\documentclass[12pt,english]{article}
\pdfoutput=1
\usepackage[T1]{fontenc}
\usepackage{graphicx,array}
\usepackage{cite}
\usepackage{color}
\usepackage{dsfont}
\usepackage{latexsym}
\usepackage{amsthm}
\usepackage{amsmath}
\usepackage{stackengine}
\usepackage[titletoc]{appendix}
\usepackage{enumitem}
\usepackage{amssymb}
\usepackage{color}
\usepackage[unicode=true,
 bookmarks=true,bookmarksnumbered=false,bookmarksopen=false,
 breaklinks=false,pdfborder={0 0 1},backref=false,colorlinks=true]
 {hyperref}
\hypersetup{pdftitle={Consistency of the Standard Model Effective Field Theory},
 pdfauthor={Grant N. Remmen, Nicholas L. Rodd},
 citecolor=black,linkcolor=black,urlcolor=black}
\usepackage{breakurl}
\usepackage[hang,flushmargin]{footmisc}

\newcolumntype{L}[1]{>{\raggedright\let\newline\\\arraybackslash\hspace{0pt}}m{#1}}
\newcolumntype{C}[1]{>{\centering\let\newline\\\arraybackslash\hspace{0pt}}m{#1}}

\def\simgt{\mathrel{\lower2.5pt\vbox{\lineskip=0pt\baselineskip=0pt
           \hbox{$>$}\hbox{$\sim$}}}}
\def\simlt{\mathrel{\lower2.5pt\vbox{\lineskip=0pt\baselineskip=0pt
           \hbox{$<$}\hbox{$\sim$}}}}

\newcommand{\be}{\begin{equation}}
\newcommand{\ee}{\end{equation}}
\newcommand{\bea}{\begin{eqnarray}}
\newcommand{\eea}{\end{eqnarray}}
\newcommand{\Eq}[1]{Eq.~(\ref{#1})}
\newcommand{\Eqs}[2]{Eqs.~(\ref{#1}) and (\ref{#2})}
\newcommand{\Sec}[1]{Sec.~\ref{#1}}

\newcommand{\Fig}[1]{Fig.~\ref{#1}}
\newcommand{\Tab}[1]{Table~\ref{#1}}
\newcommand{\App}[1]{App.~\ref{#1}}

\newcommand{\Ref}[1]{Ref.~\cite{#1}}
\newcommand{\Tr}{\textrm{Tr\,}}

\newcommand{\ket}[1]{\left| #1 \right\rangle}

\newcommand*\oline[1]{%
  \vbox{%
    \hrule height 0.5pt
    \kern0.68ex
    \hbox{%
      \kern-0.1em
      \ifmmode#1\else\ensuremath{#1}\fi
      \kern-0.1em
    }
  }
}

\definecolor{nicered}{rgb}{0.7,0.1,0.1}
\definecolor{nicegreen}{rgb}{0.1,0.5,0.1}
\hypersetup{
 colorlinks,citecolor=black,linkcolor=black,urlcolor=black}
\usepackage{xcolor}

\begin{document}

\interfootnotelinepenalty=10000
\baselineskip=18pt
\hfill

\vspace{2cm}
\thispagestyle{empty}
\begin{center}
{\LARGE \bf
Consistency of the Standard Model \\\vspace{1.5mm} Effective Field Theory
}\\
\bigskip\vspace{1cm}{
{\large Grant N. Remmen and Nicholas L. Rodd}
} \\[7mm]
{\it Center for Theoretical Physics and Department of Physics \\
     University of California, Berkeley, CA 94720, USA and \\
     Lawrence Berkeley National Laboratory, Berkeley, CA 94720, USA}
\let\thefootnote\relax\footnote{e-mail: 
\url{grant.remmen@berkeley.edu}, \url{nrodd@berkeley.edu}}
 \end{center}

\bigskip
\centerline{\large\bf Abstract}
\begin{quote} \small
We derive bounds on couplings in the standard model effective field theory (SMEFT) as a consequence of causality and the analytic structure of scattering amplitudes.
In the SMEFT, there are 64 independent operators at mass dimension eight that are quartic in bosons (either Higgs or gauge fields) and that contain four derivatives and/or field strengths, including both CP-conserving and CP-violating operators.
Using analytic dispersion relation arguments for two-to-two bosonic scattering amplitudes, we derive 27 independent bounds on the sign or magnitude of the couplings. We show that these bounds also follow as a consequence of causality of signal propagation in nonvacuum SM backgrounds.
These bounds come in two qualitative forms: i) positivity of (various linear combinations of) couplings of CP-even operators and ii) upper bounds on the magnitude of CP-odd operators in terms of (products of) CP-even couplings.
We exhibit various classes of example completions, which all satisfy our EFT bounds.
These bounds have consequences for current and future particle physics experiments, as part of the observable parameter space is inconsistent with causality and analyticity.
To demonstrate the impact of our bounds, we consider applications both to SMEFT constraints derived at colliders and to limits on the neutron electric dipole moment, highlighting the connection between such searches suggested by infrared consistency.
\end{quote}
	
\setcounter{footnote}{0}

\newpage
\tableofcontents
\newpage

\section{Introduction}
\label{sec:intro}

We are currently entering a unique era in particle physics.
With the discovery of the Higgs boson~\cite{Aad:2012tfa,Chatrchyan:2012xdj}, the Large Hadron Collider (LHC) has completed the detection of all of the fundamental constituents of the standard model (SM).
The absence as of yet of detections of supersymmetric (SUSY) partner states~\cite{Tanabashi:2018oca}---and the possible dawn of a post-naturalness era~\cite{Giudice:2017pzm,Feng:2013pwa,Cheung:2014vva}---means that the road ahead for model building is both wide open and unclear.
Without requiring a large number of SUSY states at the weak scale, possible beyond-SM (BSM) theories run the gamut from a rich structure of fields around the corner, near a TeV, to a desert up to a very high scale (see, e.g., Refs.~\cite{Dimopoulos:1990gf,Percacci:2007sz} for illustrative examples of these alternatives).
At the same time, a few clear signals of BSM physics are presently indicated, namely, dark matter, neutrino masses, and baryogenesis.

Faced with this situation, it is conceivable that the first tangible discoveries of BSM physics will come not in the form of new on-shell particle states, but instead through the detection of deviations from the SM through precision experiments, including the high luminosity/precision era at the LHC~\cite{Atlas:2019qfx,CidVidal:2018eel,Gao:2017yyd,Erler:2019hds}, Higgs factories in the form of future $e^+ e^-$ colliders (see \Ref{Reece:2016sdb} and refs. therein), measurements of the neutron and electron electric dipole moments (EDMs)~\cite{Baker:2006ts,Afach:2015sja,Graner:2016ses,Andreev:2018ayy,Baron:2013eja,Cairncross:2017fip}, and the muon $g-2$~\cite{Bennett:2006fi,Grange:2015fou}, among many others.
Such deviations from the SM, induced by new high-scale physics, can be encoded as operators of higher mass dimension in an effective field theory (EFT) of the SM~\cite{WarsawBasis,Henning:2015alf,Fonseca,Brivio:2017vri,Contino:2016jqw}.
Using current measurements from the LHC~\cite{Aaboud:2017bwk,Aaboud:2018ddq,Aaboud:2019nmv,Sirunyan:2017fvv,Sirunyan:2017ret,Sirunyan:2019ksz}, bounds have begun to be placed on the Wilson coefficients of some of the operators in the SMEFT; for a single example see \Ref{Ellis:2018gqa}.

In constructing an EFT, the Wilsonian approach is to write down all of the operators allowed, constrained only by Lorentz and gauge invariance (along with any other exact symmetries of the EFT), with arbitrary coefficients.
However, not all infrared (IR) Lagrangians can be generated by healthy ultraviolet (UV) completions.
As shown in \Ref{Adams:2006sv},\footnote{See also Refs.~\cite{Pham:1985cr,Ananthanarayan:1994hf,Pennington:1994kc} for earlier applications of such arguments to chiral perturbation theory.} causality of signal propagation, as well as quantum mechanical unitarity and the analytic properties of scattering amplitudes, constrains the Wilson coefficients in the EFT; see also Refs.~\cite{Jenkins:2006ia,Dvali:2012zc}.
This requirement of IR consistency has led to a program of bounding couplings in various EFTs of interest, including corrections to general relativity~\cite{Bellazzini:2015cra,Cheung:2016wjt,Camanho:2014apa,Gruzinov:2006ie}, nonlinear electrodynamics~\cite{Adams:2006sv}, massive gravity~\cite{Cheung:2016yqr,deRham:2017xox,Camanho:2016opx,Bellazzini:2017fep} and higher-spin states~\cite{Bellazzini:2016xrt,Bonifacio:2018vzv,Bonifacio:2016wcb,deRham:2017zjm,Hinterbichler:2017qyt,deRham:2018qqo,Bellazzini:2019bzh}, certain scalar theories~\cite{Nicolis:2009qm,Elvang:2012st,deRham:2017imi,Chandrasekaran:2018qmx,Herrero-Valea:2019hde} including the proof of the four-dimensional $a$-theorem~\cite{Komargodski:2011vj}, theories of fermion compositeness~\cite{Bellazzini:2017bkb}, electroweak chiral Lagrangian~\cite{Distler:2006if,Vecchi:2007na}, chiral perturbation theory~\cite{Pham:1985cr,Ananthanarayan:1994hf,Pennington:1994kc}, the electroweak EFT~\cite{Bellazzini:2018paj,Zhang:2018shp,Bi:2019phv}, and Einstein-Maxwell theory with applications to the Weak Gravity Conjecture~\cite{Cheung:2014ega,Cheung:2018cwt,Cheung:2019cwi,Bellazzini:2019xts,Charles:2019qqt}.
The IR consistency program, taking a bottom-up approach to constraining EFTs, has evolved in tandem with the swampland program~\cite{Vafa:2005ui,Ooguri:2006in,ArkaniHamed:2006dz}, which constrains EFTs compatible with  quantum gravity via a top-down perspective, examining the space of string theoretic UV completions.

\begin{figure}[t]
\begin{center}
\includegraphics[width=6.5cm]{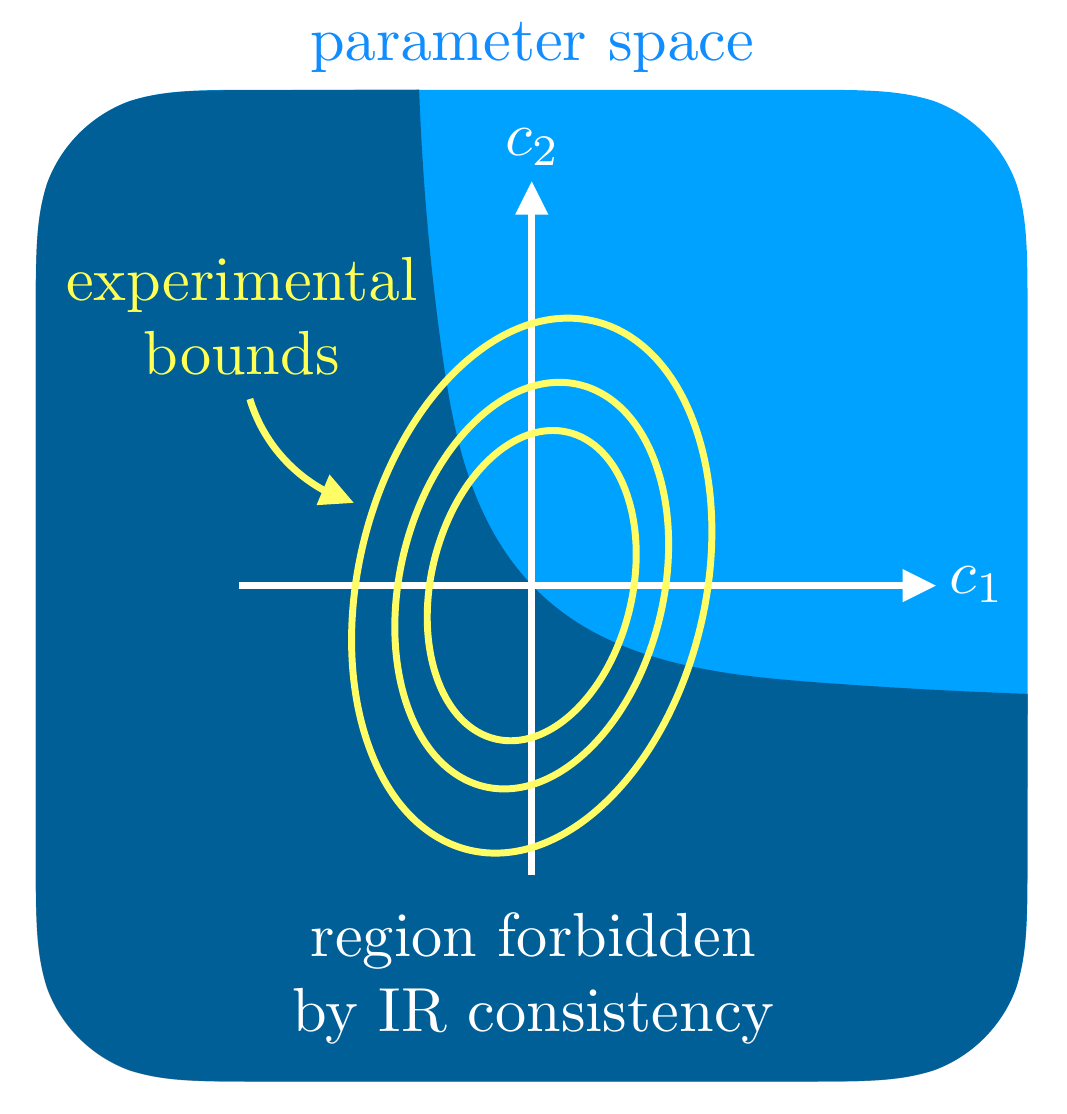}
\end{center}
\caption{Schematic depiction of bounds derived in this work.
For the example of an observable sensitive to two SMEFT operators, $\mathcal{O}_1$ and $\mathcal{O}_2$, data can weigh directly on the allowed parameter space for these operators, as shown by the yellow contours in the figure.
However, much of the parameter space is inconsistent with fundamental underlying properties of quantum field theory, and any theory with couplings in the forbidden region would depart significantly from standard assumptions about the form of UV physics (causality, unitarity, etc.).
Consequently, our bounds can be viewed as placing strong theoretical priors on the parameter space of the SMEFT.
}
\label{fig:bounds-schematic}
\end{figure}

Despite this progress, IR consistency bounds from analyticity and causality have not been systematically applied to the full SMEFT itself.
The present is an especially compelling time to do so, given that bounding and measuring SMEFT coefficients will be a focus of current and future precision particle physics experiments.
The consistency constraints can be applied to the results of such experiments in at least three ways.
First, constraining the space of possible EFT coefficients leads to enhanced statistical power for experiments sensitive to these operators, since one can incorporate IR consistency bounds into the prior probability distribution.
A schematic depiction of this interplay between the bounds derived in this work and experimental constraint is shown in \Fig{fig:bounds-schematic}.
In total we will derive 27 independent bounds; if each were as simple as requiring a certain operator coefficient to be positive, then this would represent a factor of $2^{27} \sim 10^8$ reduction in the parameter space.
This is just a rough estimate, however, because the bounds will often appear as relations between the various operators.
For example, we will see a number of examples where the size of CP-odd operators is bounded from above by certain CP-even contributions.
This implies a second possible use for our bounds: linking naively disparate experimental measurements.
We will explore an example of this in the form of connecting collider searches to measurements of the neutron EDM.
A third application of our results is that a definitive discovery of new physics, appearing in the form of nonzero SMEFT coefficients, would allow our bounds to be directly compared with data (provided that the sign of the operator coefficients can be determined).
Testing whether the resulting coefficients satisfy our bounds thereby allows for fundamental properties of the UV theory---such as causality, analyticity/locality, unitarity, and Lorentz invariance---to be tested to very high energies through experimental signatures at accessible scales.

Although we will not attempt an exhaustive exploration of the experimental consequences of our results, one example we will consider in more detail is the search for BSM contributions to electroweak boson four-point vertices, such as $W W \gamma Z$.
Such corrections are traditionally referred to as anomalous quartic gauge-boson couplings (aQGCs) and include all possible four-gauge-boson vertices in the electroweak sector of the SM (i.e., excluding gluons).
Both CMS and ATLAS are actively searching for processes sensitive to the quartic gauge-boson vertices and aQGCs; see~\Ref{Green:2016trm} for a review.
The dimension-eight SMEFT operators that induce aQGCs, first written down in~\Ref{Eboli:2006wa}, are a subset of the operators we will consider here.
As such, the bounds we derive in this work forbid couplings in certain regions of parameter space; while experimentally accessible, these regions are impossible to generate via any conventional UV completion.
We note that positivity constraints on these operators were also recently considered in Refs.~\cite{Zhang:2018shp,Bi:2019phv}, which we will discuss in \Sec{sec:aQGCs}.
In much of the experimental and phenomenological literature on aQGCs, a one-at-a-time assumption is employed, where constraints are derived by allowing only one higher-dimension operator to be nonzero; this assumption strongly limits the applicability and genericity of such bounds to arbitrary EFTs. 
We will show that even without using this simplifying assumption, we are able to derive bounds on aQGCs that are both simple and general.

In this paper, we will consider bosonic higher-dimension operators in the SMEFT. 
In order to bound operators using analyticity of scattering amplitudes or causality of particle propagation, we will specialize to gauge-invariant operators that contain terms quartic in the fields~\cite{Adams:2006sv}.
By derivative counting, we require operators with a number of derivatives (possibly inside field strengths) divisible by four in order to obtain positivity bounds from forward amplitudes or causal dispersion relations.
Hence, we will be placing positivity bounds on dimension-eight operators, which have Wilson coefficients $\sim 1/M^4$ for $M$ the cutoff scale at which new physics enters. 

A priori, one could also consider dimension-six operators, which are formally dominant, scaling as $1/M^2$ (for progress in this direction, see e.g.~Refs.~\cite{Low:2009di,Englert:2019zmt}). Indeed, generically two insertions of a dimension-six operator would compete with dimension-eight terms. In such cases, the squared contribution of the dimension-six coupling would obstruct our ability to place positivity constraints directly on dimension-eight operators in the EFT.
However, we will see in \Sec{sec:dim6} that, in the SMEFT, we can choose our scattering states to exclude the dimension-six contribution from interference with the operators of interest, leaving valid positivity bounds purely on the coefficients of dimension-eight operators.

This paper is structured as follows. 
In \Sec{sec:IR} we briefly review the IR arguments resulting in positivity bounds on EFT coefficients from analyticity and causality.
We give the basis of all dimension-eight four-boson operators in the SMEFT in \Sec{sec:basis}.
In \Sec{sec:bounds} we derive our positivity bounds on bosonic operator coefficients in the SMEFT.
As a check of our bounds, we consider broad classes of example UV completions of our operators in \Sec{sec:UV}.
We explore a subset of the phenomenological implications in \Sec{sec:pheno} and conclude in \Sec{sec:conclusions}.

\section{Infrared Consistency}\label{sec:IR}

In the interest of being self-contained, let us first briefly review how the analytic structure of scattering amplitudes and causality of signal propagation can be used to bound coefficients of operators in EFTs.
For details, see \Ref{Adams:2006sv}, as well as further discussion, extensions, and applications in Refs.~\cite{Jenkins:2006ia,Dvali:2012zc,Bellazzini:2015cra,Cheung:2016wjt,Camanho:2014apa,Gruzinov:2006ie,Cheung:2016yqr,deRham:2017xox,Camanho:2016opx,Bellazzini:2017fep,Bellazzini:2016xrt,Bonifacio:2018vzv,Bonifacio:2016wcb,deRham:2017zjm,Hinterbichler:2017qyt,deRham:2018qqo,Bellazzini:2019bzh,Nicolis:2009qm,Elvang:2012st,deRham:2017imi,Chandrasekaran:2018qmx,Herrero-Valea:2019hde,Komargodski:2011vj,Zhang:2018shp,Bi:2019phv,Cheung:2014ega,Cheung:2018cwt,Cheung:2019cwi,Bellazzini:2019xts,Charles:2019qqt}.
Throughout this section, in order to keep the discussion grounded, we will periodically return to an application of the bounds to a particularly simple example: the theory of a single massless scalar field $\phi$, invariant under a shift symmetry $\phi \to \phi + {\rm constant}$. 
The Lagrangian for this theory, including the leading-order dimension-eight interaction, is given by
\be
\mathcal{L} = -\frac{1}{2} (\partial \phi)^2 + \frac{c}{M^4} (\partial \phi)^4.
\label{eq:Lphi4}
\ee
As we will review, consistency requires $c>0$. Throughout this work, we use the mostly-plus metric signature convention, $g_{\mu\nu} = {\rm diag}(-1,+1,+1,+1)$. For our key results, however, we will always identify if there is any dependence upon convention.

In order to establish the analyticity bound, consider the scattering of massless particles $12\rightarrow 34$, with the incoming state $\ket{\psi_{\rm in}}$ describing particles 1 and 2. The two-to-two amplitude ${\cal M}(s,t)$ can be written as a function of the Mandelstam variables $s$ and $t$,\footnote{We use the all-incoming convention for momenta and define the Mandelstam variables as $s=-(p_1 + p_2)^2$, $t = -(p_1 + p_3)^2$, and $u=-(p_1 + p_4)^2$, where $s+t+u=0$ for massless states.} as well as (for a general theory) polarizations or other quantum numbers.
The forward amplitude, which we will write as ${\cal A}(s)$, is obtained by taking $t \rightarrow 0$, as well as ensuring that the outgoing state $\ket{\psi_{\rm  out}}$ is identical to the incoming one (e.g., in polarizations, flavor, species, etc.).
For our scalar field example, the two-to-two scattering amplitude is 
\be
\mathcal{M}(s,t) = \frac{2c}{M^4} ( s^2 + t^2 + u^2 ),
\ee
and so in this case, the forward amplitude is simply $\mathcal{A}(s) = 4 c s^2/M^4$.

Returning to the general case, let us promote the squared center-of-mass energy $s$ to a complex quantity. The analytic structure of scattering amplitudes as functions of complex momenta has been well studied~\cite{Froissart:1961ux,Martin:1962rt,Martin:1965jj,Mandelstam:1958xc,Lehmann1958}; in particular, ${\cal A}(s)$ is analytic everywhere in the plane except near the real $s$ axis, where simple poles correspond to on-shell tree-level exchange of massive intermediate states, and branch cuts along the axis are associated with logarithmic terms arising from loops, i.e., multiparticle on-shell production.
We can exploit this general structure as follows.
Let us expand ${\cal A}(s)$ in an EFT as $\sum_{n=0}^\infty \lambda_n s^{n}$ at small $s$.
Using this expansion and defining a small circular contour ${\cal C}$ around the origin, we can isolate the coefficient of $s^2$ in the forward amplitude using the following contour integral:
\be
\lambda_2 = \frac{1}{2\pi i} \oint_{\cal C}  \frac{{\rm d}s}{s^3} {\cal A}(s).
\label{eq:lambdacontour}
\ee

\begin{figure}[t]
\begin{center}
\includegraphics[width=7.5cm]{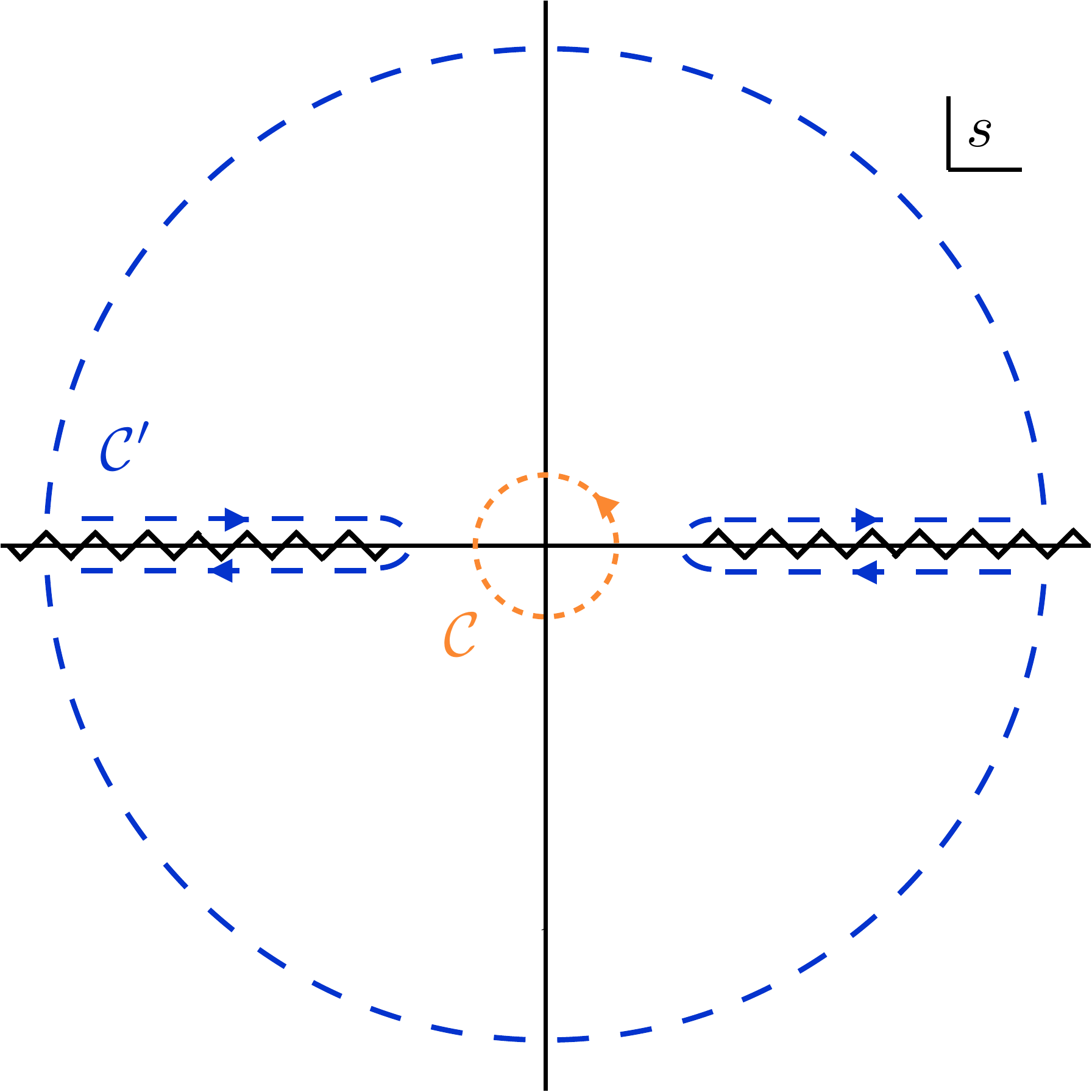}
\end{center}
\caption{A schematic depiction of the analytic structure of the amplitude in the complex $s$ plane, as well as the two contours we will use to establish our positivity bounds.
The zigzag lines denote the discontinuity in the amplitude, which can include poles and branch cuts.
} 
\label{fig:Contour}
\end{figure}

We will now establish that $\lambda_2 > 0$.\footnote{For simplicity of notation, throughout we will write all bounds on operator coefficients as strict inequalities, though this could be relaxed (i.e., $\geq$) if we are working at fixed (e.g., tree) order, cf.~Refs.~\cite{Adams:2006sv,Nicolis:2009qm,Chandrasekaran:2018qmx}.}
To do so, we exploit the analyticity of ${\cal A}(s)$ to deform ${\cal C}$ to a new contour ${\cal C}'$, which stretches out to infinity wherever it is not obstructed by the discontinuities in $\mathcal{A}(s)$, as shown in \Fig{fig:Contour}.
For simplicity, we will treat the discontinuity as beginning at $\pm s_{\rm d}$.
This choice is well motivated as it appears in simple UV completions.
For example, a tree-level completion involving the exchange of a state with mass $m$ will generate two poles on the real $s$ axis, fixing $s_{\rm d} = m^2$.
Further, if the theory is instead UV completed at one loop using a single particle of the same mass, then the associated branch cuts lead to $s_{\rm d} = 4 m^2$.
Nevertheless, in general the theory will contain poles as well as branch cuts, the exact locations of which are determined by the UV properties of the theory, and we defer to, e.g., \Ref{Adams:2006sv} for a more general discussion.\footnote{If the amplitude contains loops with massless particles, then branch cuts extend all the way to $s=0$. However, as in \Ref{Adams:2006sv} we can deform the theory in the IR by introducing a mass regulator, ensuring analyticity in a window near $s=0$, though subtleties arise in the case of gravity~\cite{Bellazzini:2015cra,Cheung:2016yqr}.}
The contribution to the integral from the boundary circle at infinity in ${\cal C}'$ will vanish, leaving \Eq{eq:lambdacontour} determined by the contribution above and below the discontinuities on the real $s$ axis.\footnote{For massive states, the vanishing of the boundary integral at infinity is a consequence of the Froissart bound~\cite{Froissart:1961ux,Martin:1962rt}. For massless states, this conclusion can instead be reached from the weak requirement that the amplitude of the UV completion scales more slowly with energy at large momentum than the EFT contribution~\cite{Chandrasekaran:2018qmx}.}
In detail, using ${\rm Disc}\, \mathcal{A}(s) = \lim_{\epsilon \to 0} [\mathcal{A}(s+i\epsilon) - \mathcal{A}(s-i\epsilon)]$, we have
\be
\frac{1}{2\pi i} \oint_{{\cal C}'}  \frac{{\rm d}s}{s^3} {\cal A}(s)
= \frac{1}{2\pi i} \left( \int_{-\infty}^{-s_{\rm d}} + \int_{s_{\rm d}}^{\infty} \right) \frac{{\rm d}s}{s^3}\,{\rm Disc} \,{\cal A}(s)
= \frac{1}{i \pi} \int_{s_{\rm d}}^{\infty} \frac{{\rm d}s}{s^3}\,{\rm Disc}\, {\cal A}(s)\label{eq:analyticity}
\ee
In the second equality in \Eq{eq:analyticity}, we sent $s \to - s$ in the first integral and subsequently invoked $1\leftrightarrow 3$ crossing symmetry---assuming real polarizations so that the forward amplitude maps onto itself under crossing---which implies $\mathcal{A}(-s) = \mathcal{A}(s)$.\footnote{While the choice of real polarizations is not strictly necessary for placing positivity bounds~\cite{Bellazzini:2016xrt}, it will be convenient for our purposes.}
Next, we can relate the discontinuity of the forward amplitude to the cross section:
\be
{\rm Disc}\, {\cal A}(s) = \lim_{\epsilon \to 0} [\mathcal{A}(s+i\epsilon) - \mathcal{A}(s-i\epsilon)]
= \lim_{\epsilon \to 0} [\mathcal{A}(s+i\epsilon) - (\mathcal{A}(s+i\epsilon))^*] = 2 i \,{\rm Im} \mathcal{A}(s),
\ee
where we used the Schwarz reflection principle, ${\cal A}(s^*) = [{\cal A}(s)]^*$.
Recalling the optical theorem ${\rm Im}\,{\cal A}(s) = s\,\sigma(s)$ (a consequence of unitarity), we find that
\be
\lambda_2 = \frac{2}{\pi} \int_{s_{\rm d}}^{\infty} \frac{{\rm d}s}{s^2}\,\sigma(s).
\ee
Finally, as cross sections are by definition positive, we conclude that $\lambda_2 > 0$.
In the case of the scalar field example, $\lambda_2 = 4 c/M^4$, and so we have demonstrated that analyticity requires $c > 0$.

Alternatively, we can arrive at $\lambda_2 > 0$ by invoking causality.
Suppose $\lambda_2$ is the coefficient of an operator ${\cal O}[\phi_1,\phi_2]$, where ${\cal O}$ is quadratic in both $\phi_1$ and $\phi_2$ (here, $\phi_1$ and $\phi_2$ may or may not be distinct fields).
Since $\lambda_2$ is defined to be the coefficient of $s^2$, there must be exactly four derivatives in ${\cal O}$.
We wish to determine the dynamics of a perturbation in $\phi_1$ in a background of nonzero $\overline{\partial \phi_2}$ condensate.
The two remaining derivatives in ${\cal O}$ will contribute to the dispersion relation quadratic in the momentum of $\phi_1$.
Thus, if $\phi_1$ is massless, the speed $v$ of the $\phi_1$ perturbation is given schematically by
\be 
v = 1 - \lambda_2 \overline{\partial \phi_2 \partial \phi_2},
\ee
where as shown in Refs.~\cite{Adams:2006sv,Cheung:2014ega,Chandrasekaran:2018qmx} the signs in the speed calculation work out such that when $\lambda_2 < 0$, $v>1$.
In that case, by taking two nonoverlapping bubbles of condensate, giving them a large relative boost (with relative speed $v_{\rm rel}$ satisfying $1-v_{\rm rel}= {\cal O}(\epsilon^2)$ when the signal speed is $v=1+\epsilon$ in the bubble \cite{Cheung:2014ega}), and sending signals back and forth between the two bubbles~\cite{Adams:2006sv}, one can form a bona fide causal paradox, with the return signal arriving at the sender before the outgoing signal was sent~\cite{Benford:1970xv,Tolman}; see Fig. 2 of \Ref{Adams:2006sv}.

\begin{figure}[t]
\begin{center}
\includegraphics[width=4cm]{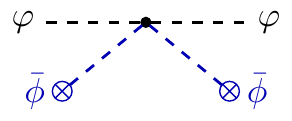}
\end{center}
\caption{A representative interaction between a small perturbation $\varphi$ and the background scalar condensate $\bar{\phi}$ in which it is propagating.
As shown in the text, in a theory where this interaction is mediated by a term $c (\partial \phi)^4$, unless $c > 0$, these interactions will lead to the superluminal propagation of $\varphi$.}
\label{fig:Dispersion}
\end{figure}

Let us demonstrate how this argument is realized in the simple scalar field example.
Here, we have ${\cal O}[\phi_1,\phi_2] = (\partial \phi)^4$, so that $\phi_1 = \phi_2 = \phi$, and the operator is quartic in $\phi$ and $\partial$ as required.
The equation of motion associated with \Eq{eq:Lphi4} is given by
\be
\Box \phi - \frac{4c}{M^4} \left[ \Box \phi (\partial \phi)^2 + 2 (\partial^{\mu} \phi) (\partial^{\nu} \phi) (\partial_{\mu} \partial_{\nu} \phi) \right] = 0.
\ee
In order to derive the bound, we must consider the dynamics of small perturbations in the scalar field around a nonzero background field.
To do so, we set $\phi = \varphi + \overline{\phi}$, with $\varphi$ the perturbation we wish to study and $\bar{\phi}$ a background condensate chosen such that $\overline{\partial_{\mu} \phi} = q_{\mu}$, with $q_{\mu}$ a constant four-vector.
The aim is to study the impact of interactions such as those in \Fig{fig:Dispersion} on the propagation of $\varphi$.
Working in this background, the linearized equation of motion for $\varphi$ becomes
\be
\left(-1+\frac{4cq^2}{M^4}\right)\Box \varphi + \frac{8 c}{M^4}q^{\mu} q^{\nu} \partial_{\mu} \partial_{\nu} \varphi = 0.
\ee
If we expand $\varphi$ in plane waves, as $\varphi \propto e^{ik \cdot x}$, this becomes a dispersion relation
\be
\left(-1+\frac{4c q^2}{M^4}\right)k^2 + \frac{8 c}{M^4} (q \cdot k)^2 = 0.\label{eq:dispex}
\ee
Writing $k_\mu = (k_0, \mathbf{k})$, the speed of propagation of $\varphi$ is $v=k_0/|{\bf k}|$, which we can compute from \Eq{eq:dispex} to be 
\be
v = \sqrt{1 - \frac{8c(q \cdot k)^2}{|{\bf k}|^2(M^4-4c q^2)}} \simeq 1 - \frac{4c(q\cdot k)^2}{M^4 k_0^2},
\ee
where in the second equality we expand to first order in the Wilson coefficient $c$.
We conclude that unless $c > 0$, $\varphi$ will experience superluminal propagation, allowing causal paradoxes to be constructed as described above.

\section{Operator Basis}\label{sec:basis}

Before proving our positivity bounds, let us first construct our basis of operators.
As discussed in \Sec{sec:intro}, we wish to consider all operators in the SMEFT that are quartic in bosonic fields and that contain four derivatives.
While the full set of operators at dimension eight in the SMEFT is vast~\cite{Henning:2015alf,Fonseca}, containing 993 terms in a minimal basis,\footnote{The figure of 993 results~\cite{Henning:2015alf} from counting operators and hermitian conjugates separately, but not counting operators corresponding to different fermion generations. In this counting, there are 84 operators at dimension six, cf. \Ref{WarsawBasis}.} the scattering processes we consider will only activate a fraction of these operators, allowing us to tractably place positivity bounds.

\subsection{Four-derivative dimension-eight bosonic operators}

We will write the gauge field strengths for ${\rm U}(1)_Y$ weak hypercharge, ${\rm SU}(2)_L$ weak isospin, and ${\rm SU}(3)_C$ color as $B_{\mu\nu}$, $W^I_{\mu\nu}$, and $G^a_{\mu\nu}$, throughout using Greek letters for Lorentz indices, $I,J,K$ for weak indices, and $a,b,c,d$ for color indices.\footnote{Written in terms of the gauge fields $B_\mu$, $W^I_\mu$, and $G^a_\mu$, the field strengths are $B_{\mu\nu} = \partial_\mu B_\nu - \partial_\nu B_\mu$, $W^I_{\mu\nu} = \partial_\mu W^I_\nu - \partial_\nu W^I_\mu + g_2 \epsilon^{IJK} W^J_\mu W^K_\nu$, and $G^a_{\mu\nu} = \partial_\mu G^a_\nu - \partial_\nu G^a_\mu + g_3 f^{abc} G^b_\mu G^c_\nu$. We write the gauge couplings of ${\rm U}(1)_Y$, ${\rm SU}(2)_L$, and ${\rm SU}(3)_C$ as $g_1$, $g_2$, and $g_3$, respectively.} We define the dual field strengths via contraction with the Levi-Civita tensor $\epsilon^{\mu\nu\rho\sigma}$: $\widetilde B^{\mu\nu} = \epsilon^{\mu\nu\rho\sigma} B_{\rho\sigma}/2$, $\widetilde W^{I\mu\nu} = \epsilon^{\mu\nu\rho\sigma} W^I_{\rho\sigma}/2$, and $\widetilde G^{a\mu\nu} = \epsilon^{\mu\nu\rho\sigma} G^a_{\rho\sigma}/2$. 

A priori, we could construct arbitrary operators that are gauge and Lorentz singlets, contracting Lorentz indices with combinations of $g^{\mu\nu}$ and $\epsilon^{\mu\nu\rho\sigma}$ and gauge indices with $\delta^{IJ}$ and $\epsilon^{IJK}$ for isospin and combinations of $\delta^{ab}$, $f^{abc}$, and $d^{abc}$ for color, where we define the structure constants via
\be 
\begin{aligned}
{} [T^a, T^b] {} &= i f^{abc}T^c \\
\{T^a,T^b\} &= \frac{1}{3}\delta^{ab}\mathds{1} + d^{abc}T^c,
\end{aligned}
\ee
for totally antisymmetric $f^{abc}$, totally symmetric $d^{abc}$, and defining the generators of ${\rm SU}(3)$ as $T^a = \lambda^a/2$ in terms of the Gell-Mann matrices $\lambda^a$.
However, many of these operators will not be linearly independent, by virtue of identities among Lorentz and color indices. The Levi-Civita identity,\footnote{We denote normalized symmetrization via $\lambda^{(a_1 \cdots a_n)}=\frac{1}{n!}(\lambda^{a_1 \cdots a_n} + \text{permutations})$ and antisymmetrization similarly with $[a_1\cdots a_n]$.}
\be 
\epsilon^{\mu\nu\rho\sigma}\epsilon_{\alpha\beta\gamma\delta} = -24\, \delta^{[\mu}_{\hphantom{[}\alpha}\, \delta^\nu_\beta\, \delta^\rho_\gamma \, \delta ^{\sigma]}_\delta,
\ee
implies, for example, 
\be
B_{\mu\nu}B^{\nu\rho}B_{\rho\sigma}B^{\sigma\mu} = \frac{1}{2} (B_{\mu\nu}B^{\mu\nu})^2 + \frac{1}{4} (B_{\mu\nu}\widetilde B^{\mu\nu})^2.\label{eq:cyclic}
\ee
The Schouten identity,
\be 
\epsilon^{[\alpha\beta\gamma\delta}g^{\mu]\nu} = 0,
\ee
can also be used to reduce the list of operators, e.g.,
\be
4\epsilon^{\mu\nu\rho\sigma} G^a_{\mu\alpha}G^{b\alpha}_\nu G^c_{\rho\beta}G^{d\beta}_\sigma =  (G^a_{\mu\nu}G^{c\mu\nu}G^b_{\rho\sigma}\widetilde G^{d\rho\sigma} + G^b_{\mu\nu}G^{d\mu\nu}G^a_{\rho\sigma}\widetilde G^{c\rho\sigma}) - (c\leftrightarrow d).
\ee

To reduce the color structures, we exploit a particularly useful ${\rm SU}(N)$ identity,
\be
f^{abe} f^{cde} = \frac{2}{N} (\delta^{ac} \delta^{bd} - \delta^{ad}\delta^{bc}) + d^{ace} d^{bde} - d^{bce} d^{ade},\label{eq:useful}
\ee
and the Jacobi identity,
\be 
f^{abe} f^{ecd} + f^{cbe} f^{aed} + f^{dbe} f^{ace} =0,
\ee
along with the additional ${\rm SU}(N)$ identities
\be
\begin{aligned}
f^{abc}f^{abd} &= N\delta^{cd} \\ 
d^{abc}d^{abd} &= \frac{N^2 - 4}{N}\delta^{cd}\\
f^{abe}d^{cde} + f^{ace}d^{bde} + f^{ade} d^{bce}& = 0\\
f^{abe} d^{cde} + f^{cbe}d^{ade} + f^{dbe} d^{ace} &= 0.
\end{aligned}
\ee
Moreover, for ${\rm SU}(3)$, there are two additional identities:
\be
\begin{aligned}
3d^{abe} d^{cde} -f^{ace} f^{bde} -f^{ade}f^{bce} &=  \delta^{ac}\delta^{bd} + \delta^{ad}\delta^{bc} - \delta^{ab}\delta^{cd} \\ 
3(d^{abe} d^{cde} + d^{ace}d^{bde} + d^{ade} d^{bce}) &= \delta^{ab}\delta^{cd} + \delta^{ac}\delta^{bd} + \delta^{ad}\delta^{bc}.
\end{aligned}\label{eq:SU3idens} 
\ee

Using these relations, one can construct a minimal basis of operators quartic in the field strength for Yang-Mills theory, as originally derived in \Ref{Morozov}. In particular, one can show that, among the $G^4$-type operators quartic in the gluon field strength of ${\rm SU}(3)_C$, the minimal basis consists of  6 CP-conserving and 3 CP-violating operators. Analogously, for the $W^4$ operators of ${\rm SU}(2)_L$, since $d^{abc}$ vanishes for this group, the minimal basis comprises 4 CP-conserving and 2 CP-violating operators, while for the $B^4$ operators of ${\rm U}(1)$, there are 2 independent CP-conserving operators and 1 CP-violating operator. The minimal basis is given in \Tab{tab:operators}.

We can similarly construct a minimal basis of cross-quartic operators. For operators of the $B^2 W^2$ form, there are 4 CP-conserving and 3 CP-violating independent operators, and similarly for operators of the $B^2 G^2$ and $W^2 G^2$ types. Moreover, there are 2 independent CP-conserving and 2 CP-violating operators of the $BG^3$ form (but there are no nonvanishing operators of the $BW^3$ form). All of these operators are also given in \Tab{tab:operators}.

Finally, we define the complex Higgs field $H$ as a fundamental ($\boldsymbol 2$) of ${\rm SU}(2)_L$, which we can write in terms of four real scalar fields $\phi_{1,2,3,4}$ as 
\be
H_i = \frac{1}{\sqrt{2}} \begin{pmatrix}
\phi_{1}+i\phi_{2}\\
\phi_{3}+i\phi_{4}
\end{pmatrix}, \label{eq:Hdef}
\ee
with conjugate
 \be 
H_{i}^{\dagger}=\frac{1}{\sqrt{2}} \begin{pmatrix}
-\phi_{3}+i\phi_{4}\\
\phi_{1}-i\phi_{2}
\end{pmatrix}.\label{eq:Hdaggerdef}
\ee
We contract fundamental ${\rm SU}(2)$ indices in two different ways. First, contractions of the form $H^{\dagger} H$ are performed using $\epsilon_{ij}$.\footnote{We follow the convention of \Ref{Hays:2018zze}, treating $H^\dagger_i = \epsilon_{ij}H^{\dagger j}$ as a $\boldsymbol{2}$ of ${\rm SU}(2)$ rather than $\boldsymbol{\bar 2}$. Here $i$ indexes the fundamental representation of ${\rm SU}(2)$ and runs over $i\in \{1,2\}$. We define $\epsilon^{12}=\epsilon_{21}=+1$, so that indices are contracted as $H_{i}=\epsilon_{ij}H^{j}$, $H^{i}=\epsilon^{ij}H_{j}$, and $H^{\dagger}H=\epsilon^{ij}H_{j}^{\dagger}H_{i}=+(\phi_{1}^{2}+\phi_{2}^{2}+\phi_{3}^{2}+\phi_{4}^{2})/2$.\label{foot:SU2}}  On occasion, we will also express the triplet resulting from the product of two doublets as, for example, $H^{\dagger} \tau^K H = H^{\dagger i} (\tau^K)_i^{\hphantom{i} j} H_j$, where $\tau^I = \sigma^I/2$, with $\sigma^I$ the usual Pauli matrices. 
Further, we define the covariant derivative $D_\mu H = (\partial_\mu + \frac{1}{2}ig_1 B_\mu + ig_2 \tau^I W^I_\mu)H$.
Then, as shown by \Ref{Hays:2018zze}, there are three linearly-independent operators of the form $(DH)^4$, as well as 10 CP-conserving and 8 CP-violating four-derivative bosonic cross-quartics containing the Higgs; see \Tab{tab:operators2}.\footnote{As confirmed by \Ref{AdamMartin}, we note that \Ref{Hays:2018zze} contains a typo in the equivalent expressions for our $\mathcal{O}_3^{H^2 W^2}$, $\widetilde{\mathcal{O}}_3^{H^2 W^2}$, and $\widetilde{\mathcal{O}}_2^{H^2 B W}$. In \Ref{Hays:2018zze}, the equivalent expressions are missing a factor of $i$ required to make them self-hermitian. (In detail, the relevant terms in their notation are $\mathcal{O}_{8,DHW3}$, $\mathcal{O}_{8,DHW3b}$, and $\mathcal{O}_{8,DH\tilde{W}B2}$ in their Table 5.)}

In total, the basis in Tables~\ref{tab:operators} and \ref{tab:operators2} contains 64 terms, of which 39 conserve and 25 violate CP.
All possible quartic, four-derivative, Lorentz and gauge invariant dimension-eight operators in the SMEFT can be written as a linear combination of operators in this basis.
We will write the SMEFT action relevant for our bounds as 
\be
\begin{aligned}
{\cal L} =&  -\frac{1}{4}B_{\mu\nu}B^{\mu\nu} - \frac{1}{4} W^I_{\mu\nu}W^{I\mu\nu} - \frac{1}{4} G^a_{\mu\nu}G^{a\mu\nu} -(D^{\mu}H)^{\dagger}D_{\mu}H -\mu^{2}H^{\dagger}H+\lambda(H^{\dagger}H)^{2}
\\&  + \frac{1}{M^4} \sum_i c_i {\cal O}_i,
\end{aligned}\label{eq:LSMEFT}
\ee
where $M$ is the scale of new physics, the sum over the ${\cal O}_i$ contains all the operators in Tables~\ref{tab:operators} and \ref{tab:operators2} (e.g., ${\cal O}_1^{G^4}$,  ${\cal O}_2^{H^2 G^2}$, $\widetilde {\cal O}_3^{B^2 W^2}$, etc.), and the $c_i$ are unitless couplings decorated analogously with their corresponding operators (e.g., $c_1^{G^4}$,  $c_2^{H^2 G^2}$, $\widetilde c_3^{B^2 W^2}$, etc.).
In all tables, each operator in the basis is self-hermitian, so the $c_i$ are real.
It is on this set of couplings $c_i$ that we will derive bounds from analyticity and causality, which will be the main result of this work.

\renewcommand{\arraystretch}{1.2}
\begin{table}[htbp]
\begin{center}
\begin{tabular}{C{2.1cm}  L{4.5cm} | C{2.1cm} L{4.5cm}}
\multicolumn{2}{c|}{$B^4$ operators} & \multicolumn{2}{c}{\hspace{-0.8cm}$F_1^2 F_2^2$/$F_1 F_2^3$ cross-quartics} \\ 
&\phantom.{}\\
${\cal O}_{1}^{B^{4}}$ & $(BB)(BB)$ & ${\cal O}_{1}^{B^{2}W^{2}}$ 
& $(BB)(W^I W^I)$ 
\\${\cal O}_{2}^{B^{4}}$ & $(B\widetilde{B})(B\widetilde{B})$ & ${\cal O}_{2}^{B^{2}W^{2}}$ & $(B\widetilde{B})(W^I\widetilde{W}^I)$
\\
$\widetilde{{\cal O}}_{1}^{B^{4}}$ & $(BB)(B\widetilde{B})$ & ${\cal O}_{3}^{B^{2}W^{2}}$ & $(B W^I)(B W^I)$
\\  & \phantom{.} & ${\cal O}_{4}^{B^{2}W^{2}}$& $(B \widetilde{W}^I)(B\widetilde{W}^I)$
\\ \cline{1-2} &\phantom{.} & $\widetilde{{\cal O}}_{1}^{B^{2}W^{2}}$ & $(B \widetilde{B})(W^I W^I)$
\\ 
\multicolumn{2}{c|}{$W^4$ operators} & $\widetilde{{\cal O}}_{2}^{B^{2}W^{2}}$ & $(B B)(W^I \widetilde{W}^I)$
\\ &\phantom{.} &$\widetilde{{\cal O}}_{3}^{B^{2}W^{2}}$ & $(B W^I)(B \widetilde{W}^I)$\\
${\cal O}_{1}^{W^{4}}$& $(W^I W^I)(W^J W^J)$ \\
${\cal O}_{2}^{W^{4}}$ & $(W^I \widetilde{W}^I)(W^J \widetilde{W}^J)$ & ${\cal O}_{1}^{B^{2}G^{2}}$& $(B B)(G^a G^a)$\\
${\cal O}_{3}^{W^{4}}$& $(W^I W^J)(W^IW^J)$ & ${\cal O}_{2}^{B^{2}G^{2}}$ & $(B \widetilde{B}) (G^a \widetilde{G}^a)$\\
${\cal O}_{4}^{W^{4}}$& $(W^I \widetilde{W}^J)(W^I \widetilde{W}^J)
$ & ${\cal O}_{3}^{B^{2}G^{2}}$& $(B G^a)(B G^a)$\\  
$\widetilde{O}_{1}^{W^{4}}$& $(W^I W^I)(W^J \widetilde{W}^J)$ & ${\cal O}_{4}^{B^{2}G^{2}}$ &$(B \widetilde{G}^a)(B \widetilde{G}^a)$ \\
$\widetilde{{\cal O}}_{2}^{W^{4}}$& $(W^I W^J)(W^I \widetilde{W}^J)$
 & $\widetilde{{\cal O}}_{1}^{B^{2}G^{2}}$& $(B \widetilde{B}) (G^a G^a)$
 \\& \phantom{.} &$\widetilde{{\cal O}}_{2}^{B^{2}G^{2}}$& $(B B)(G^a \widetilde{G}^a)$
 \\ \cline{1-2} &\phantom{.}  &$\widetilde{{\cal O}}_{3}^{B^{2}G^{2}}$ &$(B G^a)(B \widetilde{G}^a)$ \\
\multicolumn{2}{c|}{$G^4$ operators}\\ &\phantom{.} & ${\cal O}_{1}^{W^{2}G^{2}}$ & $(W^I W^I)(G^a G^a)$\\
${\cal O}_{1}^{G^{4}}$ & $(G^a G^a)(G^b G^b)$ & ${\cal O}_{2}^{W^{2}G^{2}}$& $(W^I \widetilde{W}^I)(G^a \widetilde{G}^a)$\\
${\cal O}_{2}^{G^{4}}$ & $(G^a \widetilde{G}^a)(G^b \widetilde{G}^b)$ & ${\cal O}_{3}^{W^{2}G^{2}}$ &$(W^I G^a)(W^I G^a)$\\
${\cal O}_{3}^{G^{4}}$ & $(G^a G^b)(G^a G^b)$ & ${\cal O}_{4}^{W^{2}G^{2}}$& $(W^I \widetilde{G}^a)(W^I \widetilde{G}^a)$\\
${\cal O}_{4}^{G^{4}}$ & $(G^a \widetilde{G}^b)(G^a \widetilde{G}^b)$ & $\widetilde{{\cal O}}_{1}^{W^{2}G^{2}}$& $(W^I \widetilde{W}^I) (G^a G^a)$\\
${\cal O}_{5}^{G^{4}}$ & $d^{abe}d^{cde} (G^a G^b)(G^c G^d)$ & $\widetilde{{\cal O}}_{2}^{W^{2}G^{2}}$ & $(W^I W^I)(G^a \widetilde{G}^a)$\\
${\cal O}_{6}^{G^{4}}$ & $d^{abe}d^{cde} (G^a \widetilde{G}^b)(G^c \widetilde{G}^d)$ & $\widetilde{{\cal O}}_{3}^{W^{2}G^{2}}$& $(W^I G^a)(W^I \widetilde{G}^a)$\\
$\widetilde{O}_{1}^{G^{4}}$ & $(G^a G^a)(G^b \widetilde{G}^b)$\\
$\widetilde{{\cal O}}_{2}^{G^{4}}$ & $(G^a G^b)(G^a \widetilde{G}^b)$ & ${\cal O}_{1}^{BG^{3}}$ & $d^{abc} (B G^a)(G^b G^c)$\\
$\widetilde{{\cal O}}_{3}^{G^{4}}$ & $d^{abe}d^{cde} (G^a G^b)(G^c \widetilde{G}^d)$ & ${\cal O}_{2}^{BG^{3}}$& $d^{abc} (B \widetilde{G}^a)(G^b \widetilde{G}^c)$ \\  & \phantom{.} & $\widetilde{{\cal O}}_{1}^{BG^{3}}$& $d^{abc} (B \widetilde{G}^a) (G^b G^c)$
 \\  &\phantom{.}  & $\widetilde{{\cal O}}_{2}^{BG^{3}}$ & $d^{abc} (B G^a)(G^b \widetilde{G}^c)$
\end{tabular}
\caption{Basis of independent dimension-eight operators in the SMEFT that are quartic in gauge field strengths. CP-violating operators are denoted with a tilde. Throughout the table, we have for clarity suppressed Lorentz indices, writing $(AB)$ for $A_{\mu \nu} B^{\mu \nu}$.}
\label{tab:operators}
\end{center}
\end{table}

\renewcommand{\arraystretch}{1.2}
\begin{table}[htbp]
\begin{center}
\begin{tabular}{C{2.1cm}  L{7.7cm}} 
\multicolumn{2}{c}{$(DH)^4$ operators} \\[-0.3cm] &\phantom.{}\\
${\cal O}_{1}^{H^{4}}$ & $(D_{\mu}H^{\dagger}D_{\nu}H)(D^{\nu}H^{\dagger}D^{\mu}H)$ 
\\${\cal O}_{2}^{H^{4}}$& $(D_{\mu}H^{\dagger}D_{\nu}H)(D^{\mu}H^{\dagger}D^{\nu}H)$ 
\\${\cal O}_{3}^{H^{4}}$& $(D^{\mu}H^{\dagger}D_{\mu}H)(D^{\nu}H^{\dagger}D_{\nu}H)$ 
\\  & \phantom{.} 
\\ \cline{1-2} &\phantom{.}
\\ 
\multicolumn{2}{c}{$(DH)^2 F^2$ cross-quartics}
\\[-0.3cm] &\phantom{.} 
\\ ${\cal O}_{1}^{H^{2}B^{2}}$ & $(D^{\mu}H^{\dagger}D^{\nu}H)B_{\mu\rho}B_{\nu}^{\;\;\rho}$
\\ ${\cal O}_{2}^{H^{2}B^{2}}$ & $(D^{\mu}H^{\dagger}D_{\mu}H)B_{\rho\sigma}B^{\rho\sigma}$
\\$\widetilde{{\cal O}}_{1}^{H^{2}B^{2}}$ & $(D^{\mu}H^{\dagger}D_{\mu}H)B_{\rho\sigma}\widetilde{B}^{\rho\sigma}$
\\\\${\cal O}_{1}^{H^{2}W^{2}}$ & $(D^{\mu}H^{\dagger}D^{\nu}H)W_{\mu\rho}^{I}W_{\nu}^{I\rho}$
\\${\cal O}_{2}^{H^{2}W^{2}}$ & $(D^{\mu}H^{\dagger}D_{\mu}H)W_{\rho\sigma}^{I}W^{I\rho\sigma}$
\\${\cal O}_{3}^{H^{2}W^{2}}$ & $i\,\epsilon^{IJK}(D^{\mu}H^{\dagger}\tau^{I}D^{\nu}H)W_{\mu\rho}^{J}W_{\nu}^{K\rho}$
\\$\widetilde{{\cal O}}_{1}^{H^{2}W^{2}}$ & $(D^{\mu}H^{\dagger}D_{\mu}H)W_{\rho\sigma}^{I}\widetilde{W}^{I\rho\sigma}$
\\$\widetilde{{\cal O}}_{2}^{H^{2}W^{2}}$ & $\epsilon^{IJK}(D^{\mu}H^{\dagger}\tau^{I}D^{\nu}H)(W_{\mu\rho}^{J}\widetilde{W}_{\nu}^{K\rho}-\widetilde{W}_{\mu\rho}^{J}W_{\nu}^{K\rho})$
\\$\widetilde{{\cal O}}_{3}^{H^{2}W^{2}}$ & $i\,\epsilon^{IJK}(D^{\mu}H^{\dagger}\tau^{I}D^{\nu}H)(W_{\mu\rho}^{J}\widetilde{W}_{\nu}^{K\rho}+\widetilde{W}_{\mu\rho}^{J}W_{\nu}^{K\rho})$
\\\\${\cal O}_{1}^{H^{2}G^{2}}$ & $(D^{\mu}H^{\dagger}D^{\nu}H)G_{\mu\rho}^{a}G_{\nu}^{a\rho}$
\\${\cal O}_{2}^{H^{2}G^{2}}$ & $(D^{\mu}H^{\dagger}D_{\mu}H)G_{\rho\sigma}^{a}G^{a\rho\sigma}$
\\$\widetilde{{\cal O}}_{1}^{H^{2}G^{2}}$ & $(D^{\mu}H^{\dagger}D_{\mu}H)G_{\rho\sigma}^{a}\widetilde{G}^{a\rho\sigma}$
\\  & \phantom{.} 
\\ \cline{1-2} &\phantom{.}
\\ 
\multicolumn{2}{c}{$(DH)^2 F_1 F_2$ cross-quartics}
\\[-0.3cm] &\phantom{.} 
\\${\cal O}_{1}^{H^{2}BW}$ &$(D^{\mu}H^{\dagger}\tau^{I}D_{\mu}H)B_{\rho\sigma}W^{I\rho\sigma}$ \\
${\cal O}_{2}^{H^{2}BW}$ &$i(D^{\mu}H^{\dagger}\tau^{I}D^{\nu}H)(B^{\vphantom{I}}_{\mu\rho}W_{\nu}^{I\rho}-B^{\vphantom{I}}_{\nu\rho}W_{\mu}^{I\rho})$\\
${\cal O}_{3}^{H^{2}BW}$ & $(D^{\mu}H^{\dagger}\tau^{I}D^{\nu}H)(B^{\vphantom{I}}_{\mu\rho}W_{\nu}^{I\rho}+B^{\vphantom{I}}_{\nu\rho}W_{\mu}^{I\rho})$ \\
$\widetilde{{\cal O}}_{1}^{H^{2}BW}$ & $(D^{\mu}H^{\dagger}\tau^{I}D_{\mu}H)B_{\rho\sigma}\widetilde{W}^{I\rho\sigma}$ \\  
$\widetilde{{\cal O}}_{2}^{H^{2}BW}$ & $i(D^{\mu}H^{\dagger}\tau^{I}D^{\nu}H)(B^{\vphantom{I}}_{\mu\rho}\widetilde{W}_{\nu}^{I\rho}-B^{\vphantom{I}}_{\nu\rho}\widetilde{W}_{\mu}^{I\rho})$  \\
$\widetilde{{\cal O}}_{3}^{H^{2}BW}$ & $(D^{\mu}H^{\dagger}\tau^{I}D^{\nu}H)(B^{\vphantom{I}}_{\mu\rho}\widetilde{W}_{\nu}^{I\rho}+B^{\vphantom{I}}_{\nu\rho}\widetilde{W}_{\mu}^{I\rho})$
\end{tabular}
\caption{Basis of independent dimension-eight operators in the SMEFT that are quartic in bosons and contain four derivatives and/or field strengths and at least one Higgs. CP-violating operators are denoted with a tilde. Factors of $i$ ensure that each operator is self-hermitian and thus has real Wilson coefficient.\label{tab:operators2}}
\end{center}
\end{table}

\subsection{Lower-dimension operators}\label{sec:dim6}

In a general EFT of the SM, there are additional higher-dimension operators appearing at mass dimension five, six, and seven. For a general scattering process or perturbative calculation, such operators are formally dominant over dimension-eight terms.
However, for the four-boson operators we will consider, we can exclude such operators from consideration as follows.

The only dimension-five operator in the SMEFT is the Weinberg operator~\cite{Weinberg:1979sa}, which generates Majorana neutrino masses. This operator contains fermionic fields, but we will be considering four-point scattering of bosons and the propagation of bosonic states in backgrounds that have nonzero bosonic field support but vanishing classical fermionic field backgrounds. That is, since fermion lines cannot end in Feynman diagrams, fermions cannot contribute to four-point boson scattering at leading order in perturbation theory.
Hence, the dimension-five term's contribution to the amplitude in the low-energy EFT can be neglected for the processes we consider.
By the same token, since all dimension-seven operators in the SMEFT contain fermions~\cite{Henning:2015alf}, we can exclude these from our analysis as well.

The SMEFT at dimension six is well studied; see Refs.~\cite{Weinberg:1979sa,Buchmuller:1985jz,WarsawBasis,Henning:2015alf,Jenkins:2013zja,Jenkins:2013wua,Alonso:2013hga,Jenkins:2017jig,Jenkins:2017dyc} and refs. therein. The operator basis, given in \Ref{WarsawBasis}, contains 59 independent baryon number conserving and 4 baryon number violating operators (counting hermitian conjugates as part of the original operator). 
Of these terms, 48 contain fermions, which we exclude by the reasoning given above, leaving the 15 purely bosonic operators given in \Tab{tab:dim6}.

\renewcommand{\arraystretch}{1.2}
\begin{table}[htbp]
\begin{center}
\begin{tabular}{L{1.8cm}  L{3.8cm} | L{1.8cm} L{3.8cm}}
\hspace{0.3cm}${\cal O}_{\text{dim-6}}^{H^2 B^2}$ & $H^\dagger H B_{\mu\nu}B^{\mu\nu}$ 
&\hspace{0.3cm}${\cal O}_{\text{dim-6}}^{W^3}$ & $\epsilon^{IJK} W_\mu^{I\nu} W_\nu^{J\rho}W_\rho^{K\mu}$
\\\hspace{0.3cm}$\widetilde {\cal O}_{\text{dim-6}}^{H^2 B^2}$ & $H^\dagger H B_{\mu\nu}\widetilde B^{\mu\nu}$ 
&\hspace{0.3cm}$\widetilde {\cal O}_{\text{dim-6}}^{W^3}$ & $\epsilon^{IJK} W_\mu^{I\nu} W_\nu^{J\rho}\widetilde W_\rho^{K\mu}$
\\\hspace{0.3cm}${\cal O}_{\text{dim-6}}^{H^2 W^2}$ & $H^\dagger H W^I_{\mu\nu}W^{I\mu\nu}$ 
&\hspace{0.3cm}${\cal O}_{\text{dim-6}}^{G^3}$ & $f^{abc} G_\mu^{a\nu}G_\nu^{b\rho}G_\rho^{c\mu}$
\\\hspace{0.3cm}$\widetilde {\cal O}_{\text{dim-6}}^{H^2 W^2}$ & $H^\dagger H W^I_{\mu\nu}\widetilde W^{I\mu\nu}$ 
&\hspace{0.3cm}$\widetilde {\cal O}_{\text{dim-6}}^{G^3}$ & $f^{abc} G_\mu^{a\nu}G_\nu^{b\rho}\widetilde G_\rho^{c\mu}$
\\\hspace{0.3cm}${\cal O}_{\text{dim-6}}^{H^2 G^2}$ & $H^\dagger H G^a_{\mu\nu}G^{a\mu\nu}$ 
&\hspace{0.3cm}${\cal O}_{1,\text{dim-6}}^{H^4}$ & $(H^\dagger H)\Box (H^\dagger H)$
\\\hspace{0.3cm}$\widetilde {\cal O}_{\text{dim-6}}^{H^2 G^2}$ & $H^\dagger H G^a_{\mu\nu}\widetilde G^{a\mu\nu}$ 
&\hspace{0.3cm}${\cal O}_{2,\text{dim-6}}^{H^4}$ & $(H^\dagger D^\mu H)^\star (H^\dagger D_\mu H)$
\\\hspace{0.3cm}${\cal O}_{\text{dim-6}}^{H^2 BW}$ & $H^\dagger \tau^I H B_{\mu\nu} W^{I\mu\nu}$ 
&\hspace{0.3cm}${\cal O}_{\text{dim-6}}^{H^6}$ & $(H^\dagger H)^3$
\\\hspace{0.3cm}$\widetilde {\cal O}_{\text{dim-6}}^{H^2 BW}$ & $H^\dagger \tau^I H B_{\mu\nu} \widetilde W^{I\mu\nu}$
\end{tabular}
\caption{Bosonic operators at dimension six in the SMEFT.\label{tab:dim6}}
\end{center}
\end{table}

Since the quartic operators with a Higgs in \Tab{tab:dim6}---of the schematic form $H^2 F^2$, $H^2 \Box H^2$, or $(H D H)^2$---have terms containing at most two derivatives, a single insertion of these operators contributes at lower order in momenta in the four-point amplitude; hence, such terms will give no contribution to the $s^2$ coefficient in the forward amplitude and will thus drop out of the contour integral in \Eq{eq:lambdacontour} by the residue theorem. 
Again, we need not consider multiple insertions of quartic operators, since we are computing amplitudes to leading order low-energy EFT.
In the causality calculation, such operators also do not contribute to the first-order change to the speed of propagation, since the two derivatives in these operators become two insertions of the zeroth-order momentum $k$, and one can see by \Tab{tab:dim6} that these two $k$ momenta will always both be contracted into each other (yielding zero contribution to the first-order dispersion relation), into a polarization (yielding zero by the Lorenz gauge condition), or into a Levi-Civita tensor (again yielding zero).
Hence, we can drop all the operators in the first column of \Tab{tab:dim6}, as well as ${\cal O}_{1,\text{dim-6}}^{H^4}$ and ${\cal O}_{2,\text{dim-6}}^{H^4}$. We can also drop the six-point term ${\cal O}_{\text{dim-6}}^{H^6}$, since it cannot contribute to the leading-order four-point amplitudes in the EFT.

We are thus left with only four operators to consider at dimension six: ${\cal O}^{W^3}_{\text{dim-6}}$, $\widetilde {\cal O}^{W^3}_{\text{dim-6}}$, ${\cal O}^{G^3}_{\text{dim-6}}$, and $\widetilde {\cal O}^{G^3}_{\text{dim-6}}$.
Expanding in gauge fields, these operators can generate both cubic and quartic vertices, which can contribute to four-point $W$ boson or gluon scattering.
While a single insertion of the quartic will be of lower order in derivatives, exchange of a gauge boson through two insertions of the dimension-six operator will give a contribution independent of the sign of the coupling and at the same order in derivatives as a single insertion of a four-point dimension-eight vertex, thus obstructing the bounds we wish to place.
We can avoid this obstacle by simply choosing to scatter incoming states with commuting colors.
Writing $u_1^a$ and $u_2^a$ for the two color vectors of the incoming gluons (and similarly for the $W$ bosons), choosing $u_{1,2}$ such that  
\be 
f^{abc}u_1^b u_2^c = 0
\label{eq:T12commute}
\ee
guarantees that the leading-order four-point forward amplitude induced by ${\cal O}^{G^3}_{\text{dim-6}}$, and $\widetilde {\cal O}^{G^3}_{\text{dim-6}}$ vanishes.
Similarly, choosing \Eq{eq:T12commute}, where now $u_{1,2}$ are the colors of the background and perturbation, guarantees that these operators do not change the speed of propagation. A similar conclusion holds for the $W^3$ operators.
The requirement in \Eq{eq:T12commute} also excises the $t$-channel singularity, which scales as $s/t$, from the forward Yang-Mills amplitude.
By requiring states with commuting colors according to \Eq{eq:T12commute}, we are choosing gluons within the abelian Cartan subalgebra describing the diagonal generators of ${\rm SU}(N)$; the Cartan subalgebra has $N-1$ generators, so we will derive $N-1$ ${\rm U}(1)$-like bounds from ${\rm SU}(N)$ scattering.

Finally, we should consider whether contributions from the unperturbed SM itself interfere with our bounds.
Given the paucity of direct new physics detections at the LHC and elsewhere, we make the well-motivated assumption that the scale $M$ of new physics is above the weak scale.
We can therefore treat all of the bosons in our thought experiments as massless to good approximation.
Above the weak scale, the SM obeys perturbative unitarity, so amplitudes from purely SM operators are well behaved at large momenta.
In contrast, the addition of four-derivative (dimension-eight) EFT operators induces amplitudes that diverge $s^{2}$. 
As before, since the contour integral in $s$ extracts the $s^2$ part of the amplitude, we can ignore at leading order any contribution from the uncorrected SM.
Accordingly, in all amplitudes we subsequently write, we will suppress these contributions, retaining only the terms induced by the operators in Tables~\ref{tab:operators} and \ref{tab:operators2}.
Moreover, the condition in \Eq{eq:T12commute} (along with its ${\rm SU}(2)$ analogue) ensures that a gauge field background of one color does not interfere with the null propagation of gauge fields of the other color, and similarly the Yang-Mills contribution to the forward amplitude vanishes.

Thus, provided we take $M$ above the weak scale and choose colors and isospins according to \Eq{eq:T12commute}, we are able to apply the analyticity and causality arguments reviewed in \Sec{sec:IR} to four-point contact terms in the EFT in order to bound the dimension-eight operators without being affected by operators of lower mass dimension.

\section{Bounds}\label{sec:bounds}

Using the causality and analyticity techniques of \Sec{sec:IR}, we now derive bounds on the EFT coefficients of the operators in Tables~\ref{tab:operators} and \ref{tab:operators2}. In total we will derive 27 independent bounds, and we will do so by considering different classes of operators separately.

\subsection{Single field strength quartics}\label{sec:singlegaugebounds}

Let us first consider bounding the operators in \Tab{tab:operators} that contain a single type of gauge field. It will be useful to first generalize to gauge group ${\rm SU}(N)$ for $N$ arbitrary, writing the gauge field as $A^a_\mu$ and field strength at $F^a_{\mu\nu}$. 
As shown in \Ref{Morozov}, there are eight CP-conserving and four CP-violating operators of the form $F^4$, given in \Tab{tab:operatorsN}; the larger number of operators, compared to the ${\rm SU}(3)$ operators in \Tab{tab:operators}, is due to the fact that the identities in \Eq{eq:SU3idens} do not apply for general $N$.

\renewcommand{\arraystretch}{1.2}
\begin{table}[htbp]
\begin{center}
\begin{tabular}{C{1.5cm}  L{4.5cm}}
${\cal O}_{1}^{F^4}$ & $(F^a F^a)( F^b F^b)$
\\${\cal O}_{2}^{F^4}$ & $(F^a \widetilde{F}^a)(F^b \widetilde{F}^b)$
\\${\cal O}_{3}^{F^4}$ & $(F^a F^b)(F^a F^b)$
\\${\cal O}_{4}^{F^4}$ & $(F^a \widetilde{F}^b)(F^a \widetilde{F}^b)$
\\${\cal O}_{5}^{F^4}$ & $d^{abe}d^{cde} (F^a F^b)(F^c F^d)$
\\${\cal O}_{6}^{F^4}$ & $d^{abe}d^{cde} (F^a \widetilde{F}^b)(F^c \widetilde{F}^d)$
\\${\cal O}_{7}^{F^4}$ & $d^{ace}d^{bde} (F^a F^b)(F^c F^d)$
\\${\cal O}_{8}^{F^4}$ & $d^{ace}d^{bde} (F^a \widetilde{F}^b)(F^c \widetilde{F}^d)$
\\$\widetilde{{\cal O}}_{1}^{F^4}$ & $(F^a F^a)(F^b \widetilde{F}^b)$
\\$ \widetilde{{\cal O}}_{2}^{F^4}$ & $(F^a F^b)(F^a \widetilde{F}^b)$
\\$\widetilde{{\cal O}}_{3}^{F^4}$ & $d^{abe}d^{cde} (F^a F^b)(F^c \widetilde{F}^d)$
\\$\widetilde{{\cal O}}_{4}^{F^4}$ & $d^{ace}d^{bde} (F^a F^b)(F^c \widetilde{F}^d)$
\end{tabular}
\caption{Basis of independent operators for ${\rm SU}(N)$ Yang-Mills theory that are quartic in gauge field strengths. Notation is as in \Tab{tab:operators}.}
\label{tab:operatorsN}
\end{center}
\end{table}

We first consider the constraint that causality imposes on this more general set of operators. Taking the Lagrangian $-\frac{1}{4}F^a_{\mu\nu}F^{a\mu\nu} + \frac{1}{M^4} \sum_i c_i {\cal O}_i$, for the ${\cal O}_i$ in \Tab{tab:operatorsN}, the leading-order (i.e., uncorrected Yang-Mills) equation of motion is
\be
D^\mu F^a_{\mu\nu} = \partial^\mu F^a_{\mu\nu} + g\, f^{abc} A^{\mu b} F^c_{\mu\nu} = 0, \label{eq:EOM}
\ee
where $g$ is the gauge coupling. This equation of motion is solved by a constant $F^a_{\mu\nu}$, so we take the background gauge field to be 
\be 
\overline A^a_\mu = u_1^a \epsilon_{1\mu} w,\label{eq:simple}
\ee 
where $u_1$ is a constant real vector in color space, $\epsilon_1$ is a constant four-vector of arbitrary signature, and $w$ is an arbitrary Cartesian coordinate in spacetime, i.e., $\partial_\mu w = \ell_\mu$, where $\ell_\mu$ is a constant four-vector; we make no assumptions regarding the direction or signature of $\ell$ or $\epsilon_1$.\footnote{For the dispersion relation calculation, we work in the geometric optics limit, so that we can approximate any solution with constant $F^a_{\mu\nu}$ provided that we  ultimately consider perturbations on length scales small compared to the characteristic scale of the background.}
Note that $u_1$ is {\it not} a generator of ${\rm SU}(N)$; $u_1$ is not matrix-valued, but is simply a constant vector of real numbers.
We then expand around this background, writing $A^a_\mu = \overline A^a_\mu + \delta A^a_\mu$, where the fluctuation is taken to be a plane wave,
\be 
\delta A^a_\mu = u_2^a \epsilon_{2\mu} e^{ik \cdot x}, \label{eq:planewave}
\ee
choosing an arbitrary color vector $u_2$ and spacelike polarization $\epsilon_2$. In vacuum, where $\overline{A}_{\mu}^{a}=0$, \Eq{eq:planewave} is a solution of the equations of motion~\eqref{eq:EOM} provided $k^{2}=k\cdot \epsilon_2=0$. However, for general backgrounds, this does not solve the unperturbed equations of motion, due to the self-interaction of the gluon. If we take the background $\overline{A}_{\mu}^{a}$ to be given by the constant solution in \Eq{eq:simple}, then we require $u_{1,2}$ to satisfy \Eq{eq:T12commute} in order for the null plane wave perturbation to solve its equation of motion (i.e., to get $k^{2}=0$ as a solution of the dispersion relation at zeroth order in the $c_{i}$). In spite of \Eq{eq:T12commute}, there are additional color combinations we can form using $d^{abc}$, which we define as $U^{a}=d^{abc}u_1^{b}u_2^{c}$, $V^a = d^{abc} u_1^b u_1^c$, and $W^a = d^{abc} u_2^b u_2^c$. The identity in \Eq{eq:useful} relates these as
\be
VW = U^2 + \frac{2}{N}[ (u_1 u_2)^2 - u_1^2 u_2^2],
\ee
where we contract color indices using $\delta^{ab}$.

Within this framework, we now compute the speed $v$ of propagation of our fluctuation to leading order in the $c_i$, as in \Sec{sec:IR} writing $k_\mu = (k_0,{\bf k})$ and identifying the speed $v$ as $k_0/|{\bf k}|$. To our given order in perturbation theory, we find
\be 
\begin{aligned}
v &= 1 - \frac{8}{M^4 \epsilon_2^2 u_2^2 k_0^2 N}\left\{[(\epsilon_1 \cdot k)(\epsilon_2 \cdot \ell) -(k\cdot \ell)(\epsilon_1 \cdot \epsilon_2)]^2 A + (\epsilon_1^\mu \epsilon_2^\nu  k^\rho \ell^\sigma \epsilon_{\mu\nu\rho\sigma})^2 B\right. \\& \left. \hspace{3.5cm} - [(\epsilon_1 \cdot  k)(\epsilon_2 \cdot \ell) -(k\cdot \ell)(\epsilon_1 \cdot \epsilon_2)]\epsilon_1^\mu \epsilon_2^\nu  k^\rho \ell^\sigma \epsilon_{\mu\nu\rho\sigma}C \right\},
\end{aligned}\label{eq:vcorr}
\ee 
where $A$, $B$, and $C$ are combinations of the operator coefficients defined as
\be
\begin{aligned}
A & =N\left[(2c_{1}+c_{3})(u_{1}u_{2})^{2}+c_{3}u_{1}^{2}u_{2}^{2}+2(c_{5}+c_{7})U^{2}\right]+2c_{7}\left[(u_{1}u_{2})^{2}-u_{1}^{2}u_{2}^{2}\right]\\
B & =N\left[(2c_{2}+c_{4})(u_{1}u_{2})^{2}+c_{4}u_{1}^{2}u_{2}^{2}+2(c_{6}+c_{8})U^{2}\right]+2c_{8}\left[(u_{1}u_{2})^{2}-u_{1}^{2}u_{2}^{2}\right]\\
C & =N\left[(2\widetilde{c}_{1}+\widetilde{c}_{2})(u_{1}u_{2})^{2}+\widetilde{c}_{2}u_{1}^{2}u_{2}^{2}+2(\widetilde{c}_{3}+\widetilde{c}_{4})U^{2}\right]+2\widetilde{c}_{4}\left[(u_{1}u_{2})^{2}-u_{1}^{2}u_{2}^{2}\right].
\end{aligned} \label{eq:ABC}
\ee
Normalizing the color vectors such that $u_1^2 = u_2^2 = 1$ will allow us to define $u_1 u_2 = \cos\zeta$. For concreteness, we can use Lorentz invariance to fix the direction of ${\bf k}$ and the direction of the (spatial vector) $\epsilon_2^{\mu}$ orthogonal to $k$. Let us take ${\bf k}$ to point in the $z$-direction and $\epsilon_2^{\mu}$ to be a unit vector pointing in the $x$-direction. Then let us define parameters $\theta_{1,2,3}$, $\phi_{1,2,3}$, and $E$ via
\be 
\begin{aligned}
\ell_{\mu} & =(\cos\theta_1,\sin\theta_1\cos\theta_2,\sin\theta_1\sin\theta_2\cos\theta_3,\sin\theta_1\sin\theta_2\sin\theta_3)\\
\epsilon_1^{\mu} & =E(\cos\phi_1,\sin\phi_1\cos\phi_2,\sin\phi_1\sin\phi_2\cos\phi_3,\sin\phi_1\sin\phi_2\sin\phi_3).
\end{aligned}\label{eq:le1def}
\ee
Note that we do not need to specify a separate overall scale for both $\ell$ and $\epsilon_1$; since each term in the speed correction in \Eq{eq:vcorr} is quadratic in each of these vectors individually, we can absorb the normalization into the constant $E$ in \Eq{eq:le1def}.
Then the speed $v$ is
\be 
v=1-\frac{8E^2}{M^4 N}\left(AX^{2}+BY^{2}+CXY\right),
\ee
where
\be
\left( \begin{matrix} 
X \\
Y 
\end{matrix} \right)
= \left(\begin{matrix}\sin \theta_1 \cos \theta_2 & \sin\phi_1 \cos\phi_2 \\ 
\sin\theta_1\sin\theta_2\cos\theta_3 & \sin\phi_1\sin\phi_2\cos\phi_3
\end{matrix} \right)
\left( \begin{matrix} 
\cos\phi_1+\sin\phi_1\sin\phi_2\sin\phi_3 \\
\cos\theta_1- \sin\theta_1\sin\theta_2\sin\theta_3
\end{matrix} \right).\label{eq:XY}
\ee
Writing $X=Z\cos\psi$ and $Y=Z\sin\psi$ for some $Z>0$, the causality bound becomes
\be 
A\cos^{2}\psi+B\sin^{2}\psi+C\cos\psi\sin\psi>0.\label{eq:boundpre}
\ee

We can alternately obtain the constraint in \Eq{eq:boundpre} from analyticity and unitarity of scattering amplitudes. To do so, we compute the two-to-two gluon scattering amplitude for gluons\footnote{Even though gluons in realistic scattering experiments would be confined into nonperturbative objects like hadrons, it is still well-defined to consider the perturbative S-matrix for asymptotic states of gluons and analyze its analytic structure. Indeed, the properties of gluon amplitudes are among the central objects of study in the modern amplitudes program~\cite{Elvang:2013cua,Britto:2004ap,Britto:2005fq,ArkaniHamed:2012nw,Cheung:2017yef,Kawai:1985xq}. Our analyticity constraints therefore derive from considering such an idealized thought experiment for gluon scattering.} with incoming momenta $k_{1,2,3,4}$, polarizations $\epsilon_{1,2,3,4}$, and color vectors $u_{1,2,3,4}$.
Let us choose forward kinematics ($k_3 = -k_1$, $k_4 = -k_2$, $\epsilon_3 = \epsilon_1$, $\epsilon_4 = \epsilon_2$, $u_3 = u_1$, $u_4 = u_2$) and require that $u_i$ and $\epsilon_i$ be real vectors. As discussed in \Sec{sec:dim6}, by requiring that $u_{1,2}$ be chosen such that $f^{abc} u_1^a u_2^b = 0$ as in the causality bound, we ensure that contributions from the Yang-Mills cubic interaction and any dimension-six cubics vanish.
The forward amplitude is then
\be 
\begin{aligned}
{\cal A}_{F^{4}}(s) &= \frac{8}{M^4 N} \left\{A(\epsilon_1\cdot \epsilon_2)^2 s^2 + B [\epsilon_1^2 \epsilon_2^2 - (\epsilon_1 \cdot \epsilon_2)^2] s^2 + 2C (\epsilon_1 \cdot \epsilon_2)\epsilon_1^\mu \epsilon_2^\nu k_1^\rho k_2^\sigma \epsilon_{\mu\nu\rho\sigma}s\right\}
\\&=\frac{8s^{2}}{M^4 N}\left[A\cos^{2}\psi+B\sin^{2}\psi+C\cos\psi\sin\psi \right],
\end{aligned}
\ee
where in the second line $\psi$ is now the angle between the unit vectors $\epsilon_1$ and $\epsilon_2$, defined with sign such that the cross product of $\epsilon_1$ and $\epsilon_2$ points along the spatial direction of $k_1$.
Thus, analyticity of gluon-gluon scattering requires the same bound in \Eq{eq:boundpre} as implied by causality.
Marginalizing over $\psi$, we have the requirements:\footnote{In particular, one obtains $A>0$ by setting $\psi = 0$, $B>0$ by setting $\psi = \pi/2$, and $C^2 < 4 AB$ by setting $\psi = \pm \arctan \sqrt{A/B}$. One does not find the third bound on the CP-violating term by considering strictly parallel or perpendicular polarizations alone; equivalently, it would not be obtained by considering positivity bounds from fixed-helicity amplitudes, but only from superpositions thereof.}
\be
A>0,\qquad B>0, \qquad \text{and} \qquad C^2 < 4AB,\label{eq:bound1}
\ee
which must be satisfied for all $u_{1,2}$ for which $f^{abc} u_1^a u_2^b = 0$.
Note that if all the CP-conserving terms vanish (so that $A=B=0$), then the CP-violating terms are forced to vanish as well, since all our Wilson coefficients are real; we will discuss this feature further in \Sec{sec:square}.

\subsubsection{${\rm SU}(3)$}

Let us determine the consequences of the bound in \Eq{eq:bound1} for ${\rm SU}(3)_C$. For $N=3$, we can rewrite ${\cal O}_{7}$, ${\cal O}_8$, and $\widetilde {\cal O}_4$ in \Tab{tab:operatorsN} in terms of the other operators, resulting in the basis for ${\cal O}_i^{G^4}$, $\widetilde {\cal O}_i^{G^4}$ in \Tab{tab:operators}. Moreover, for ${\rm SU}(3)$ the identities in \Eqs{eq:useful}{eq:SU3idens} imply that $U^2 = u_1^2 u_2^2/3$, so  we can redefine $A,B,C$ as
\be 
\begin{aligned}
A & =3c_{3}^{G^4}+2c_{5}^{G^4}+3(2c_{1}^{G^4}+c_{3}^{G^4})\cos^{2}\zeta\\
B & =3c_{4}^{G^4}+2c_{6}^{G^4}+3(2c_{2}^{G^4}+c_{4}^{G^4})\cos^{2}\zeta\\
C & =3\widetilde{c}_{2}^{G^4}+2\widetilde{c}_{3}^{G^4}+3(2\widetilde{c}_{1}^{G^4}+\widetilde{c}_{2}^{G^4})\cos^{2}\zeta.
\end{aligned}
\qquad[\text{for \ensuremath{{\rm SU}(3)}}]
\label{eq:QCDbounds}
\ee
Since the bound in \Eq{eq:bound1} is now linear in $\cos^2\zeta$, we obtain a basis of bounds by considering the two cases of $\cos^2 \zeta = 0$ or $1$, so the independent bounds are
\be 
\begin{aligned}
3c_{1}^{G^{4}}+3c_{3}^{G^{4}}+c_{5}^{G^{4}}&>0\\3c_{3}^{G^{4}}+2c_{5}^{G^{4}}&>0\\3c_{2}^{G^{4}}+3c_{4}^{G^{4}}+c_{6}^{G^{4}}&>0\\3c_{4}^{G^{4}}+2c_{6}^{G^{4}}&>0\\(3\widetilde{c}_{1}^{G^{4}}+3\widetilde{c}_{2}^{G^{4}}+\widetilde{c}_{3}^{G^{4}})^{2}&<4(3c_{1}^{G^{4}}+3c_{3}^{G^{4}}+c_{5}^{G^{4}})(3c_{2}^{G^{4}}+3c_{4}^{G^{4}}+c_{6}^{G^{4}})\\(3\widetilde{c}_{2}^{G^{4}}+2\widetilde{c}_{3}^{G^{4}})^{2}&<4(3c_{3}^{G^{4}}+2c_{5}^{G^{4}})(3c_{4}^{G^{4}}+2c_{6}^{G^{4}}).
\end{aligned}
\qquad[\text{for \ensuremath{{\rm SU}(3)}}]\label{eq:SU3bounds}
\ee

\subsubsection{${\rm SU}(2)$}

For the case of ${\rm SU(2)}_L$, the basis and bounds simplify further. The $d^{abc}$ coefficients all vanish for ${\rm SU}(2)$, so we reduce to the basis of ${\cal O}_i^{W^4}$, $\widetilde {\cal O}_i^{W^4}$ in \Tab{tab:operators} and have $U^2 = 0$. Moreover, since the structure constants for ${\rm SU}(2)$ are simply $\epsilon^{IJK}$, the requirement in \Eq{eq:T12commute} implies that we must have $u_1 = \pm u_2$. Thus, we can now redefine $A,B,C$ for ${\rm SU}(2)$ as
\be 
\begin{aligned}
A & =4(c_{1}^{W^4}+c_3^{W^4})\\
B & =4(c_{2}^{W^4}+c_4^{W^4})\\
C & =4(\widetilde c_{1}^{W^4}+\widetilde c_2^{W^4}),
\end{aligned}
\qquad[\text{for \ensuremath{{\rm SU}(2)}}]
\ee
so the independent bounds become
\be 
\begin{aligned}
c_{1}^{W^{4}}+c_{3}^{W^{4}}&>0
\\c_{2}^{W^{4}}+c_{4}^{W^{4}}&>0\\
(\widetilde{c}_{1}^{W^{4}}+\widetilde{c}_{2}^{W^{4}})^{2}&<4(c_{1}^{W^{4}}+c_{3}^{W^{4}})(c_{2}^{W^{4}}+c_{4}^{W^{4}}).
\end{aligned}
\qquad[\text{for \ensuremath{{\rm SU}(2)}}]\label{eq:SU2bound}
\ee

\subsubsection{${\rm U}(1)$}

For the ${\rm U}(1)_Y$ case, all the color structure disappears, and we are left with only the three $B^4$ operators in \Tab{tab:operators}. We can write
\be 
\begin{aligned}
A & =2c_{1}^{B^4}\\
B & =2c_{2}^{B^4}\\
C & =2\widetilde{c}_{1}^{B^4},
\end{aligned}
\qquad[\text{for \ensuremath{{\rm U}(1)}}]
\ee
so the independent bounds become
\be 
\begin{aligned}
c_{1}^{B^4} & >0\\
c_{2}^{B^4} & >0\\
(\widetilde{c}_{1}^{B^4})^{2} & <4c_{1}^{B^4} c_{2}^{B^4}.
\end{aligned}
\qquad[\text{for \ensuremath{{\rm U}(1)}}]\label{eq:U1bound}
\ee
The first two lines of \Eq{eq:U1bound} match the results derived in \Ref{Adams:2006sv}, and the third line can further be obtained by demanding index of refraction greater than unity in the vacuum birefringence calculation of \Ref{Rebhan:2017zdx}; see also Refs.~\cite{Gibbons:2000xe,Fouche:2016qqj}. Discussion of experimental limits on the photonic component of $c_i^{B^4}$ can be found in \Ref{Abalos:2015gha}.

That the SU(2) and SU(3) results in \Eq{eq:SU2bound} and \Eq{eq:SU3bounds} take the form of single and pairwise copies of the U(1) bounds in \Eq{eq:U1bound} is not by accident. As commented above, it is due to our insistence on imposing \Eq{eq:T12commute}, which restricts us to the Cartan subalgebra in each case. For SU($N$) the Cartan subalgebra consists of $N-1$ copies of U(1), explaining this pattern between the bounds.

\subsection{Field strength cross-quartics}

Let us next bound the cross-quartic operators among the gauge field strengths appearing in \Tab{tab:operators}. For a Yang-Mills theory with gauge group ${\rm SU}(N)\otimes {\rm SU}(n)$ for $N,n$ arbitrary with respective gauge fields $A_\mu^a$ and $a^A_\mu$ and field strengths $F^a_{\mu\nu}$ and $f^A_{\mu\nu}$, let us again choose a background for $A^a_{\mu}$ as in \Eq{eq:simple}. The ansatz for the perturbation of the ${\rm SU}(n)$ gauge field, $\delta a^A_\mu$, will be written as in \Eq{eq:planewave}, $u_2^A \epsilon_{2\mu}e^{ik \cdot x}$. We define the coordinate system for $\epsilon_{1,2}$ and $\partial_\mu w$ as in \Sec{sec:singlegaugebounds}.

Writing the basis of cross-quartic operators analogously with those in \Tab{tab:operators} as
\be 
\begin{aligned}
{\cal O}_1 &= F_{\mu\nu}^a F^{a\mu\nu} f^A_{\rho\sigma}f^{A\rho\sigma}\\
{\cal O}_2 &= F_{\mu\nu}^a \widetilde F^{a\mu\nu} f^A_{\rho\sigma} \widetilde f^{A\rho\sigma}\\
{\cal O}_3 &= F_{\mu\nu}^a f^{A\mu\nu} F^a_{\rho\sigma}  f^{A\rho\sigma}\\
{\cal O}_4 &= F_{\mu\nu}^a \widetilde f^{A\mu\nu} \widetilde F^a_{\rho\sigma}  f^{A\rho\sigma}\\
\widetilde {\cal O}_1 &= F_{\mu\nu}^a \widetilde F^{a\mu\nu} f^A_{\rho\sigma}f^{A\rho\sigma}\\
\widetilde {\cal O}_2 &= F_{\mu\nu}^a F^{a\mu\nu} f^A_{\rho\sigma} \widetilde f^{A\rho\sigma}\\
\widetilde {\cal O}_3 &= F_{\mu\nu}^a f^{A\mu\nu} \widetilde F^a_{\rho\sigma}  f^{A\rho\sigma},
\end{aligned}
\ee
with Wilson coefficients $c_i/M^4,\widetilde c_i/M^4$, we find that the speed of propagation of the $\delta a^A_\mu$ fluctuations is
\be 
v = 1- \frac{4E^2}{M^4} (c_3 X^2 + c_4 Y^2 + \widetilde c_3 XY),
\ee
for $X,Y$ defined in \Eq{eq:XY}. Similarly, the four-point amplitude becomes
\be
{\cal A}_{F^2 f^2}(s) = \frac{4s^2}{M^4}\left[c_3 \cos^{2}\psi+c_4 \sin^{2}\psi+\widetilde c_3\cos\psi\sin\psi \right],
\ee
so the bounds from analyticity and causality are identical and, marginalizing over $\psi$, can be written simply as the requirements that $c_3>0$, $c_4>0$, and $\widetilde c_3^2 < 4 c_3 c_4$.

Applying this result to the operators in \Tab{tab:operators}, we obtain the bounds on the SMEFT:
\begingroup
\allowdisplaybreaks
\begin{align}
c_{3}^{B^{2}W^{2}}&>0 \nonumber \\
c_{4}^{B^{2}W^{2}}&>0 \nonumber \\
(\widetilde{c}_{3}^{B^{2}W^{2}})^{2}&<4c_{3}^{B^{2}W^{2}}c_{4}^{B^{2}W^{2}} \nonumber \\
c_{3}^{B^{2}G^{2}}&>0 \nonumber \\
c_{4}^{B^{2}G^{2}}&>0\\
(\widetilde{c}_{3}^{B^{2}G^{2}})^{2}&<4c_{3}^{B^{2}G^{2}}c_{4}^{B^{2}G^{2}} \nonumber \\
c_{3}^{W^{2}G^{2}}&>0 \nonumber \\
c_{4}^{W^{2}G^{2}}&>0 \nonumber \\
(\widetilde{c}_{3}^{W^{2}G^{2}})^{2}&<4c_{3}^{W^{2}G^{2}}c_{4}^{W^{2}G^{2}}.\nonumber
\end{align}\label{eq:crossbound}

\subsection{Higgs quartics}\label{sec:H4}

Let us now bound the dimension-eight operators quartic in the Higgs in \Tab{tab:operators2}. We will be considering the change to the speed of Higgs propagation in nonzero Higgs backgrounds, as well as Higgs four-point scattering. 
As we reviewed in \Sec{sec:IR}, a positivity bound on the coefficient of $(\partial \phi)^4$ was proven for a real massless scalar $\phi$ in \Ref{Adams:2006sv}; for the ${\cal O}_i^{H^4}$, however, the ${\rm SU}(2)$ doublet nature of the SM Higgs field leads to more interesting structure in the bounds.
Since we are considering a completion scale $M$ above the weak scale, we can approximate the Higgs as effectively massless in our dispersion relation calculation. 

Expanding the covariant derivatives and considering backgrounds with zero gauge field, the ${\cal O}^{H^4}_i$ operators become $(\partial H)^4$.  Expanding in the $\phi_{1,2,3,4}$ defined in \Eq{eq:Hdef}, we have 
\begingroup
\allowdisplaybreaks
\begin{align}
{\cal O}^{H^4}_{1}  \rightarrow &\,\frac{1}{4}\sum_{i=1}^{4}(\partial\phi_{i})^{4}+\frac{1}{2}\left[(\partial_{\mu}\phi_{1}\partial^{\mu}\phi_{3})^{2}+(\partial_{\mu}\phi_{1}\partial^{\mu}\phi_{4})^{2}+(\partial_{\mu}\phi_{2}\partial^{\mu}\phi_{3})^{2}+(\partial_{\mu}\phi_{2}\partial^{\mu}\phi_{4})^{2}\right] \nonumber \\
 & +\frac{1}{2}\left[(\partial\phi_{1})^{2}(\partial\phi_{2})^{2}+(\partial\phi_{3})^{2}(\partial\phi_{4})^{2}\right]-\partial_{\mu}\phi_{1}\partial^{\mu}\phi_{4}\partial_{\nu}\phi_{2}\partial^{\nu}\phi_{3}+\partial_{\mu}\phi_{1}\partial^{\mu}\phi_{3}\partial_{\nu}\phi_{2}\partial^{\nu}\phi_{4} \nonumber \\
\nonumber \\
{\cal O}^{H^4}_{2}  \rightarrow &\,\frac{1}{4}\sum_{i=1}^{4}(\partial\phi_{i})^{4}+\frac{1}{2}\left[(\partial_{\mu}\phi_{1}\partial^{\mu}\phi_{3})^{2}+(\partial_{\mu}\phi_{1}\partial^{\mu}\phi_{4})^{2}+(\partial_{\mu}\phi_{2}\partial^{\mu}\phi_{3})^{2}+(\partial_{\mu}\phi_{2}\partial^{\mu}\phi_{4})^{2}\right] \nonumber \\
 & -\frac{1}{2}\left[(\partial\phi_{1})^{2}(\partial\phi_{2})^{2}+(\partial\phi_{3})^{2}(\partial\phi_{4})^{2}\right]+(\partial_{\mu}\phi_{1}\partial^{\mu}\phi_{2})^{2}+(\partial_{\mu}\phi_{3}\partial^{\mu}\phi_{4})^{2} \label{eq:OH4simp} \\
 & +\partial_{\mu}\phi_{1}\partial^{\mu}\phi_{4}\partial_{\nu}\phi_{2}\partial^{\nu}\phi_{3}-\partial_{\mu}\phi_{1}\partial^{\mu}\phi_{3}\partial_{\nu}\phi_{2}\partial^{\nu}\phi_{4} \nonumber \\ 
{\cal O}^{H^4}_{3}  \rightarrow &\, \frac{1}{4} \sum_{i=1}^{4}(\partial\phi_{i})^{4}+\frac{1}{2}\left[(\partial\phi_{1})^{2}(\partial\phi_{2})^{2}+(\partial\phi_{1})^{2}(\partial\phi_{3})^{2}+(\partial\phi_{2})^{2}(\partial\phi_{3})^{2}\right] \nonumber \\
 & +\frac{1}{2}\left[(\partial\phi_{1})^{2}(\partial\phi_{4})^{2}+(\partial\phi_{2})^{2}(\partial\phi_{4})^{2}+(\partial\phi_{3})^{2}(\partial\phi_{4})^{2}\right]. \nonumber
\end{align}
\endgroup

Let us now compute the effect of the operators in \Eq{eq:OH4simp} on the speed of Higgs propagation. In a background of nonzero $\overline \phi_{i}$ for $i\in\{1,2,3,4\}$, for perturbations $\delta\phi_{i} \propto e^{ik \cdot x}$ with wavelengths much smaller than the characteristic scale of the background, $\overline \phi_i$ can be expanded as a constant vacuum expectation value plus a constant derivative $\overline{\partial_{\mu}\phi_{i}}=j_{\mu}$. 
We choose this background to be smaller than the weak scale, so that it does not induce an effective Higgs mass via the Higgs quartic or higher-dimension operators.
We then find a speed of fluctuations equal to
\be 
v=1-(c^{H^4}_{1}+c^{H^4}_{2}+c^{H^4}_{3})\frac{(j\cdot k)^{2}}{k_0^2 M^4},
\ee
so we must have $c^{H^4}_{1}+c^{H^4}_{2}+c^{H^4}_{3}>0$. 

Similarly, if we take a background of $\phi_{1}$ or $\phi_{2}$, and consider perturbations in $\phi_{3}$ or $\phi_{4}$ (or vice versa), we have
\be
v=1-(c^{H^4}_{1}+c^{H^4}_{2})\frac{(j\cdot k)^{2}}{2k_0^2 M^4},
\ee
so we must have $c^{H^4}_{1}+c^{H^4}_{2}>0$. 
Finally, if we take a background of $\phi_{1}$ (or $\phi_{3}$) and consider perturbations in $\phi_{2}$ (or $\phi_{4}$, respectively), or vice versa, we obtain
\be 
v=1-c^{H^4}_{2}\frac{(j\cdot k)^{2}}{k_0^2 M^4},
\ee
so we must have $c^{H^4}_{2}>0$.

We can arrive at these same bounds from analyticity of two-to-two Higgs scattering. 
Let us compute an amplitude for particles I and II scattering to particles III and IV.\footnote{We use $\ket{\rm I}\ket{\rm II}\rightarrow \ket{\rm III}\ket{\rm IV}$ here, instead of our usual $12\rightarrow 34$ notation, to avoid ambiguity with the four Higgs fields, which we have already labeled $1234$.} For each of particles I, II, III, IV, let us consider an arbitrary superposition of $\phi_{i}$ states:
\be 
\begin{aligned}
\ket{\rm I} & =\sum_{i=1}^{4}\alpha_{i}|\phi_{i}\rangle\qquad &\ket{\rm III} & =\sum_{i=1}^{4}\gamma_{i}|\phi_{i}\rangle\\
\ket{\rm II} & =\sum_{i=1}^{4}\beta_{i}|\phi_{i}\rangle &\ket{\rm IV} & =\sum_{i=1}^{4}\delta_{i}|\phi_{i}\rangle,
\end{aligned}
\ee
where $\sum_i |\alpha_i|^2 = \sum_i |\beta_i|^2 = \sum_i |\gamma_i|^2 = \sum_i |\delta_i|^2 = 1$. To apply the optical theorem, we must consider forward scattering, which means $k_{\rm I}=-k_{\rm III}$ and $k_{\rm II}=-k_{\rm IV}$ in the all-incoming convention and $\alpha_{i}=\gamma_{i}^{*}$ and $\beta_{i}=\delta_{i}^{*}$. Schematically, the $s^2$ part of the forward amplitude coming from the ${\cal O}_i^{H^4}$ is then
\be 
{\cal A}_{H^4}(s)=\sum_{ijkl} K_{ijkl}\alpha_{i}\beta_{j}\alpha_{k}^{*}\beta_{l}^{*}\frac{s^{2}}{M^4},
\ee
where $K_{ijkl}$ is a rank-four tensor, with each entry proportional to a sum of $c^{H^4}_{1,2,3}$. 
Analogously with polarizations, to ensure for simplicity that ${\cal A}_{H^4}(s)$ maps onto itself under crossing, we choose $\alpha_i$ and $\beta_i$ to be real. 
Analyticity and unitarity then require 
\be 
\sum_{ijkl} K_{ijkl}\alpha_{i}\beta_{j}\alpha_{k} \beta_{l}>0\label{eq:quarticform}
\ee
for all $\alpha_{i},\beta_{i}$. Numerical analysis shows that the analyticity and unitarity conditions in \Eq{eq:quarticform} can be reduced to
\be 
\begin{aligned}
c^{H^4}_{1}+c^{H^4}_{2}+c^{H^4}_{3} & >0\\
c^{H^4}_{1}+c^{H^4}_{2} & >0\\
c^{H^4}_{2} & >0
\end{aligned}\label{eq:H4bounds}
\ee
and that these conditions are unchanged even if we let $\alpha_i$ and $\beta_i$ be complex (i.e., marginalize over $K_{ijkl}\alpha_i \beta_j \alpha^*_k \beta^*_l$). These are precisely the same conditions we obtained from causality.
The first bound in \Eq{eq:H4bounds} can be obtained from scattering $\phi_{i}\phi_{i}\rightarrow\phi_{i}\phi_{i}$ (all the same state) for any $i\in\{1,2,3,4\}$, in which case ${\cal A}_{H^4}=(c^{H^4}_{1}+c^{H^4}_{2}+c^{H^4}_{3})s^{2}/M^4$. The second bound can be obtained from scattering $\phi_{1}\phi_{3}\rightarrow\phi_{1}\phi_{3}$, $\phi_{1}\phi_{4}\rightarrow\phi_{1}\phi_{4}$, $\phi_{2}\phi_{3}\rightarrow\phi_{2}\phi_{3}$, or $\phi_{2}\phi_{4}\rightarrow\phi_{2}\phi_{4}$, in which case ${\cal A}_{H^4}=(c^{H^4}_{1}+c^{H^4}_{2})s^{2}/ 2M^4$. Finally, the third bound can be obtained from scattering $\phi_{1}\phi_{2}\rightarrow\phi_{1}\phi_{2}$ or $\phi_{3}\phi_{4}\rightarrow\phi_{3}\phi_{4}$, in which case ${\cal A}_{H^4}=c^{H^4}_{2}s^{2}/M^4$.

\subsection{Higgs/field strength cross-quartics}

Let us now bound the coefficients of operators in \Tab{tab:operators2} that give cross-quartic couplings between the Higgs and gauge bosons.
Since we will be considering the speed of fluctuations and scattering amplitudes for bosons of definite type (i.e., $H$, $B$, $W$, or $G$), we will be bounding operators of the type ${\cal O}_i^{H^2 F^2}$, for $F=B,W,G$.

Within a Higgs background $\overline{\partial_\mu \phi_i}$ as described in \Sec{sec:H4}, a $B$, $W$, or $G$ boson's speed of propagation is corrected by the operators in \Tab{tab:operators2} to
\be 
\begin{aligned}
v &= 1-c_1^{H^2 B^2} \frac{(j\cdot k)^{2}}{2k_0^2 M^4}\\
v &= 1-c_1^{H^2 W^2} \frac{(j\cdot k)^{2}}{2k_0^2 M^4}\\
\text{or}\;\; v &= 1-c_1^{H^2 G^2} \frac{(j\cdot k)^{2}}{2k_0^2 M^4},
\end{aligned}
\ee
respectively. 
If we instead have a gauge field background of $B$, $W$, or $G$ as defined in \Sec{sec:singlegaugebounds} and compute the speed of propagation for a perturbation of the Higgs field, we have
\be 
\begin{aligned}
v &= 1 - \frac{E^2}{2M^4} c_1^{H^2 B^2}(X^2 + Y^2)\\
v &= 1 - \frac{E^2}{2M^4} c_1^{H^2 W^2}(X^2 + Y^2)\\
\text{or}\;\;v &= 1 - \frac{E^2}{2M^4} c_1^{H^2 G^2}(X^2 + Y^2),
\end{aligned}
\ee
respectively. Thus, causality requires\footnote{We note that in the mostly-minus metric convention, the opposite of what we use in this work, these bounds all become $c_1^{H^2 F^2} < 0$. This sign change in the bound arises as the associated operators involve an odd number of Lorentz contractions and therefore also change signs between signatures. These are the only bounds presented in this work that vary with signature convention.}
\be 
\begin{aligned}
c_1^{H^2 B^2}&>0 \\
c_1^{H^2 W^2}&>0 \\
c_1^{H^2 G^2}&>0.
\end{aligned}\label{eq:H2F2bounds}
\ee

We can check that analyticity of scattering amplitudes gives the same bounds. Scattering $HB \rightarrow HB$ in the forward limit, we have, at ${\cal O}(s^2)$,
\be 
{\cal A}(s)_{H^2 B^2} =  c_1^{H^2 B^2} \frac{s^2}{2M^4}.
\ee
As in \Sec{sec:H4}, we can ignore any SM contribution to this process, since the contour integral discussed in \Sec{sec:IR} will extract the $s^2$ term in the forward amplitude. 
Similarly for forward $HW\rightarrow HW$ and $HG\rightarrow HG$ scattering, we obtain, at ${\cal O}(s^2)$,
\be
\begin{aligned}
{\cal A}(s)_{H^2 W^2} &=  c_1^{H^2 W^2} \frac{s^2}{2M^4}\\
{\cal A}(s)_{H^2 G^2} &= c_1^{H^2 G^2} \frac{s^2}{2M^4}.
\end{aligned}
\ee
Thus, analyticity implies the same bounds as causality for Higgs/field strength cross-quartics, given in \Eq{eq:H2F2bounds}.

\section{UV Completions}\label{sec:UV}

It is instructive to check whether the EFT coefficients in example theories satisfy the positivity bounds we derived in \Sec{sec:bounds}.
We should expect the EFT coefficients arising from any well defined field theoretic completion to automatically obey these bounds.
By verifying that the inequalities are manifestly satisfied for a large class of completions, we thus perform a nontrivial consistency check of our calculation.

In all of the examples considered in this section, we will analyze the operator coefficients generated at the scale where the new physics is integrated out. This qualifier is important: at lower energy scales, the coefficients will generically run and mix under renormalization-group evolution. While the form of our bounds is independent of these details, if one wanted to convert the operator coefficients appearing in our bounds at a given scale into either the parameters of the UV theory, or the parameters of the EFT at the matching scale, these complications must be accounted for. For the analysis in this section, we could require that the UV theory is sufficiently weakly coupled that, at the scale we imagine performing our scattering experiments, the matching coefficients remain the only relevant contributions to the dimension-8 operators.

\subsection{One-loop completions of gauge field operators}\label{sec:oneloop}

Let us consider introducing some extra massive field $\Phi$, with mass $M$, charged under the SM gauge group.\footnote{For example, the well known Euler-Heisenberg coefficients for ${\rm U}(1)$ electrodynamics, $c_{1,2}^{B^4}$, are both positive when the electron is integrated out~\cite{Heisenberg:1935qt,Schwinger:1951nm,Weisskopf:1996bu}, satisfying the positivity bounds of \Ref{Adams:2006sv}; we wish to substantially generalize this check.}
We will let $\Phi$ be a complex scalar, Dirac fermion, or complex vector field, in an arbitrary irreducible representation of ${\rm SU}(3)_C \otimes {\rm SU}(2)_L \otimes {\rm U}(1)_Y$. Let us write the representation under ${\rm SU}(3)_C$ as ${\bf R}_3$, the representation under ${\rm SU}(2)_L$ as ${\bf R}_2$, and the charge under ${\rm U}(1)_Y$ as $Q$. 

Integrating out $\Phi$ at one loop generates quartic terms among the gauge field strengths of the form given in \Tab{tab:operators}, specifically, the CP-conserving operators.
The corresponding Feynman diagram is shown in \Fig{fig:F4loop}, where in the EFT below $M$, the loop is replaced by an effective four-point vertex, giving rise to the higher-dimension operators we consider.
In \Ref{Quevillon:2018mfl}, the Wilson coefficients for these generalized Euler-Heisenberg terms in the EFT were computed.
To translate from the basis of operators in \Ref{Quevillon:2018mfl} to that in \Tab{tab:operators}, the following identities are useful: for ${\rm SU}(3)$,
\be 
\begin{aligned}
f^{abe}f^{cde}G_{\mu\nu}^{a}G^{c\mu\nu}G_{\rho\sigma}^{b}G^{d\rho\sigma}&=\frac{1}{2}{\cal O}_{1}^{G^{4}}-{\cal O}_{3}^{G^{4}}+\frac{3}{2}{\cal O}_{5}^{G^{4}}\\f^{abe}f^{cde}G_{\mu\nu}^{a}\widetilde{G}^{c\mu\nu}G_{\rho\sigma}^{b}\widetilde{G}^{d\rho\sigma}&=\frac{1}{2}{\cal O}_{2}^{G^{4}}-{\cal O}_{4}^{G^{4}}+\frac{3}{2}{\cal O}_{6}^{G^{4}},
\end{aligned}
\ee
and for ${\rm SU}(2)$,
\be 
\begin{aligned}
\epsilon^{IJM}\epsilon^{KLM}W_{\mu\nu}^{I}W^{K\mu\nu}W_{\rho\sigma}^{J}W^{L\rho\sigma}&={\cal O}_{1}^{W^{4}}-{\cal O}_{3}^{W^{4}}\\
\epsilon^{IJM}\epsilon^{KLM}W_{\mu\nu}^{I}\widetilde{W}^{K\mu\nu}W_{\rho\sigma}^{J}\widetilde{W}^{L\rho\sigma}&={\cal O}_{2}^{W^{4}}-{\cal O}_{4}^{W^{4}}.
\end{aligned}
\ee

\begin{figure}[t]
\begin{center}
\includegraphics[width=6cm]{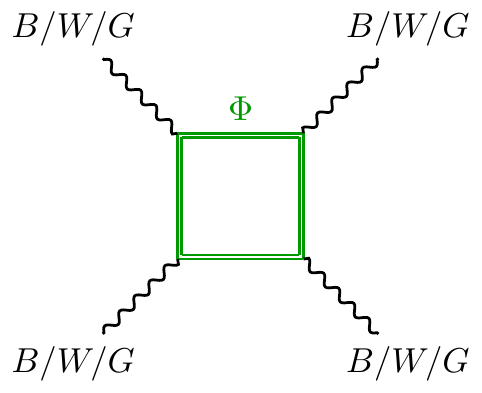}
\end{center}
\caption{One-loop diagram involving a heavy state $\Phi$, which when integrated out will generate a set of CP-even $F^4$ operators, the coefficients of which must obey our bounds.
Here $\Phi$ can be a complex scalar, Dirac fermion, or complex vector field.}
\label{fig:F4loop}
\end{figure}

We can specify the irreducible representations of ${\rm SU}(2)$ and ${\rm SU}(3)$ as ${\bf R}_2 = (r)$ and ${\bf R}_3 = (p,q)$, labeled by their Dynkin indices.\footnote{The number of states in the ${\rm SU}(2)$ multiplet is $r+1$. For the ${\rm SU}(3)$ multiplet, there are $(p+1)(q+1)(p+q+2)/2$ particles in the representation, and one can think of the $(p,q)$ of ${\rm SU}(3)$ as the number of quarks and antiquarks, respectively, in the representation, corresponding to a Young tableau with $p$ columns of one box and $q$ columns of two boxes (or, equivalently, to the number of steps across the top and bottom, respectively, of the multiplet diagram)~\cite{Tanabashi2}.}
Let us define the quadratic invariants for the representations,
\be
\begin{aligned}
I_2(r) &= \frac{1}{12}r(r+1)(r+2)\\
I_2(p,q) &= \frac{1}{48}(p+1)(q+1)(p+q+2)[p(p+3)+q(q+3)+pq],
\end{aligned}\label{eq:I2}
\ee
the cubic anomaly invariant~\cite{Banks:1976yg}, which vanishes for SU(2),
\be
\begin{aligned}
I_3(p,q) = \frac{1}{120}(p+1)(q+1)(p+q+2)(p-q)(p+2q+3)(q+2p+3),
\end{aligned} 
\ee
and the $\Lambda$ constants
\be
\begin{aligned}
\Lambda(r)&=\frac{1}{720}r(r+1)(r+2)(3r^{2}+6r-4)\\
\Lambda(p,q)&=\frac{1}{480}(p+1)(q+1)(p+q+2)[p(p+3)+q(q+3)+pq]\times \\&\qquad \times\left\{ \frac{1}{3}[p(p+3)+q(q+3)+pq]-\frac{1}{2}\right\} .
\end{aligned} \label{eq:Lambdas}
\ee
In terms of these quantities, the $c_i$ as computed for this class of completions are given in \Tab{tab:cs}.
For detailed definitions of these group theoretic quantities, see \App{app:GeneralN}.

\renewcommand{\arraystretch}{1.3}
\begin{table}[htbp]
\noindent \begin{centering}
\begin{tabular}{L{1.1cm} | C{4.8cm} | C{4.8cm} | C{4.8cm}}
 & scalar & fermion & vector\tabularnewline
 \hline
$c_{1}^{B^{4}}$ & $\frac{7}{32}g_{1}^{4}Q^{4}$ & $\frac{1}{2}g_{1}^{4}Q^{4}$ & $\frac{261}{32}g_{1}^{4}Q^{4}$\tabularnewline
  
$c_{2}^{B^{4}}$ & $\frac{1}{32}g_{1}^{4}Q^{4}$ & $\frac{7}{8}g_{1}^{4}Q^{4}$ & $\frac{243}{32}g_{1}^{4}Q^{4}$\tabularnewline
  
$c_{1}^{W^{4}}$ & $g_{2}^{4}\left[\frac{7}{32}\Lambda(\mathbf{R}_{2})+\frac{1}{48}I_{2}(\mathbf{R}_{2})\right]$ & $g_{2}^{4}\left[\frac{1}{2}\Lambda(\mathbf{R}_{2})+\frac{1}{48}I_{2}(\mathbf{R}_{2})\right]$ & $g_{2}^{4}\left[\frac{261}{32}\Lambda(\mathbf{R}_{2})-\frac{3}{16}I_{2}(\mathbf{R}_{2})\right]$\tabularnewline
  
$c_{2}^{W^{4}}$ & $g_{2}^{4}\left[\frac{1}{32}\Lambda(\mathbf{R}_{2})+\frac{1}{336}I_{2}(\mathbf{R}_{2})\right]$ & $g_{2}^{4}\left[\frac{7}{8}\Lambda(\mathbf{R}_{2})+\frac{19}{336}I_{2}(\mathbf{R}_{2})\right]$ & $g_{2}^{4}\left[\frac{243}{32}\Lambda(\mathbf{R}_{2})-\frac{27}{112}I_{2}(\mathbf{R}_{2})\right]$\tabularnewline
  
$c_{3}^{W^{4}}$ & $g_{2}^{4}\left[\frac{7}{16}\Lambda(\mathbf{R}_{2})-\frac{1}{48}I_{2}(\mathbf{R}_{2})\right]$ & $g_{2}^{4}\left[\Lambda(\mathbf{R}_{2})-\frac{1}{48}I_{2}(\mathbf{R}_{2})\right]$ & $g_{2}^{4}\left[\frac{261}{16}\Lambda(\mathbf{R}_{2})+\frac{3}{16}I_{2}(\mathbf{R}_{2})\right]$\tabularnewline
  
$c_{4}^{W^{4}}$ & $g_{2}^{4}\left[\frac{1}{16}\Lambda(\mathbf{R}_{2})-\frac{1}{336}I_{2}(\mathbf{R}_{2})\right]$ & $g_{2}^{4}\left[\frac{7}{4}\Lambda(\mathbf{R}_{2})-\frac{19}{336}I_{2}(\mathbf{R}_{2})\right]$ & $g_{2}^{4}\left[\frac{243}{16}\Lambda(\mathbf{R}_{2})+\frac{27}{112}I_{2}(\mathbf{R}_{2})\right]$\tabularnewline
  
$c_{1}^{G^{4}}$ & $g_{3}^{4}\left[\frac{7}{32}\Lambda({\bf R}_{3})+\frac{1}{96}I_{2}(\mathbf{R}_{3})\right]$ & $g_{3}^{4}\left[\frac{1}{2}\Lambda({\bf R}_{3})+\frac{1}{96}I_{2}(\mathbf{R}_{3})\right]$ & $g_{3}^{4}\left[\frac{261}{32}\Lambda({\bf R}_{3})-\frac{3}{32}I_{2}(\mathbf{R}_{3})\right]$\tabularnewline
  
$c_{2}^{G^{4}}$ & $g_{3}^{4}\left[\frac{1}{32}\Lambda({\bf R}_{3})+\frac{1}{672}I_{2}(\mathbf{R}_{3})\right]$ & $g_{3}^{4}\left[\frac{7}{8}\Lambda({\bf R}_{3})+\frac{19}{672}I_{2}(\mathbf{R}_{3})\right]$ & $g_{3}^{4}\left[\frac{243}{32}\Lambda({\bf R}_{3})-\frac{27}{224}I_{2}(\mathbf{R}_{3})\right]$\tabularnewline
  
$c_{3}^{G^{4}}$ & $g_{3}^{4}\left[\frac{7}{16}\Lambda({\bf R}_{3})-\frac{1}{48}I_{2}(\mathbf{R}_{3})\right]$ & $g_{3}^{4}\left[\Lambda({\bf R}_{3})-\frac{1}{48}I_{2}(\mathbf{R}_{3})\right]$ & $g_{3}^{4}\left[\frac{261}{16}\Lambda({\bf R}_{3})+\frac{3}{16}I_{2}(\mathbf{R}_{3})\right]$\tabularnewline
  
$c_{4}^{G^{4}}$ & $g_{3}^{4}\left[\frac{1}{16}\Lambda({\bf R}_{3})-\frac{1}{336}I_{2}(\mathbf{R}_{3})\right]$ & $g_{3}^{4}\left[\frac{7}{4}\Lambda({\bf R}_{3})-\frac{19}{336}I_{2}(\mathbf{R}_{3})\right]$ & $g_{3}^{4}\left[\frac{243}{16}\Lambda({\bf R}_{3})+\frac{27}{112}I_{2}(\mathbf{R}_{3})\right]$\tabularnewline
  
$c_{5}^{G^{4}}$ & $\frac{1}{32}g_{3}^{4}I_{2}(\mathbf{R}_{3})$ & $\frac{1}{32}g_{3}^{4}I_{2}(\mathbf{R}_{3})$ & $-\frac{9}{32}g_{3}^{4}I_{2}(\mathbf{R}_{3})$\tabularnewline
  
$c_{6}^{G^{4}}$ & $\frac{1}{224}g_{3}^{4}I_{2}(\mathbf{R}_{3})$ & $\frac{19}{224}g_{3}^{4}I_{2}(\mathbf{R}_{3})$ & $-\frac{81}{224}g_{3}^{4}I_{2}(\mathbf{R}_{3})$\tabularnewline
  
$c_{1}^{B^{2}W^{2}}$ & $\frac{7}{16}g_{1}^{2}g_{2}^{2}Q^{2}I_{2}(\mathbf{R}_{2})$ & $g_{1}^{2}g_{2}^{2}Q^{2}I_{2}(\mathbf{R}_{2})$ & $\frac{261}{16}g_{1}^{2}g_{2}^{2}Q^{2}I_{2}(\mathbf{R}_{2})$\tabularnewline
  
$c_{2}^{B^{2}W^{2}}$ & $\frac{1}{16}g_{1}^{2}g_{2}^{2}Q^{2}I_{2}(\mathbf{R}_{2})$ & $\frac{7}{4}g_{1}^{2}g_{2}^{2}Q^{2}I_{2}(\mathbf{R}_{2})$ & $\frac{243}{16}g_{1}^{2}g_{2}^{2}Q^{2}I_{2}(\mathbf{R}_{2})$\tabularnewline
  
$c_{3}^{B^{2}W^{2}}$ & $\frac{7}{8}g_{1}^{2}g_{2}^{2}Q^{2}I_{2}(\mathbf{R}_{2})$ & $2g_{1}^{2}g_{2}^{2}Q^{2}I_{2}(\mathbf{R}_{2})$ & $\frac{261}{8}g_{1}^{2}g_{2}^{2}Q^{2}I_{2}(\mathbf{R}_{2})$\tabularnewline
  
$c_{4}^{B^{2}W^{2}}$ & $\frac{1}{8}g_{1}^{2}g_{2}^{2}Q^{2}I_{2}(\mathbf{R}_{2})$ & $\frac{7}{2}g_{1}^{2}g_{2}^{2}Q^{2}I_{2}(\mathbf{R}_{2})$ & $\frac{243}{8}g_{1}^{2}g_{2}^{2}Q^{2}I_{2}(\mathbf{R}_{2})$\tabularnewline
  
$c_{1}^{B^{2}G^{2}}$ & $\frac{7}{16}g_{1}^{2}g_{3}^{2}Q^{2}I_{2}(\mathbf{R}_{3})$ & $g_{1}^{2}g_{3}^{2}Q^{2}I_{2}(\mathbf{R}_{3})$ & $\frac{261}{16}g_{1}^{2}g_{3}^{2}Q^{2}I_{2}(\mathbf{R}_{3})$\tabularnewline
  
$c_{2}^{B^{2}G^{2}}$ & $\frac{1}{16}g_{1}^{2}g_{3}^{2}Q^{2}I_{2}(\mathbf{R}_{3})$ & $\frac{7}{4}g_{1}^{2}g_{3}^{2}Q^{2}I_{2}(\mathbf{R}_{3})$ & $\frac{243}{16}g_{1}^{2}g_{3}^{2}Q^{2}I_{2}(\mathbf{R}_{3})$\tabularnewline
  
$c_{3}^{B^{2}G^{2}}$ & $\frac{7}{8}g_{1}^{2}g_{3}^{2}Q^{2}I_{2}(\mathbf{R}_{3})$ & $2g_{1}^{2}g_{3}^{2}Q^{2}I_{2}(\mathbf{R}_{3})$ & $\frac{261}{8}g_{1}^{2}g_{3}^{2}Q^{2}I_{2}(\mathbf{R}_{3})$\tabularnewline
  
$c_{4}^{B^{2}G^{2}}$ & $\frac{1}{8}g_{1}^{2}g_{3}^{2}Q^{2}I_{2}(\mathbf{R}_{3})$ & $\frac{7}{2}g_{1}^{2}g_{3}^{2}Q^{2}I_{2}(\mathbf{R}_{3})$ & $\frac{243}{8}g_{1}^{2}g_{3}^{2}Q^{2}I_{2}(\mathbf{R}_{3})$\tabularnewline
  
$c_{1}^{W^{2}G^{2}}$ & $\frac{7}{16}g_{2}^{2}g_{3}^{2}I_{2}(\mathbf{R}_{2})I_{2}(\mathbf{R}_{3})$ & $g_{2}^{2}g_{3}^{2}I_{2}(\mathbf{R}_{2})I_{2}(\mathbf{R}_{3})$ & $\frac{261}{16}g_{2}^{2}g_{3}^{2}I_{2}(\mathbf{R}_{2})I_{2}(\mathbf{R}_{3})$\tabularnewline
  
$c_{2}^{W^{2}G^{2}}$ & $\frac{1}{16}g_{2}^{2}g_{3}^{2}I_{2}(\mathbf{R}_{2})I_{2}(\mathbf{R}_{3})$ & $\frac{7}{4}g_{2}^{2}g_{3}^{2}I_{2}(\mathbf{R}_{2})I_{2}(\mathbf{R}_{3})$ & $\frac{243}{16}g_{2}^{2}g_{3}^{2}I_{2}(\mathbf{R}_{2})I_{2}(\mathbf{R}_{3})$\tabularnewline
  
$c_{3}^{W^{2}G^{2}}$ & $\frac{7}{8}g_{2}^{2}g_{3}^{2}I_{2}(\mathbf{R}_{2})I_{2}(\mathbf{R}_{3})$ & $2g_{2}^{2}g_{3}^{2}I_{2}(\mathbf{R}_{2})I_{2}(\mathbf{R}_{3})$ & $\frac{261}{8}g_{2}^{2}g_{3}^{2}I_{2}(\mathbf{R}_{2})I_{2}(\mathbf{R}_{3})$\tabularnewline
  
$c_{4}^{W^{2}G^{2}}$ & $\frac{1}{8}g_{2}^{2}g_{3}^{2}I_{2}(\mathbf{R}_{2})I_{2}(\mathbf{R}_{3})$ & $\frac{7}{2}g_{2}^{2}g_{3}^{2}I_{2}(\mathbf{R}_{2})I_{2}(\mathbf{R}_{3})$ & $\frac{243}{8}g_{2}^{2}g_{3}^{2}I_{2}(\mathbf{R}_{2})I_{2}(\mathbf{R}_{3})$\tabularnewline
  
$c_{1}^{BG^{3}}$ & $\frac{7}{32}g_{1}g_{3}^{3}QI_{3}(\mathbf{R}_{3})$ & $\frac{1}{2}g_{1}g_{3}^{3}QI_{3}(\mathbf{R}_{3})$ & $\frac{261}{32}g_{1}g_{3}^{3}QI_{3}(\mathbf{R}_{3})$\tabularnewline
  
$c_{2}^{BG^{3}}$ & $\frac{1}{32}g_{1}g_{3}^{3}QI_{3}(\mathbf{R}_{3})$ & $\frac{7}{8}g_{1}g_{3}^{3}QI_{3}(\mathbf{R}_{3})$ & $\frac{243}{32}g_{1}g_{3}^{3}QI_{3}(\mathbf{R}_{3})$\tabularnewline
\end{tabular}
\par\end{centering}
\caption{Wilson coefficients, from \Ref{Quevillon:2018mfl}, for the generalized Euler-Heisenberg operators in \Tab{tab:operators} generated by integrating out a massive complex scalar, Dirac fermion, or complex vector charged under the SM gauge group as described in text. All $c_i$ have been multiplied by $6!\pi^2$, and all $c_i$ not given in this table vanish at one loop for these completions. \label{tab:cs}}
\end{table}

The relevant positivity bounds from \Sec{sec:bounds}, which we repeat here for convenience, are
\be
\begin{aligned}
c_{1}^{B^{4}} & >0&\qquad &&3c_{4}^{G^{4}}+2c_{6}^{G^{4}} & >0\\
c_{2}^{B^{4}} & >0 & &&c_{3}^{B^{2}W^{2}} & >0\\
c_{1}^{W^{4}}+c_{3}^{W^{4}} & >0 && &c_{4}^{B^{2}W^{2}} & >0 \\
c_{2}^{W^{4}}+c_{4}^{W^{4}} & >0 && &c_{3}^{B^{2}G^{2}} & >0\\
3c_{1}^{G^{4}}+3c_{3}^{G^{4}}+c_{5}^{G^{4}} & >0 & &&c_{4}^{B^{2}G^{2}} & >0\\
3c_{3}^{G^{4}}+2c_{5}^{G^{4}} & >0& &&c_{3}^{W^{2}G^{2}} & >0&\\
3c_{2}^{G^{4}}+3c_{4}^{G^{4}}+c_{6}^{G^{4}} & >0& &&c_{4}^{W^{2}G^{2}} & >0.\\
\end{aligned} \label{eq:boundsubset}
\ee
Plugging the $c_i$ from \Tab{tab:cs} for the scalar, fermion, and vector cases into  \Eq{eq:boundsubset}, we find that the left-hand sides of the inequalities each become a positive numerical coefficient times one of the following:
\be 
g_1^4 Q^4,\; g_2^4 \Lambda(\mathbf{R}_2),\; g_3^4 \Lambda(\mathbf{R}_3),\; g_1^2 g_2^2 Q^2 I_2(\mathbf{R}_2),\; g_1^2 g_3^2 Q^2 I_2(\mathbf{R}_3), \;\text{or}\;g_2^2 g_3^2 I_2(\mathbf{R}_2)I_2(\mathbf{R}_3).
\ee
Since $I_2({\bf R}_2)$, $I_2({\bf R}_3)$, $\Lambda({\bf R}_2)$, and $\Lambda({\bf R}_3)$ are all  nonnegative per Eqs.~\eqref{eq:I2} and \eqref{eq:Lambdas}, our bounds are satisfied by this class of completions.

\subsection{Born-Infeld}\label{sec:BI}
Another well-motivated extension of the quartic gauge field operators in \Tab{tab:operators} is the Born-Infeld (BI) action, in which field strengths have an upper limit. Originally formulated in the context of nonlinear Maxwell theory~\cite{Born:1934gh}, the BI action appears in string theory models in which the gauge fields are coupled to brane-localized matter (see Refs.~\cite{Fradkin:1985qd,Tseytlin:1999dj} and refs. therein). The BI action moreover yields amplitudes with structures connected to the gauge/gravity double copy~\cite{Cheung:2017ems,Cheung:2017yef} and possessing special soft limits~\cite{Cheung:2018oki}.

If the SM gauge group is extended to a BI model with universal BI parameter $M^2$, then the action is~\cite{Ellis:2018cos}
\be
 {\cal L}=M^4 \left[ 1- \sqrt{1+\frac{1}{2M^4} F_{\mu\nu}^A F^{A\mu\nu} - \frac{1}{16M^8} (F^A_{\mu\nu}\widetilde F^{A\mu\nu})^2} \right],
\ee
where $A$ runs over all twelve generators of ${\rm SU}(3)\otimes {\rm SU}(2)\otimes{\rm U}(1)$, i.e., $F^A_{\mu\nu} = (B_{\mu\nu},W^I_{\mu\nu},G^a_{\mu\nu})$.
Expanding for small field values, we have the EFT up to dimension-eight operators:
\be 
\begin{aligned}
{\cal L}  =&-\frac{1}{4}B_{\mu\nu}B^{\mu\nu}-\frac{1}{4}W_{\mu\nu}^{I}W^{I\mu\nu}-\frac{1}{4}G_{\mu\nu}^{a}G^{a\mu\nu}\\
 &+\frac{1}{32M^4}\left({\cal O}_{1}^{B^{4}}+{\cal O}_{2}^{B^{4}}+{\cal O}_{1}^{W^{4}}+{\cal O}_{2}^{W^{4}}+{\cal O}_{1}^{G^{4}}+{\cal O}_{2}^{G^{4}}\right)\\
 &+\frac{1}{16M^4}\left({\cal O}_{1}^{B^{2}W^{2}}+{\cal O}_{2}^{B^{2}W^{2}}+{\cal O}_{1}^{B^{2}G^{2}}+{\cal O}_{2}^{B^{2}G^{2}}+{\cal O}_{1}^{W^{2}G^{2}}+{\cal O}_{2}^{W^{2}G^{2}}\right).
\end{aligned}
\ee
Reading off the $c_i$, we see that the bounds in \Eq{eq:boundsubset} are manifestly satisfied.

We could also consider a different BI action, with different BI parameters $M_i^2$,
\be 
\begin{aligned}
{\cal L}  =&\phantom{+}\;\,M_1^4\left[1-\sqrt{1+\frac{1}{2M_1^4}B_{\mu\nu}B^{\mu\nu}-\frac{1}{16M_1^8}\left(B_{\mu\nu}\widetilde{B}^{\mu\nu}\right)^{2}}\right]\\
 &+M_2^4\left[1-\sqrt{1+\frac{1}{2M_2^4}W_{\mu\nu}^{I}W^{I\mu\nu}-\frac{1}{16M_2^8}\left(W_{\mu\nu}^{I}\widetilde{W}^{I\mu\nu}\right)^{2}}\right]\\
 &+M_3^4\left[1-\sqrt{1+\frac{1}{2M_3^4}G_{\mu\nu}^{a}G^{a\mu\nu}-\frac{1}{16M_3^8}\left(G_{\mu\nu}^{a}\widetilde{G}^{a\mu\nu}\right)^{2}}\right],
\end{aligned}
\ee
for which the EFT becomes
\be 
\begin{aligned}
{\cal L}  =&-\frac{1}{4}B_{\mu\nu}B^{\mu\nu}-\frac{1}{4}W_{\mu\nu}^{I}W^{I\mu\nu}-\frac{1}{4}G_{\mu\nu}^{a}G^{a\mu\nu}\\
 &+\frac{1}{32M_1^4}{\cal O}_{1}^{B^{4}}+\frac{1}{32M_1^4}{\cal O}_{2}^{B^{4}}+\frac{1}{32M_2^4}{\cal O}_{1}^{W^{4}}+\frac{1}{32M_2^4}{\cal O}_{2}^{W^{4}}+\frac{1}{32M_3^4}{\cal O}_{1}^{G^{4}}+\frac{1}{32M_3^4}{\cal O}_{2}^{G^{4}},
\end{aligned}
\ee
which also clearly satisfies the bounds in \Eq{eq:boundsubset}.

\subsection{CP-violating terms and completing the square}\label{sec:square}

A striking feature of the bounds derived in \Sec{sec:bounds} is that they come in two distinct forms, either $c>0$, for some CP-even term with coefficient $c$, or $\widetilde c^2 < 4 c_1 c_2$, where $\widetilde c$ is the coefficient of a CP-odd term and $c_{1,2}$ are the positive coefficients of CP-even terms. 
In other words, the bounds take the form of a cone, $\widetilde c^2 + c_-^2 < c_+^2$ for $c_+ >0$, where $c_\pm = c_1 \pm c_2$.
That is, the overall size of CP-odd terms tends to be upper-bounded by the size of CP-even terms appearing in the action; see \Fig{fig:cone}.

\begin{figure}[ht]
\begin{center}
\includegraphics[width=5cm]{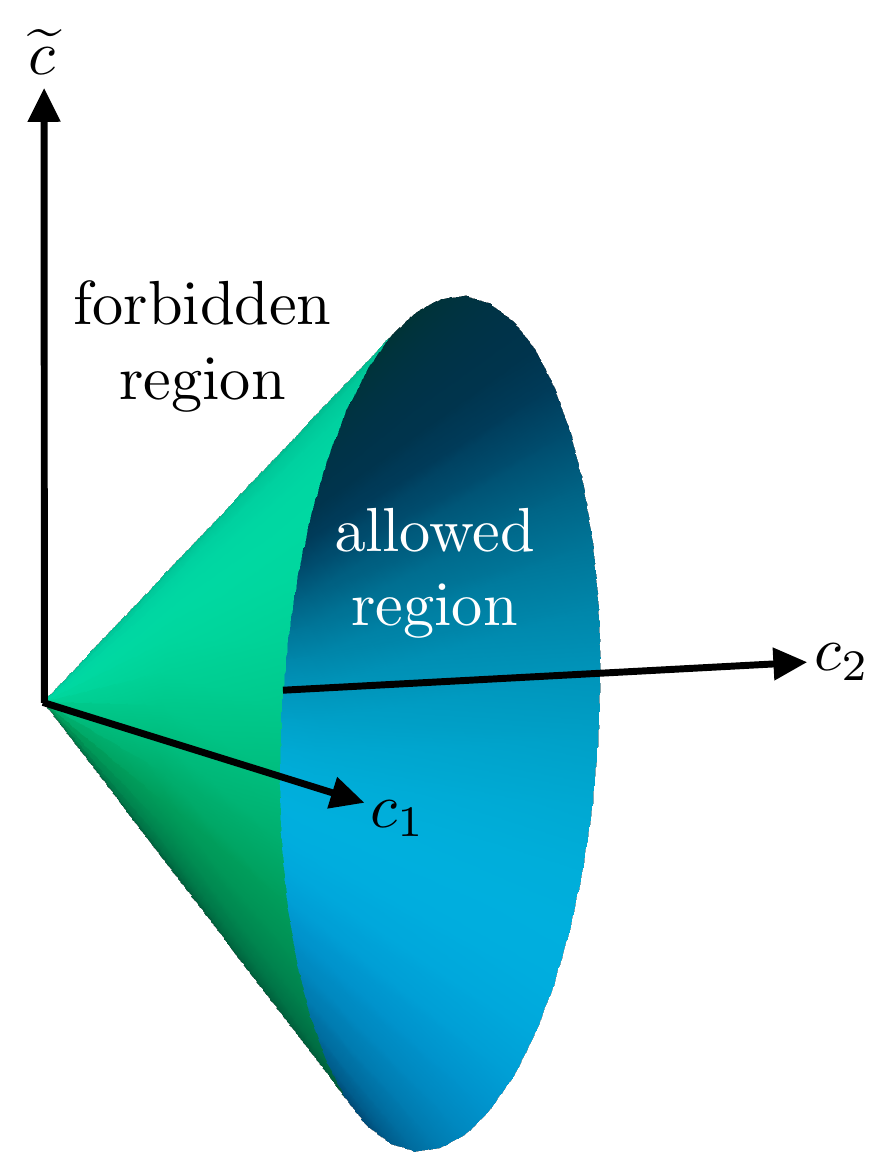}
\end{center}
\caption{General form of the bounds derived in \Sec{sec:bounds}. CP-conserving terms (or linear combinations thereof) have coefficients $c_1$ and $c_2$ bounded to be positive, while the corresponding CP-breaking term has coefficient $\widetilde c$ satisfying $\widetilde c^2 < 4c_1 c_2$.}
\label{fig:cone}
\end{figure}

Let us restrict our attention to the ${\cal O}_i^{B^4}$ terms containing four ${\rm U}(1)$ field strengths. The bounds in \Eq{eq:U1bound}, $c_{1,2}^{B^4}>0$ and $(\widetilde c_1^{B^4})^2 < 4c_1^{B^4} c_2^{B^4}$, imply that it is possible to rewrite the Lagrangian corrections
\be 
\Delta{\cal L}=\frac{1}{M^4}\left[c^{B^4}_{1}(B_{\mu\nu}B^{\mu\nu})^{2}+c^{B^4}_{2}(B_{\mu\nu}\widetilde{B}^{\mu\nu})^{2}+\widetilde{c}^{B^4}_{1}B_{\mu\nu}B^{\mu\nu}B_{\rho\sigma}\widetilde{B}^{\rho\sigma}\right]
\ee
as a sum of perfect squares,
\be 
\Delta{\cal L}=\frac{\alpha^2}{2M^4}\left[(B_{\mu\nu}B^{\mu\nu} + \beta B_{\mu\nu}\widetilde{B}^{\mu\nu})^{2}+\gamma^2 (B_{\mu\nu}B^{\mu\nu} - \beta B_{\mu\nu}\widetilde{B}^{\mu\nu})^{2}\right],\label{eq:U1act}
\ee
where $\alpha,\beta,\gamma$ are real constants chosen such that $\alpha^2(1+\gamma^2) = 2c_1^{B^4}$, $\alpha^2 \beta^2 (1+\gamma^2) = 2c_2^{B^4}$, and $\alpha^2 \beta (1-\gamma^2) = \widetilde c_1^{B^4}$, so $(\widetilde c_1^{B^4})^2/4 c_1^{B^4}c_2^{B^4} = [(1-\gamma^2)/(1+\gamma^2)]^2$ and \Eq{eq:U1bound} guarantees that such a choice of $\alpha,\beta,\gamma \in \mathbb{R}$ exists.

\begin{figure}[t]
\begin{center}
\includegraphics[width=5.2cm]{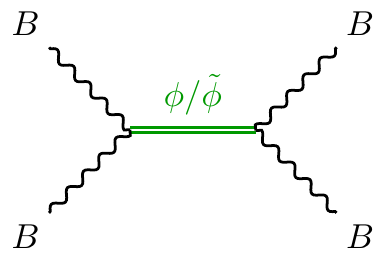}
\end{center}
\caption{Tree-level interaction between four U(1)$_Y$ gauge bosons, mediated by a CP-even scalar $\phi$, a CP-odd scalar $\widetilde{\phi}$, and their mixing.
When these states are integrated out, CP even and odd $B^4$ operators are generated, and their coefficients satisfy our bound on the CP-odd coefficient given in \Eq{eq:U1bound}, in detail $(\tilde{c}_1^{B^4})^2 < 4 c_1^{B^4} c_2^{B^4}$.
}
\label{fig:F4tree}
\end{figure}

The sum-of-squares form of the action in \Eq{eq:U1act} suggests how it can be generated at tree level through mass mixing of a massive dilaton and axion,\footnote{Such a model may be motivated by string theory, where the universal axion and dilaton can be treated as real and imaginary parts of a complex modular parameter\cite{Schwarz:1992tn,Kallosh:2002wj,Conlon:2006tq}, though difficulties remain in engineering mass splitting~\cite{Gao:2015nra}.} coupling a CP-even scalar $\phi$ and CP-odd scalar $\widetilde \phi$ to $B_{\mu\nu}B^{\mu\nu}$ and $B_{\mu\nu}\widetilde B^{\mu\nu}$, respectively:\footnote{As this Lagrangian involves dimension-five operators, it is not a complete UV theory.
However, it is still a UV extension of the theory written in terms of dimension-eight operators, in that its amplitudes will grow more slowly with momentum and its cutoff will be higher if the theory is weakly coupled; focusing on the $\alpha$ coefficient, the cutoff of the original EFT in \Eq{eq:U1act} is at $\sim M/\sqrt{\alpha}$, while that of the theory in \Eq{eq:phitildephi} is at $\sim M/\alpha$.  The latter action can be itself completed in any of the ways theories of axions and dilatons traditionally are without impacting the arguments in this section.
A similar comment applies for the completions considered in Sec.~\ref{sec:treeDH4}.}
\be 
{\cal L} \supset -\frac{M^2}{2} (\phi + \widetilde\phi)^2 - \frac{M^2}{2\gamma^2}(\phi - \widetilde \phi)^2  + \frac{2\alpha}{M}\phi B_{\mu\nu}B^{\mu\nu} + \frac{2\alpha\beta}{M} \widetilde \phi B_{\mu\nu}\widetilde B^{\mu\nu},\label{eq:phitildephi}
\ee
where $\phi +\widetilde \phi$ and $\phi - \widetilde \phi$ are the mass eigenstates, with masses $M$ and $M/|\gamma|$, respectively. (When $\gamma = 1$ the two masses are degenerate, the mass mixing between $\phi$ and $\widetilde \phi$ vanishes, and the CP-odd coefficient $\widetilde c_1^{B^4} = 0$.)
Integrating out $\phi$ and $\widetilde\phi$, effectively replacing the diagram in \Fig{fig:F4tree} by a quartic $B$ vertex, we obtain the interactions in \Eq{eq:U1act}.
Indeed, any scalar tree-level completion of ${\cal O}_{1,2}^{B^4}$ and $\widetilde {\cal O}_1^{B^4}$ will manifestly obey the bounds in \Eq{eq:U1bound}: $c_{1,2}^{B^4}$ will both be positive when integrating out a nontachyonic state coupled at tree level to $B_{\mu\nu}B^{\mu\nu}$ and $B_{\mu\nu}\widetilde B^{\mu\nu}$, respectively, while $\widetilde c_1^{B^4}$ can be generated by a scalar at tree level only through a state that breaks CP by simultaneously coupling to $B_{\mu\nu}B^{\mu\nu}$ and $B_{\mu\nu}\widetilde B^{\mu\nu}$, generating $\widetilde O_1^{B^4}$ through the cross term but also generating $O_{1,2}^{B^4}$.
Beyond the example completion in \Eq{eq:phitildephi}, it would be interesting to understand, from a UV perspective and for more general completions, why CP-violating terms in the EFT are constrained to always come in the form of sums of perfect squares with CP-even terms.
For example, a similar pattern of bounds was found from causality for CP-even and -odd quartic Riemann operators in \Ref{Endlich:2017tqa}.
We leave such investigations to future work.

\subsection{Tree-level completions of $(DH)^4$ operators}\label{sec:treeDH4}

Finally, let us consider the three $(DH)^4$ operators ${\cal O}_{1,2,3}^{H^4}$ given in \Tab{tab:operators2},
\be 
\begin{aligned}
{\cal O}_1^{H^4} &= (D_\mu H^\dagger D_\nu H)(D^\nu H^\dagger D^\mu H)\\
{\cal O}_2^{H^4} &= (D_\mu H^\dagger D_\nu H)(D^\mu H^\dagger D^\nu H)\\
{\cal O}_3^{H^4} &= (D^\mu H^\dagger D_\mu H)(D^\nu H^\dagger D_\nu H).
\end{aligned}
\ee
We will construct a few representative completions of these operators via the tree-level exchange of massive states, as depicted in \Fig{fig:H4tree}, and verify that they obey the three bounds derived in \Eq{eq:H4bounds}, $c_2^{H^4}>0$, $c_1^{H^4} + c_2^{H^4}>0$, and $c_1^{H^4}+c_2^{H^4}+c_3^{H^4}>0$.

\begin{figure}[t]
\begin{center}
\includegraphics[width=5.2cm]{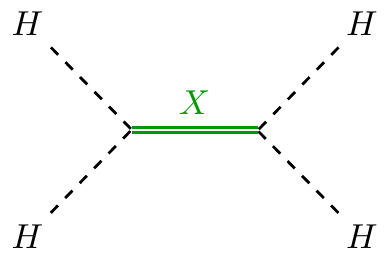}
\end{center}
\caption{Tree-level completion of a $(DH)^4$ operator, where $X$ is a heavy field transforming under a specific representation of SU(2)$_L$.
In the text, we consider a number of examples for $X$ and show in all cases that, when it is integrated out, the coefficients of the $(DH)^4$ operators generated always satisfy our bounds.
}
\label{fig:H4tree}
\end{figure}

First, the simplest example tree completion of a $(DH)^4$ operator is a massive real scalar $\phi$ transforming as a singlet under ${\rm SU}(2)$, with UV action
\be 
{\cal L}\supset-D^{\mu}H^{\dagger}D_{\mu}H-\frac{1}{2}(\partial\phi)^{2}-\frac{1}{2}m^{2}\phi^{2}+\frac{\alpha}{M}\phi(D^{\mu}H^{\dagger}D_{\mu}H),
\ee
where $\alpha$ is an arbitrary real coupling. 
Integrating out $\phi$ generates the Wilson coefficients
\be 
(c_{1}^{H^{4}},c_{2}^{H^{4}},c_{3}^{H^{4}})=\frac{\alpha^{2}}{2m^{2}M^{2}}(0,0,1).\qquad [\text{singlet scalar}]\label{eq:cH41}
\ee

Second, let us consider completion via a massive state $\pi^I$ that transforms as an ${\rm SU}(2)$ triplet:
\be 
{\cal L}\supset-D^{\mu}H^{\dagger}D_{\mu}H-\frac{1}{2}\partial^{\mu}\pi^{I}\partial_{\mu}\pi^{I}-\frac{1}{2}m^{2}\pi^{I}\pi^{I}+\frac{\alpha}{M}\pi^I (D^{\mu}H^{\dagger}\tau^{I}D_{\mu}H).
\ee
Using the Pauli matrix completeness relation,
\be 
(\sigma^{I})_{i}^{\;\;j}(\sigma^{I})_{k}^{\;\;l}=2\delta_{i}^{l}\delta_{k}^{j}-\delta_{i}^{j}\delta_{k}^{l},
\ee
one finds that at low energies $\pi^I$ generates the coefficients
\be 
(c_{1}^{H^{4}},c_{2}^{H^{4}},c_{3}^{H^{4}})=\frac{\alpha^{2}}{8m^{2}M^{2}}(2,0,-1).\qquad[\text{triplet scalar}]\label{eq:cH42}
\ee

Third, let us consider a complex bifundamental scalar $\rho_{ij}$ transforming as a $\boldsymbol{2}\otimes\boldsymbol{2}$ of ${\rm SU}(2)$. In analogy with $H$ and $H^\dagger$ in Eqs.~\eqref{eq:Hdef} and \eqref{eq:Hdaggerdef}, we can write out $\rho_{ij}$ and $\rho^\dagger_{ij}$ in terms of real fields $\sigma_{1,\ldots,8}$,\footnote{As in footnote~\ref{foot:SU2}, we again use the ${\rm SU}(2)$ notation conventions of \Ref{Hays:2018zze}, where each lowered index of $\rho^\dagger_{ij}$ is treated as a $\boldsymbol{2}$ rather than a $\boldsymbol{\bar 2}$.}
\be 
\rho_{ij} = \frac{1}{\sqrt{2}} \begin{pmatrix}
\sigma_{1}+i\sigma_{2} & \sigma_3 +i\sigma_4\\
\sigma_5 +i\sigma_6 & \sigma_7 +i\sigma_8
\end{pmatrix}
 \qquad \text{and}\qquad
\rho^\dagger_{ij} = \frac{1}{\sqrt{2}} \begin{pmatrix}
\sigma_7 - i \sigma_8 & -\sigma_5 + i\sigma_6 \\
-\sigma_3 + i \sigma_4 & \sigma_1 - i \sigma_2
\end{pmatrix},
\ee
and define $\rho^{ij} = \epsilon^{ik} \epsilon^{jl} \rho_{kl}$ and $\rho^{\dagger ij} = \epsilon^{ik} \epsilon^{jl} \rho^\dagger_{kl}$.
Let us then consider the action
\be
{\cal L} \supset - D^\mu H^\dagger D_\mu H - \partial^\mu \rho^\dagger_{ij}\partial_\mu \rho^{ij} - m^2 \rho^\dagger_{ij} \rho^{ij} + \frac{1}{M}(\alpha D^\mu H^\dagger_i \rho^{ij} D_\mu H^\dagger_j + \alpha^* D^\mu H^i \rho^\dagger_{ij} D_\mu H^j ).
\ee
Since $\rho_{ij}^\dagger \rho^{ij} = +\frac{1}{2} \sum_{n=1}^8 \sigma_n^2$, the action contains no ghosts or tachyons.
Here, $\alpha$ is an arbitrary complex unitless coupling, and one van verify that the interaction term is self-hermitian.
Integrating out $\rho_{ij}$, we obtain the EFT coefficients
\be 
(c_{1}^{H^{4}},c_{2}^{H^{4}},c_{3}^{H^{4}})=\frac{|\alpha|^2}{m^{2}M^{2}}(0,1,0).\qquad[\text{bifundamental scalar}]\label{eq:cH43}
\ee

\begin{figure}[t]
\begin{center}
\includegraphics[width=8cm]{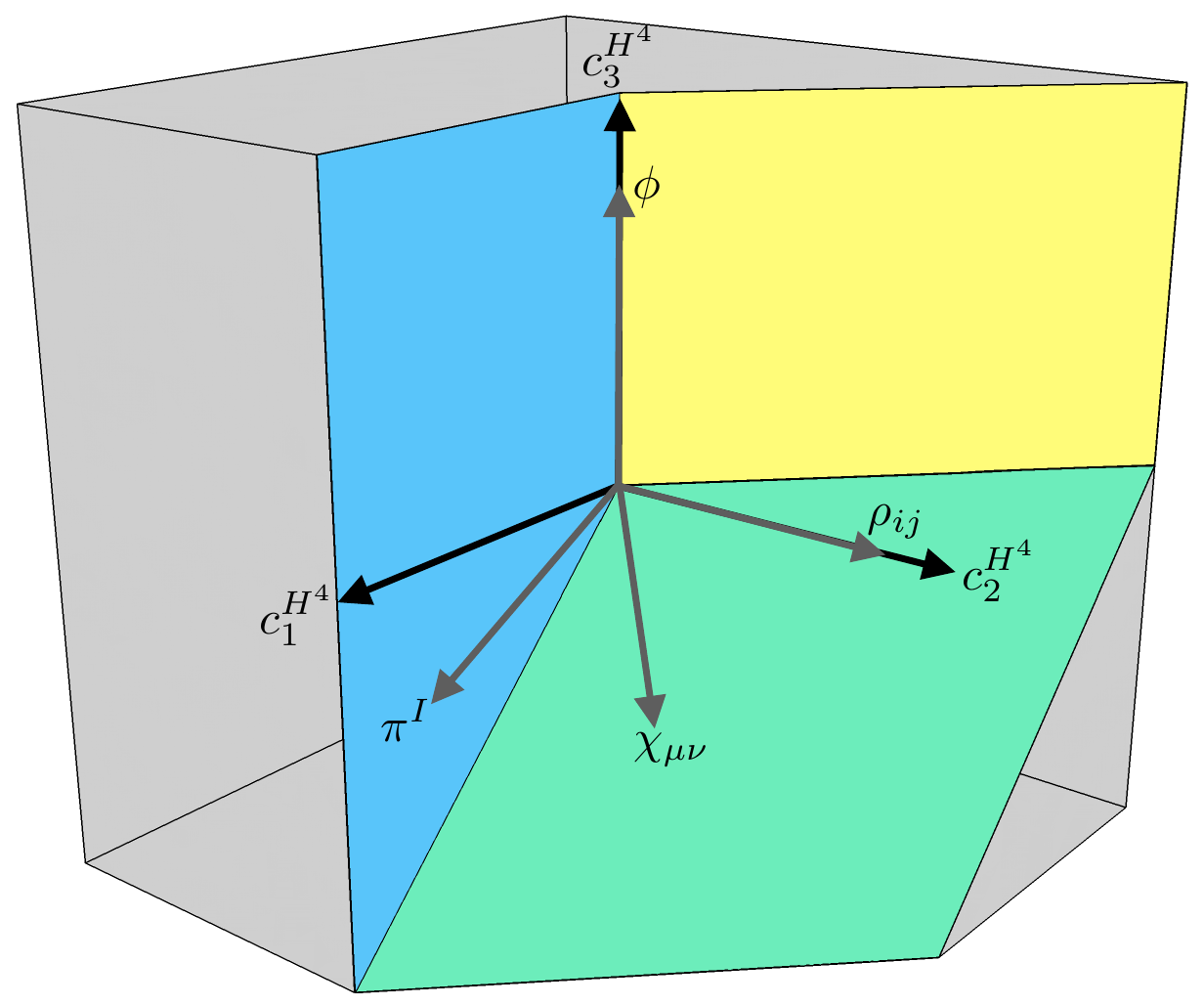}
\end{center}
\caption{Bounds on the three $(DH)^{4}$ operator coefficients from \Eq{eq:H4bounds}: $c^{H^4}_{2}>0$ (blue), $c^{H^4}_{1}+c^{H^4}_{2}>0$ (yellow), and $c^{H^4}_{1}+c^{H^4}_{2}+c^{H^4}_{3}>0$ (green). Gray arrows indicate the vectors of Wilson coefficients generated in example tree-level completions, all lying in the interior or on the boundary of the allowed space, with various massive states coupling to $(DH)^2$: a scalar singlet $\phi$, scalar triplet $\pi^I$, bifundamental scalar $\rho_{ij}$, and symmetric spin-two field $\chi_{\mu\nu}$. The forbidden region is shaded in gray.}
\label{fig:Higgscone}
\end{figure}

A final well-motivated example completion is a massive symmetric spin-two
state $\chi_{\mu\nu}$, coupled to the Higgs stress tensor $T_{\mu\nu}$
akin to a massive graviton (as would naturally occur in, e.g., a Kaluza-Klein
model):
\be 
{\cal L}\supset-D^{\mu}H^{\dagger}D_{\mu}H+{\cal L}_{{\rm FP}}+\frac{\alpha}{M}\chi^{\mu\nu}T_{\mu\nu},
\ee
where ${\cal L}_{{\rm FP}}$ is the Fierz-Pauli action~\cite{Hinterbichler:2011tt}, $\alpha$ is real, and we can
take $T_{\mu\nu}=D_{\mu}H^{\dagger}D_{\nu}H+D_{\nu}H^{\dagger}D_{\mu}H-g_{\mu\nu}(D^{\rho}H^{\dagger}D_{\rho}H)$, neglecting Higgs mass and potential terms as we are interested
in specifically generating ${\cal O}_{1,2,3}^{H^{4}}$ enumerated
above. 
Integrating out $\chi_{\mu\nu}$, we obtain the EFT operators
with coefficients
\be 
(c_{1}^{H^{4}},c_{2}^{H^{4}},c_{3}^{H^{4}})=\frac{\alpha^{2}}{3m^{2}M^{2}}(3,3,-2).\qquad[\text{symmetric tensor}]\label{eq:cH44}
\ee

The bounds in \Eq{eq:H4bounds} delineate a the triangular cone defining the allowed space of $c_{1,2,3}^{H^4}$; see \Fig{fig:Higgscone}.
We find, as required, that the sets of Wilson coefficients explored in the examples in Eqs.~\eqref{eq:cH41}, \eqref{eq:cH42}, \eqref{eq:cH43}, and \eqref{eq:cH44} correspond to vectors lying either on the edge of or interior to this cone.

\section{Phenomenological Consequences}\label{sec:pheno}

Searching for deviations from the SM is one of the driving forces for all of particle physics.
The bounds derived in this work demarcate the space of experimentally accessible deviations into those that can be associated with well-defined UV completions and those that cannot.
Accordingly, the bounds represent significantly constraining theoretical priors on the SMEFT parameter space, which would be particularly appropriate in the context of global analyses~\cite{Hartland:2019bjb,vanBeek:2019evb}.
Conversely, any conclusively-measured experimental violation of our bounds would have powerful implications for fundamental assumptions about physics, falsifying low-energy causality/locality, Lorentz invariance, and, by extension, both low-energy quantum field theory and perturbative string theory~\cite{ArkaniHamed:2006dz}.
Performing such a direct test of our bounds is a challenging endeavor.
Not only does it require a discovery of new physics generated by the SMEFT, for many of our bounds a definitive determination of the sign as well as the magnitude is required.\footnote{Extracting the sign of the operator requires an observable linear in the operator coefficient.
Some observables, such as EDMs, are automatically linear in the coefficient, but for others extracting the sign will require interference with an associated SM contribution.}
In spite of the challenges, the possibility to test the fundamental pillars of modern physics represents a unique motivation for experiments sensitive to the operators we bound.

The full reach of experimental effects associated with the operators considered in this work is vast.
Given this scope, in this section we will restrict ourselves to a brief exploration of searches where our bounds have immediate application.
One example of such a search is the measurement of aQGCs at the LHC.
Such searches are expected to represent a viable discovery channel for several of the operators we have considered, making our bounds immediately relevant.
Beyond aQGCs, we also outline how our bounds are likely to have consequences for additional collider searches, as well as other precision probes such as neutron EDM measurements.

As demonstrated in Sec.~\ref{sec:UV}, a generic UV completion will satisfy our bounds.
For this reason, we do not see our results as immediate probes of the viable BSM model building space.
To take an example, a theory where dark matter is made up of bound states of color octet fermions~\cite{DeLuca:2018mzn} will automatically satisfy our bounds, as shown in Sec.~\ref{sec:oneloop}.
Instead, our bounds are most applicable in the context of the SMEFT, which will be the focus of this section.

Before presenting explicit examples, we mention a caveat that must be kept in mind whenever constraints on the SMEFT are being considered.
Whenever a limit on the scale $M$ of higher-dimension operators is quoted, one must ensure that the limit was derived within the regime of validity for the EFT.
In particular, if the measurement was performed at an invariant mass scale $\sqrt{s}$, then unless $\sqrt{s} < M$, the EFT limit was derived in a region where the theory violated perturbative unitarity, and can no longer be trusted.
This point is often relevant for SMEFT constraints at the LHC---for a few examples see Refs.~\cite{Eboli:2000ad,Abazov:2013opa,Chatrchyan:2013akv,Khachatryan:2016mud,Aad:2016sau}---but must always be considered when EFT constraints are discussed.

\subsection{Anomalous quartic gauge-boson couplings}\label{sec:aQGCs}

Through a subset of the operators we have considered, BSM physics can induce anomalous corrections to the tree-level SM quartic gauge-boson couplings $WWWW$, $WWZZ$, $WW \gamma \gamma$, and $WW Z \gamma $ or even induce vertices not observed in the SM at tree level such as $ZZZZ$.
The set of all such processes are collectively referred to as aQGCs.\footnote{Additional quartic gauge couplings that are not always considered within the aQGC framework can also be probed at the LHC.
For example, constraints on dimension-eight operators associated with $\gamma \gamma \to \gamma \gamma$ and $gg \to \gamma \gamma$ have been considered in the literature~\cite{Ellis:2017edi,Ellis:2018cos}, although we will not discuss them further.}
Importantly, the LHC can probe these vertices and, accordingly, the corrections that dimension-eight operators induce.
For example, the $WWWW$ and $WWZZ$ vertices can be probed using channels such as $q q \to qq WW$ and $qq \to qq ZZ$.
This is illustrated in \Fig{fig:Feynman} for an example Feynman diagram contributing to these processes, where dimension-eight operators can provide corrections to the highlighted four-point vertex.

\begin{figure}[t]
\begin{center}
\includegraphics[width=6cm]{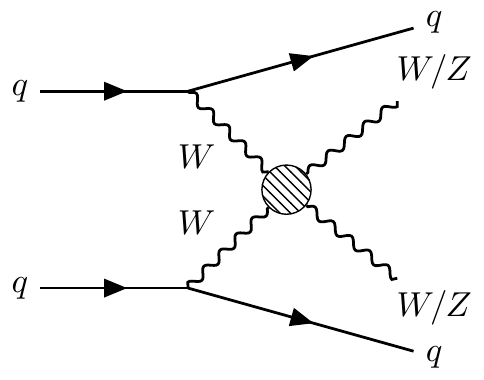}
\end{center}
\caption{Example Feynman diagram for the contribution of aQGCs to the $qq\rightarrow qqWW$ or $qq\rightarrow qqZZ$ channel.}
\label{fig:Feynman}
\end{figure}

As pointed out in \Ref{Eboli:2006wa}, aQGC searches represent a powerful channel for probing a number of dimension-eight SMEFT operators.
Naively, one may worry that any effects will be swamped by the contribution from the dimension-six bosonic operators shown in Table~\ref{tab:dim6}, which, when squared, will also contribute to aQGCs.
Yet a subset of these dimension-six operators also induce anomalous triple gauge-boson couplings, on which the LHC has significantly extended the constraints from LEP~\cite{Butter:2016cvz} (see also \Ref{Henning:2018kys}), potentially leaving room for aQGCs from dimension-eight operators.
Deviations from the SM in these channels have been searched for at both ATLAS~\cite{Aaboud:2017bwk,Aaboud:2018ddq,Aaboud:2019nmv} and CMS~\cite{Sirunyan:2017fvv,Sirunyan:2017ret,Sirunyan:2019ksz}; for more detail see~Refs.~\cite{Degrande:2013rea,Green:2016trm,Rauch:2016pai}.
At the moment no anomalies have been observed, and assuming $\mathcal{O}(1)$ couplings, these searches are currently probing new physics scales from the hundreds of GeV to TeV range.
Importantly, however, if a deviation from the SM were to emerge in these channels, it could represent a sign of new physics first appearing through dimension-eight operators.

In addition to pointing out the importance of aQGCs for dimension-eight operators, \Ref{Eboli:2006wa} also introduced the commonly-used set of operators for these effects.
This list, which has since received several corrections (see, for example,~\Ref{Rauch:2016pai} for details), contains 18 CP-even operators, all of which are related to operators we considered above.
Conventionally, the 18 are divided into three types of operators:
\begin{enumerate}
\item 3 scalar or S-type operators of the form $(DH)^4$,
\item 7 mixed or M-type operators, schematically $(DH)^2 F^2$, and
\item 8 tensor or T-type operators, such as $F^4$.
\end{enumerate}
As aQGCs exclude gluonic vertices, in the above $F$ can be $B$ or $W$.
A complete mapping between this notation and the basis of operators used in the present work is possible and will be provided below.
We note in passing that our basis for such processes includes three additional independent operators that are missing from the standard aQGC list.
In detail, while we also have three operators of the form $(DH)^4$, as shown in Table~\ref{tab:operators2} we have 8 M-type operators such as $(DH)^2 B^2$, $(DH)^2 W^2$, or $(DH)^2 (BW)$, and we have 10 T-type operators as can be seen in Table~\ref{tab:operators}.
One can show that all the operators in our basis are indeed independent~\cite{Morozov}, so the aQGC operator basis commonly used in the literature is incomplete.
A complete basis is crucial, since an incomplete operator basis could in principle lead to conflation and/or overlooking of distinct signals of new physics.

Before converting our bounds into the conventional aQGC language, let us comment on the related work in Refs.~\cite{Zhang:2018shp,Bi:2019phv}, which also considered the question of positivity constraints on these operators.
Those works determined the restrictions associated with the forward limit of all possible $2 \to 2$ electroweak boson scatterings, such as $\gamma \gamma \to \gamma \gamma$ or $\gamma W^{\pm} \to \gamma W^{\pm}$.
This can be contrasted with the approach we have taken in this work, where we have restricted our attention to the scattering of bosons in the limit of unbroken electroweak symmetry; this choice is natural for theories with completions above the electroweak scale, and we accordingly organized our SMEFT operators in the unbroken basis using $B$ and $W^I$.
In \Ref{Zhang:2018shp}, the authors considered 18 aQGC operators, with the results in their Table 1 stated using a one-at-a-time approach, where limits are derived on a single operator assuming all others vanish.
The work of \Ref{Zhang:2018shp} also considered the more general case where multiple operators at a time are allowed to be nonzero, the results of which were then expanded upon significantly in \Ref{Bi:2019phv}.
Given the different starting points used to derive our bounds, as contrasted to those in \Ref{Zhang:2018shp}, instead of comparing each of our full expressions, we will consider just results restricted to a single operator and show that our bounds are generally in agreement.

We caution that the one-at-a-time approach can lead to results that must be interpreted carefully.
Indeed, unlike our bounds, which we exhaustively verified in Sec.~\ref{sec:UV} to be consistent with a large suite of UV completions, bounds derived in a one-at-a-time approach will generically be violated by these checks.
We will provide several explicit examples of this below.
The loophole is that the bounds are always violated in a context where at least one additional operator is also generated, implying a contradiction with the basic one-at-a-time assumption.
This highlights that one-at-a-time bounds are not generic conditions that must be satisfied by any UV completion, but rather conditions that can only possibly apply to particular high-scale theories that, when the UV degrees of freedom are integrated out, generate very simple IR EFTs. 
As we saw in Sec.~\ref{sec:UV}, such completions are highly nongeneric.
Even for those theories where this assumption is satisfied, however, below the matching scale renormalization-group evolution will also generically induce operator mixing, and therefore, at the relevant experimental scale, invalidate the single-operator assumption.
For such reasons, one-at-a-time results must be interpreted carefully, and often have a narrow range of applicability.

Such theoretical considerations do not immediately address the practical consideration underlying the one-at-a-time approach: it is not straightforward to determine, and then present in a useful manner, limits on the 18-dimensional space spanned by the aQGC operators.
Fundamentally, it is a challenge to find enough observables with sufficient statistics to disentangle the various operators.
Motivated by this concern, experimental limits are almost exclusively quoted on single operators, although there have also been efforts to extend this to two operators; see for example~\Ref{Sirunyan:2019ksz}.
In spite of this, theoretical consistency only requires that the complete basis of appropriate operators be considered when deriving the bounds; as we have seen many times above, individual bounds themselves usually only apply to a significantly smaller subset of operators.
The same will be true when our bounds are converted to the space of aQGC operators.
We will find many examples of bounds that apply to a single operators, or to small numbers of operators, and, unlike for one-at-a-time results, our bounds apply to all possible UV completions.

With the above considerations in mind, let us return to converting our operators and bounds into the conventional notation of aQGCs, considering the S-type, M-type, and T-type operators in turn.
To begin with, the three S-type operators in the notation of Refs.~\cite{Rauch:2016pai,Zhang:2018shp,Bi:2019phv} map directly onto our $(DH)^4$ operators given in Table~\ref{tab:operators2} as follows:
\be
\mathcal{O}_{S,0} = \mathcal{O}^{H^4}_2,\qquad
\mathcal{O}_{S,1} = \mathcal{O}^{H^4}_3,\qquad\text{and}\qquad
\mathcal{O}_{S,2} = \mathcal{O}^{H^4}_1.
\ee
Using this mapping, our constraints on these operators given in \Eq{eq:H4bounds} become
\be\begin{aligned}
c_{S,0} + c_{S,1} + c_{S,2} &> 0 \\
c_{S,0} + c_{S,2} &> 0 \\
c_{S,0} &> 0.\label{eq:aQGCSbounds}
\end{aligned}\ee
These results can be contrasted with the one-at-a-time bounds derived in Table 1 of \Ref{Zhang:2018shp}, which required that each of these coefficients be individually positive.
Looking back at the results in Sec.~\ref{sec:treeDH4}, we can see that two of the four example completions discussed in that section have one of these coefficients negative.
These same two completions also generate multiple operators, so these provide examples of the inconsistency associated with interpreting operator bounds on a one-at-a-time basis.

Turning to the M-type operators, our bounds impact three of the seven operators in the commonly-used partial aQGC basis.
Those operators, in the notation of Refs.~\cite{Rauch:2016pai,Zhang:2018shp,Bi:2019phv}, can be converted to our basis in Table~\ref{tab:operators2} using
\begingroup
\allowdisplaybreaks
\begin{align}
\mathcal{O}_{M,0} &= - \frac{g_2^2}{2} \, \mathcal{O}^{H^2 W^2}_2 \nonumber \\
\mathcal{O}_{M,1} &= \frac{g_2^2}{2} \, \mathcal{O}^{H^2 W^2}_1 \nonumber  \\
\mathcal{O}_{M,2} &= - \frac{g_1^2}{4}\, \mathcal{O}^{H^2 B^2}_2 \nonumber  \\
\mathcal{O}_{M,3} &= \frac{g_1^2}{4}\, \mathcal{O}^{H^2 B^2}_1 \\
\mathcal{O}_{M,4} &= - \frac{g_1 g_2}{2} \, \mathcal{O}_1^{H^2 BW} \nonumber  \\
\mathcal{O}_{M,5} &= \frac{g_1g_2}{4} ( i\, \mathcal{O}_2^{H^2 BW} - \mathcal{O}_3^{H^2 BW} ) \nonumber  \\
\mathcal{O}_{M,7} &= \frac{g_2^2}{4} ( 2\mathcal{O}_3^{H^2 W^2} - \mathcal{O}_1^{H^2 W^2}), \nonumber 
\end{align}
\endgroup
where $g_{1,2}$ are the U(1)$_Y$ and SU(2)$_L$ couplings, respectively.
Accordingly, our two relevant bounds in \Eq{eq:H2F2bounds} convert to a requirement that
\be\begin{aligned}
2c_{M,1} - c_{M,7}&> 0 \\
c_{M,3}&> 0.\label{eq:aQGCMbounds}
\end{aligned}\ee
Recall that the overall sign of these operators, and hence the bounds, varies between metric conventions.
After accounting for the fact we use the mostly-plus convention in this work, these bounds are also consistent with those in \Ref{Zhang:2018shp}.

Finally, for the T-type operators, we will make use of the following conversion from the notation of Refs.~\cite{Rauch:2016pai,Zhang:2018shp,Bi:2019phv} to a subset of the operators in our basis, as listed in Table~\ref{tab:operators}:
\be\begin{aligned}
\mathcal{O}_{T,0} &= \frac{g_2^4}{4}\, \mathcal{O}^{W^4}_1 \\
\mathcal{O}_{T,1} &= \frac{g_2^4}{4}\, \mathcal{O}^{W^4}_3 \\
\mathcal{O}_{T,2} &= \frac{g_2^4}{16} (\mathcal{O}^{W^4}_1 + \mathcal{O}^{W^4}_3 + \mathcal{O}^{W^4}_4 ) \\
\mathcal{O}_{T,5} &= \frac{g_1^2 g_2^2}{8}\, \mathcal{O}_1^{B^2 W^2} \\
\mathcal{O}_{T,6} &= \frac{g_1^2 g_2^2}{8}\, \mathcal{O}_3^{B^2 W^2} \\
\mathcal{O}_{T,7} &= \frac{g_1^2 g_2^2}{32} (\mathcal{O}_1^{B^2 W^2} + \mathcal{O}_3^{B^2 W^2} + \mathcal{O}_4^{B^2 W^2} ) \\
\mathcal{O}_{T,8} &= \frac{g_1^4}{16}\,  \mathcal{O}_1^{B^4} \\
\mathcal{O}_{T,9} &=  \frac{g_1^4}{64} (2 \mathcal{O}_1^{B^4} + \mathcal{O}_2^{B^4} ). 
\end{aligned}\ee
Our constraints given in Eqs.~\eqref{eq:SU2bound}, \eqref{eq:U1bound}, and \eqref{eq:crossbound} then imply
\be\begin{aligned}
2 c_{T,0} + 2c_{T,1} + c_{T,2} &> 0 \\
4 c_{T,6} + c_{T,7} &> 0 \\
c_{T,7} &> 0 \\
2 c_{T,8} + c_{T,9} &> 0 \\
c_{T,9} &> 0.\label{eq:aQGCTbounds}
\end{aligned}\ee

Moreover, our bound $c_2^{W^4} + c_4^{W^4} >0$ from \Eq{eq:SU2bound} cannot be converted into aQGC notation, as the conventional aQGC basis is incomplete, missing the operator $\mathcal{O}_2^{W^4}$.
The majority of the T-type bounds in \Eq{eq:aQGCTbounds} are consistent with those in Table 1 of \Ref{Zhang:2018shp}.
The only exception is that we require $c_{T,7} > 0$, whereas when those authors restricted their results to a single operator, they found $c_{T,7} = 0$. 
In that work, the authors also required that the coefficient of $\mathcal{O}_{T,5}$, or equivalently our $\mathcal{O}_1^{B^2 W^2}$, vanished.
Looking back to the relevant row in Table~\ref{tab:cs}, we see that any one-loop completion of these operators that involves a scalar, fermion, or vector that is charged under both SU(2)$_L$ and U(1)$_Y$ will generate a nonzero coefficient for both ${\cal O}_{T,5}$ and ${\cal O}_{T,7}$, in conflict with the one-at-a-time bounds in \Ref{Zhang:2018shp}.
These completions are standard quantum field theories and are analogous to the completion of the Euler-Heisenberg action by the electron loop in quantum electrodynamics~\cite{Quevillon:2018mfl}, so these models should not be forbidden by any consistent set of positivity bounds, providing another explicit example of the limitations of one-at-a-time results.

\begin{figure}[t]
\begin{center}
\includegraphics[width=7.0cm]{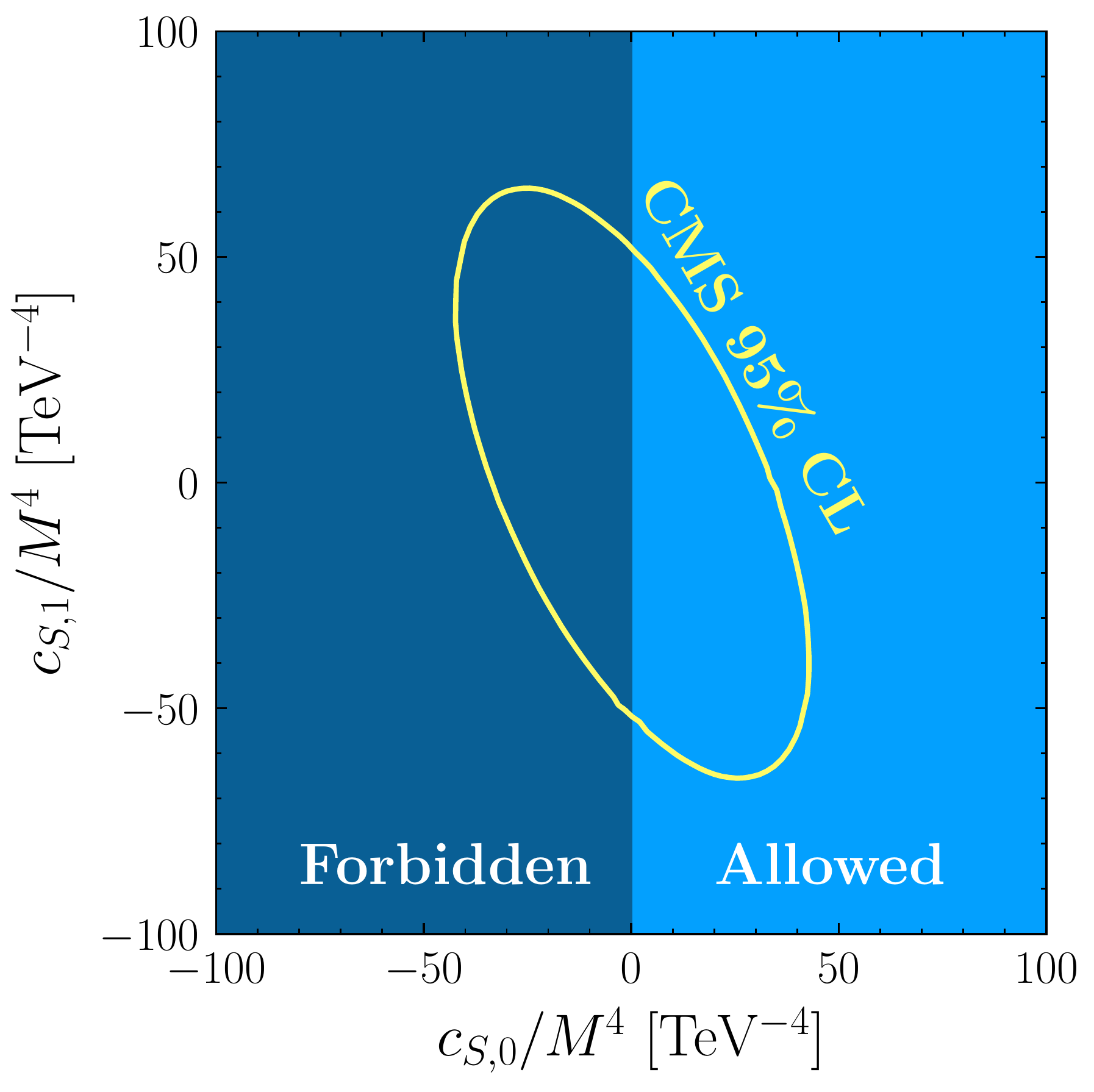} 
\hspace{1cm}
\includegraphics[width=7.0cm]{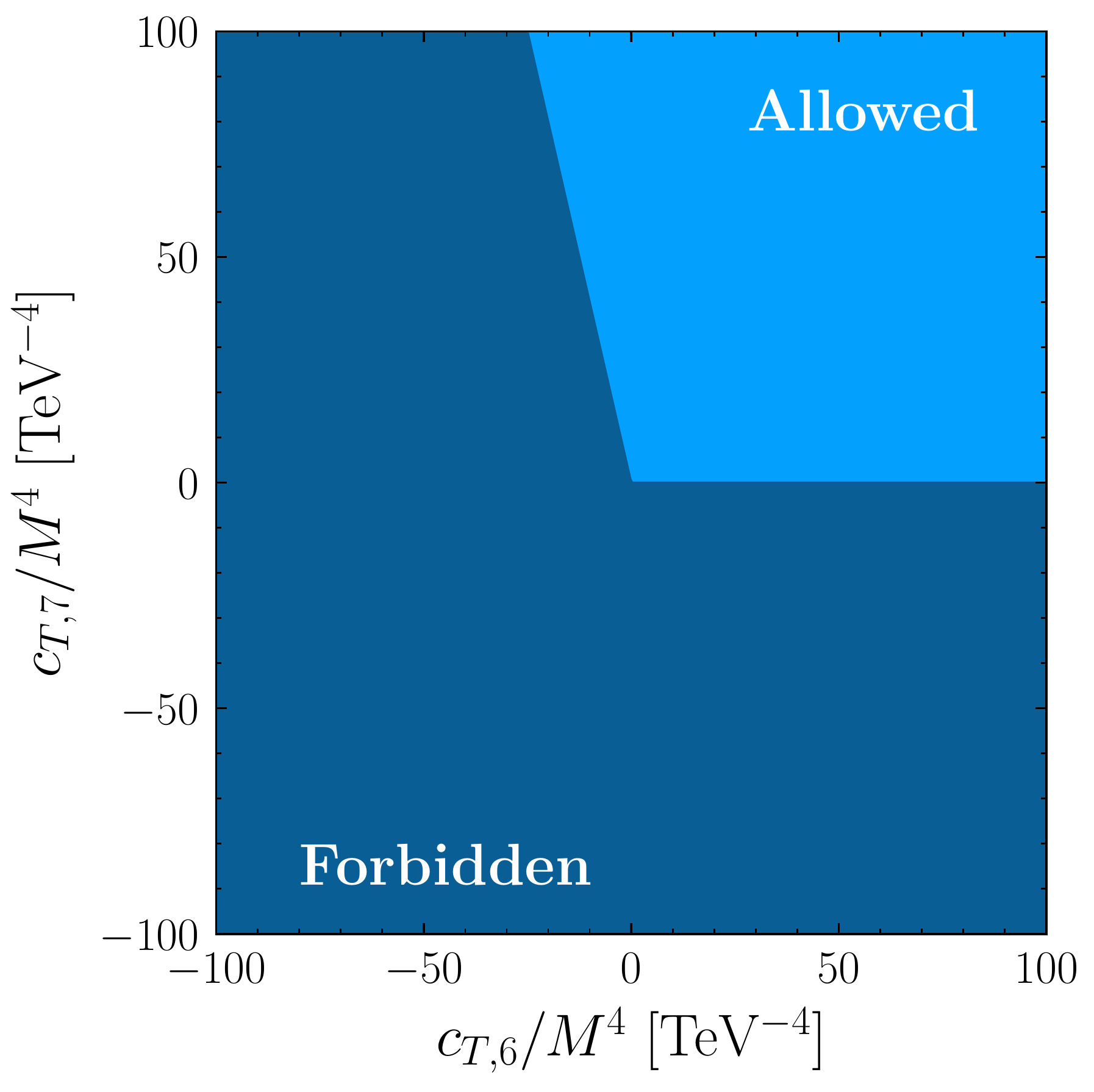}
\end{center}
\caption{Two examples of the allowed and forbidden aQGC parameter space, as demarcated by our bounds.
In each figure we have marginalized over the operators that are not shown.
On the left we show the allowed parameter space in the $c_{S,0}$ and $c_{S,1}$ plane, applying the bounds in \Eq{eq:aQGCSbounds}.
Note that $c_{S,0}+c_{S,1}+c_{S,2}>0$ provides no additional information in this plane once we marginalize over $c_{S,2}$.
In this space, we show the 95\% confidence level (CL) limits derived by CMS~\cite{Sirunyan:2019ksz} under the assumption that $c_{S,2}$ precisely vanishes.
On the right, we provide an example of two parameters relevant for aQGCs where our bounds are more restrictive, $c_{T,6}$ and $c_{T,7}$, as determined from \Eq{eq:aQGCTbounds}.
These figures represent an explicit realization of the schematic goal of our work shown in Fig.~\ref{fig:bounds-schematic}.
}
\label{fig:bounds-aQGC}
\end{figure}

Having converted our bounds into the aQGC notation, we see from Eqs.~\eqref{eq:aQGCSbounds}, \eqref{eq:aQGCMbounds}, and \eqref{eq:aQGCTbounds} that we are able to demand the positivity of four of the 18 commonly-used aQGC couplings individually, with a further five coefficients entering bounds containing two operators and an additional four coefficients entering only bounds containing three operators.
Two depictions of the allowed space are shown in Fig.~\ref{fig:bounds-aQGC} and represent two-dimensional views of the parameter space where the unseen parameters have been marginalized over.
As we derived the bounds while considering all possible four-field-strength operators, these constraints must be satisfied by any consistent UV completion.
Nevertheless, while our bounds apply to every UV completion, not every UV completion will generate only the combinations of operator coefficients that appear in our bounds.
This point is manifest in Table~\ref{tab:cs}, where we see that introducing a new field at one loop that is charged under the SM will generically yield a large number of higher-dimension operators.

At this stage, we have ten bounds that constrain 13 of the aQGC operators (plus one additional bound that cannot be translated as the standard aQGC basis is incomplete).
This can be contrasted with Refs.~\cite{Zhang:2018shp} and \cite{Bi:2019phv}, where the authors gave bounds on 16 and 18 operators, respectively.
One point of difference is that in the present work we have restricted ourselves to the scattering of individual gauge-bosons in the unbroken phase, with colors chosen subject to the condition in \Eq{eq:T12commute}.
A possibility for going beyond this is to consider scattering a superpositions of gauge bosons, which is particularly well motivated given that the SM $\gamma$, $Z$, and $W^{\pm}$ are themselves superpositions of the unbroken fields.
Fully mapping out in detail the space of all positivity bounds on the operators in Tables~\ref{tab:operators} and \ref{tab:operators2} arising from arbitrary superpositions of $B$, $W$, $G$, and $H$ fields is beyond the scope of our present work, but would proceed similarly to the approach we used for scalars in Sec.~\ref{sec:H4}.
To provide a flavor of the additional constraints that can be derived from such an approach, however, let us consider a simple example of scattering states of the form $(B + W^3)/\sqrt{2}$, which would be the photon in a theory with $\theta_{\scriptscriptstyle W} = \pi/4$.
Doing so, we arrive at the following additional bounds on the operators in Table~\ref{tab:operators}:
\be
\begin{aligned}
&c_1^{B^4} + c_1^{W^4} + c_3^{W^4} + c_1^{B^2 W^2} + c_3^{B^2 W^2} > 0 \\
&c_2^{B^4} + c_2^{W^4} + c_4^{W^4} + c_2^{B^2 W^2} + c_4^{B^2 W^2} > 0 \\[3mm]
&(\tilde{c}_1^{B^4} + \tilde{c}_1^{W^4} + \tilde{c}_2^{W^4} + \tilde{c}_1^{B^2 W^2} + \tilde{c}_2^{B^2 W^2} + \tilde{c}_3^{B^2 W^2})^2 \\ 
&\qquad < 4 ( c_1^{B^4} + c_1^{W^4} + c_3^{W^4} + c_1^{B^2 W^2} + c_3^{B^2 W^2})( c_2^{B^4} + c_2^{W^4} + c_4^{W^4} + c_2^{B^2 W^2} + c_4^{B^2 W^2}).
\end{aligned}\ee
The first of these can be converted completely into the aQGC notation, and requires 
\be
4 g_2^4 \left( 2 c_{T,0} + 2 c_{T,1} + c_{T,2} \right) + 2g_1^2 g_2^2 \left( 2 c_{T,5} + 2 c_{T,6} + c_{T,7} \right) + g_1^4 \left( 2 c_{T,8} + c_{T,9} \right) > 0.
\ee
This inequality now provides a constraint on the coefficient of $\mathcal{O}_{T,5}$, although in a more complicated form than the bounds already presented.
More generally, additional superpositions will allow the construction of further bounds than those already considered, but we postpone a full exploration to future work.

Before moving on, let us comment on additional collider searches where our bounds are likely to be relevant.
The discussion here will be brief, as it would be appropriate to have the complete set of superposition bounds before determining the full scope of applicability.
Fundamentally, any such search must be sensitive to the effect of dimension-eight operators, preferably in a channel where the dimension-six effects can be suppressed.
Of course, even if both types of operators are present, it is not true that all UV theories predict that the dimension-six operators dominate.
For example, in certain theories that invoke strong dynamics at the TeV scale, it has been shown that dimension-eight operators can sometimes dominate the corresponding LHC observables~\cite{Liu:2016idz,Liu:2019vid}.
Even in situations where the dimension-six operators are expected to dominate, there are still experimental handles to expose the higher-dimension effects.
In the case of aQGCs, as already mentioned, there are separate channels where the dimension-six operators can be independently constrained.
Another possibility would involve observables where the leading contribution occurs at dimension eight.
Such a possibility is provided by neutral triple gauge boson couplings~\cite{Degrande:2013kka}; future $e^+ e^-$ colliders will be particularly sensitive to a possible $ZZ \gamma$ vertex~\cite{Ellis:2019zex}.

More generically, the contribution from higher-dimension operators is expected to fall less quickly with $s$, and thereby can be isolated by studying the kinematic tails of distributions.
A classic example of this, discussed in Refs.~\cite{Simmons:1989zs,Simmons:1990dh}, is that if the Superconducting Super Collider had been realized, the effect of dimension-eight pure gluon operators of the form $G^4$ could be observed in the high $p_T$ tail of the inclusive differential jet cross section, $pp \to {\rm jet} + X$.
This effect can also manifest at the LHC.
In \Ref{Hays:2018zze}, the authors considered the systematic error induced by neglecting dimension-eight operators when extracting coefficients for the dimension-six corrections to $pp \to h W^+$.
They found that, depending on the measured deviation from the SM for $\sigma (pp \to h W^+)$, the associated error was generally on the order of several percent.
Nevertheless, if the analysis were repeated with an invariant mass cut of $m_{hW} > 500$ GeV, the error increased to as large as 50\%, highlighting the increased relevance of dimension-eight operators in the tails.
This approach can also be pursued to probe SMEFT corrections to the Higgs couplings~\cite{Henning:2018kys}.
Despite these possibilities, there are disadvantages inherent in focusing on the tails, such as the inherent reduction in statistics and further the enhanced PDF uncertainties (for an example see Ref~\cite{Czakon:2017wor}), although this latter uncertainty can potentially be reduced by considering cross section ratios instead.\footnote{We thank Jesse Thaler for this suggestion.}

\subsection{Neutron EDM}

The form in which CP-odd operators enter our results, always being bounded by the associated CP-even contributions, raises the intriguing possibility of connecting potentially disparate experimental measurements.
A comprehensive analysis of this possibility is beyond our present scope, but in this section we demonstrate a proof-of-principle example: the neutron EDM.

The coefficient of the dimension-four operator that would generate a neutron EDM, $G \widetilde{G}$, is famously small, which raises the possibility of a dominant contribution arising from higher-dimension operators.
Indeed, as first observed in \Ref{Weinberg:1989dx}, in extensions to the SM involving multiple Higgs, at low energies the neutron EDM receives a large contribution from the following dimension-six operator from Table~\ref{tab:dim6}:
\be
\widetilde{\cal O}^{G^3}_{\text{dim-6}} = f^{abc} G^{a\nu}_{\mu} G^{b\rho}_{\nu} \widetilde{G}^{c\mu}_{\rho}.
\ee
The low energy SMEFT associated with these models also includes several of the dimension-eight CP-odd operators considered in this work, namely $\widetilde{\mathcal{O}}^{G^4}_{1-3}$.
In \Ref{Chemtob:1992au}, it was shown that in these same multi-Higgs scenarios, it is possible for the dimension-eight operators to actually dominate over the dimension-six in the contribution to the neutron EDM, highlighting this as an observable potentially sensitive to CP-odd operators that are subject to our bounds.
As another example, in \Ref{Chang:1991ry} it was demonstrated that in a theory containing a heavy quark with a chromoelectric dipole moment, the corresponding EFT at low energies generates both $\widetilde{\mathcal{O}}^{G^4}_1$ and $\widetilde{\mathcal{O}}^{G^4}_3$. (It can be verified that in this case the EFT coefficients explicitly satisfy our bounds.)

Recall that the bounds we derived on these operators in \Eq{eq:SU3bounds}, absent any fine tuning, are schematically of the form $(\widetilde{c}^{G^4})^2 < (c^{G^4})^2$; heuristically, the CP-even contribution bounds the CP-odd.
Yet as seen above, experiment should also bound the CP-odd contribution in this case.
To estimate the impact of the dimension-eight operators on the neutron EDM, following \Ref{Weinberg:1989dx}, we can use naive dimensional analysis~\cite{Manohar:1983md,Georgi:1986kr} applied to an operator of the form $(c/M^4) G^3 \widetilde{G}$.
Doing so, we obtain
\bea
|d_n| \sim \frac{e\,c}{M^4} \frac{\Lambda_{\chi {\rm SB}}^3}{(4\pi)^2} \approx c \left( \frac{1~{\rm TeV}}{M} \right)^4 \times 10^{-28}~e~{\rm cm}.
\eea
Here $\Lambda_{\chi {\rm SB}} = 4 \pi F_{\pi} \approx 1190$ MeV is the scale associated with chiral symmetry breaking.
Taking $c \sim \mathcal{O}(1)$ and the current experimental bound of $|d_n| \lesssim 10^{-26}~e~{\rm cm}$~\cite{Baker:2006ts,Afach:2015sja,Graner:2016ses}, we see these measurements are already probing the scale associated with $\widetilde{\mathcal{O}}^{G^4}_{1-3}$ to several hundred GeV.

The hope would then be to compare this estimate to the limits on the associated CP-even contribution.
Unfortunately, bounds on CP-even dimension-eight gluon operators, for example from multijet events at colliders, have not been comprehensively considered in the literature; as such, we cannot explore this possibility here.
We note that there are clear challenges inherent in such an analysis.
Most importantly there are significant uncertainties associated with measurements of multijet cross sections at the LHC (see, for example, Refs.~\cite{Aaboud:2017wsi,Sirunyan:2017skj}), on top of which a small correction from dimension-eight SMEFT operators would be difficult to constrain.
Nevertheless, this example is a proof of principle for how our CP-odd bounds could be utilized.
Another possible consequence of our bounds, in this case, is if a definitive neutron EDM were discovered, one way to test if it had a dimension-eight origin would be to search for the associated CP-even effect, the magnitude of which can be estimated from the same bound.

\section{Conclusions}\label{sec:conclusions}

In this paper, we have placed bounds on operator coefficients in the SMEFT.
For the 64 quartic bosonic operators with four derivatives at mass dimension eight enumerated in \Sec{sec:basis} (see Tables~\ref{tab:operators} and \ref{tab:operators2}), we derived 27 independent bounds on their Wilson coefficients in \Sec{sec:bounds}, which we restate here for convenience:\footnote{We reiterate that in this work we have used the mostly-plus metric convention.
Working instead in the mostly-minus convention, these results are identical except for three sign changes, resulting in $c_1^{H^2 B^2} < 0$, $c_1^{H^2 W^2} < 0$, and $c_1^{H^2 G^2} < 0$.}
\begingroup
\allowdisplaybreaks
\begin{align}
3c_{1}^{G^{4}}+3c_{3}^{G^{4}}+c_{5}^{G^{4}}&>0 \qquad & c_{3}^{B^{2}W^{2}}&>0 \nonumber \\
3c_{3}^{G^{4}}+2c_{5}^{G^{4}}&>0 & c_{4}^{B^{2}W^{2}}&>0 \nonumber \\
3c_{2}^{G^{4}}+3c_{4}^{G^{4}}+c_{6}^{G^{4}}&>0 & c_{3}^{B^{2}G^{2}}&>0 \nonumber\\
3c_{4}^{G^{4}}+2c_{6}^{G^{4}}&>0 & c_{4}^{B^{2}G^{2}}&>0 \nonumber\\
c_{1}^{W^{4}}+c_{3}^{W^{4}}&>0 & c_{3}^{W^{2}G^{2}}&>0 \nonumber\\
c_{2}^{W^{4}}+c_{4}^{W^{4}}&>0 & c_{4}^{W^{2}G^{2}}&>0 \\
c_{1}^{B^4} & >0 & c_1^{H^2 B^2}&>0 \nonumber\\
c_{2}^{B^4} & >0 & c_1^{H^2 W^2}&>0 \nonumber\\
c^{H^4}_{1}+c^{H^4}_{2}+c^{H^4}_{3} & >0 & c_1^{H^2 G^2}&>0 \nonumber\\
c^{H^4}_{1}+c^{H^4}_{2} & >0 \nonumber\\
c^{H^4}_{2} & >0 \nonumber
\end{align}
and 
\be 
\begin{aligned}
(3\widetilde{c}_{1}^{G^{4}}+3\widetilde{c}_{2}^{G^{4}}+\widetilde{c}_{3}^{G^{4}})^{2}&<4(3c_{1}^{G^{4}}+3c_{3}^{G^{4}}+c_{5}^{G^{4}})(3c_{2}^{G^{4}}+3c_{4}^{G^{4}}+c_{6}^{G^{4}}) \\
(3\widetilde{c}_{2}^{G^{4}}+2\widetilde{c}_{3}^{G^{4}})^{2}&<4(3c_{3}^{G^{4}}+2c_{5}^{G^{4}})(3c_{4}^{G^{4}}+2c_{6}^{G^{4}})\\
(\widetilde{c}_{1}^{W^{4}}+\widetilde{c}_{2}^{W^{4}})^{2}&<4(c_{1}^{W^{4}}+c_{3}^{W^{4}})(c_{2}^{W^{4}}+c_{4}^{W^{4}})\\
(\widetilde{c}_{1}^{B^4})^{2} & <4c_{1}^{B^4} c_{2}^{B^4}\\
(\widetilde{c}_{3}^{B^{2}W^{2}})^{2}&<4c_{3}^{B^{2}W^{2}}c_{4}^{B^{2}W^{2}}\\
 (\widetilde{c}_{3}^{B^{2}G^{2}})^{2}&<4c_{3}^{B^{2}G^{2}}c_{4}^{B^{2}G^{2}}\\
 (\widetilde{c}_{3}^{W^{2}G^{2}})^{2}&<4c_{3}^{W^{2}G^{2}}c_{4}^{W^{2}G^{2}}.
\end{aligned}
\ee
As discussed in Secs.~\ref{sec:IR} and \ref{sec:bounds}, these bounds follow from amplitudes and causality arguments analogous to those of \Ref{Adams:2006sv}: analyticity and unitarity of two-to-two scattering amplitudes, along with subluminal signal propagation in the EFT in nontrivial backgrounds.
While IR consistency techniques have been previously used to constrain various EFTs of interest~\cite{Adams:2006sv,Jenkins:2006ia,Dvali:2012zc,Bellazzini:2015cra,Cheung:2016wjt,Camanho:2014apa,Gruzinov:2006ie,Cheung:2016yqr,deRham:2017xox,Camanho:2016opx,Bellazzini:2017fep,Bellazzini:2016xrt,Bonifacio:2018vzv,Bonifacio:2016wcb,deRham:2017zjm,Hinterbichler:2017qyt,deRham:2018qqo,Bellazzini:2019bzh,Nicolis:2009qm,Elvang:2012st,deRham:2017imi,Chandrasekaran:2018qmx,Herrero-Valea:2019hde,Komargodski:2011vj,Cheung:2014ega,Cheung:2018cwt,Cheung:2019cwi,Bellazzini:2019xts,Charles:2019qqt,Zhang:2018shp,Bi:2019phv}, they hitherto have not been systematically applied to the full SMEFT itself.
As a test of our bounds, in \Sec{sec:UV} we exhibited a litany of UV completions of these operators and found that our inequalities were always satisfied.
The IR requirements of analyticity and causality are relatively agnostic of the details of the UV completion and therefore provide a robust means of bounding the coefficients of deformations of the low-energy EFT. They are hence highly relevant to the current situation in particle physics.
Indeed, as discussed in \Sec{sec:pheno}, the bounds derived in this paper will be useful in any search that can be cast in terms of the operators we constrain.
As two explicit examples, we demonstrated the immediate applicability of our results to aQGC searches at the LHC and measurements of the neutron EDM.

The results of this paper suggest many well-motivated directions for future study.
While in the present work we have restricted ourselves to scattering amplitudes for boson states of definite type ($H$, $B$, $W$, or $G$), it would be very interesting to apply the optical theorem to incoming states that are superpositions of different bosons (e.g., a superposition of a Higgs and a gluon), which would result in yet more positivity bounds on the EFT coefficients.
However, this presents a mathematical challenge, since marginalizing over the space of superposition coefficients to find the minimal set of required bounds amounts to determining positivity of a quartic form, a problem known to be analytically strongly NP-hard~\cite{Ahmadi} (see also \Ref{Cheung:2016yqr}).
Beyond this, there is significant scope to explore the experimental consequences of our existing (and potentially superposition-extended) bounds.
In particular, the possibility of finding additional connections between a CP-even and -odd measurement is an especially compelling direction.
Another exciting future direction would be to derive bounds on operators that include fermion fields in the SMEFT.
Finally, we could consider constraints on operator coefficients arising from analyticity and dispersion relations beyond the forward limit~\cite{Nicolis:2009qm}---including the beyond-positivity methods of \Ref{Bellazzini:2017fep} and the EFThedron construction of \Ref{EFThedron}---which could provide yet more bounds on Wilson coefficients.
We leave these interesting questions of generalization and application of our SMEFT bounds to future work. 

\vspace{5mm}
 
\begin{center} 
{\bf Acknowledgments}
\end{center}
\noindent 
We thank Brando Bellazzini, Cliff Cheung, Nathaniel Craig, Markus Luty, Adam Martin, Tom Melia, Christopher Murphy, Francesco Riva, Dean Robinson, David Shih, Yotam Soreq, Dave Sutherland, Jesse Thaler, Cen Zhang, Zhengkang Zhang, and Shuang-Yong Zhou for useful discussions and comments.  
G.N.R. and N.L.R. are supported by the Miller Institute for 
Basic Research in Science at the University of California, Berkeley.
The work of G.N.R. was performed in part at the Aspen Center for Physics, which is supported by National Science Foundation grant PHY-1607611.

\appendix

\section{Group Theory Definitions}\label{app:GeneralN}

In \Sec{sec:oneloop}, we considered completions of the CP-conserving operators in \Tab{tab:operators} by integrating out at one loop a massive state $\Phi$ charged under an arbitrary irreducible representation of the SM gauge group. Accordingly, the Wilson coefficients in \Tab{tab:cs} depend on various group theoretic quantities. In this appendix, we define these quantities in more detail.
 Throughout, we adopt the conventions of \Ref{Quevillon:2018mfl}. 
For an arbitrary irreducible representation ${\bf R}$ of gauge group ${\rm SU}(N)$ for arbitrary $N$, we write $T_{\bf R}$ for the generators of ${\bf R}$ and define the structure constants as
\be 
{}[T_{\bf R}^a,T_{\bf R}^b]{} = i f^{abc} T_{\bf R}^c.
\ee
We can define the quadratic, cubic, and quartic invariants $I_{2,3,4}$ via
\be
\begin{aligned}
\Tr(T_{\bf R}^{(a} T_{\bf R}^{b)}) &= I_2({\bf R})\delta^{ab} \\
\Tr(T_{\bf R}^{(a} T_{\bf R}^b T_{\bf R}^{c)}) &= \frac{1}{4}I_3({\bf R}) d^{abc} \\
\Tr(T_{\bf R}^{(a} T_{\bf R}^b T_{\bf R}^c T_{\bf R}^{d)}) &= I_4({\bf R}) d^{abcd} + \Lambda({\bf R}) (\delta^{ab}\delta^{cd} + \delta^{ac}\delta^{bd} + \delta^{ad}\delta^{bc}).\label{eq:I234}
\end{aligned} 
\ee
The generators are normalized such that $I_2({\bf F})=1/2$, where ${\bf F}$ denotes the fundamental representation, i.e., the representation of dimension $N$, and the $d^{abc}$ are normalized such that $I_3({\bf F}) = 1$.
The quartic symbol $d^{abcd}$ in \Eq{eq:I234}, which vanishes for $N=2,3$, is given by
\be
d^{abcd}=\frac{1}{24}(d^{abe}d^{cde}+d^{ace}d^{bde}+d^{ade}d^{bce})-\frac{(N+2)(N-2)}{12N(N^{2}+1)}(\delta^{ab}\delta^{cd}+\delta^{ac}\delta^{bd}+\delta^{ad}\delta^{bc})
\ee
and the constant $\Lambda({\bf R})$ in \Eq{eq:I234} is defined as
\be 
\Lambda(\mathbf{R})=\left[\frac{D(\mathbf{A})I_{2}(\mathbf{R})}{D(\mathbf{R})}-\frac{I_{2}(\mathbf{A})}{6}\right]\frac{I_{2}(\mathbf{R})}{2+D(\mathbf{A})},\label{eq:Lambdadef}
\ee
where $D({\bf R})$ denotes the dimension of the representation and ${\bf A}$ denotes the adjoint representation, for which $D({\bf A}) = N^2 - 1$ and $I_2({\bf A}) = N$. Thus, $I_4({\bf F}) = 1$ and $\Lambda({\bf F}) = (2N^2 - 3)/12 N(N^2+1)$. The definition in \Eq{eq:Lambdadef}, along with the third line of \Eq{eq:I234}, ensures that $I_4(\mathbf{R}) = 0$ for $N=2,3$.

In order to evaluate these quantities for specific representations, we make use of the fact that an arbitrary irreducible representation ${\bf R}$ of ${\rm SU}(N)$ can be specified by its Dynkin indices $(a_1,a_2,\ldots,a_{N-1})$, an ordered vector of $N-1$ nonnegative integers. The fundamental, antifundamental, and singlet representations are written as $(1,0,\ldots,0)$, $(0,\ldots,0,1)$, and $(0,\ldots,0)$, respectively.
The dimension of the representation is \cite{Dietrich:2006cm}
\be
\begin{aligned}
D(\mathbf{R}) &= \prod_{p=1}^{N-1}\left\{\frac{1}{p!}\prod_{q=p}^{N-1}\left[\sum_{r=q-p+1}^p (1+a_r) \right] \right\}
\\& =\left(1+a_{1}\right)\left(1+a_{2}\right)\cdots\left(1+a_{N-1}\right)\times\\
 & \qquad\times\left(1+\frac{a_{1}+a_{2}}{2}\right)\left(1+\frac{a_{2}+a_{3}}{2}\right)\cdots\left(1+\frac{a_{N-2}+a_{N-1}}{2}\right)\times\\
 & \qquad\times\left(1+\frac{a_{1}+a_{2}+a_{3}}{3}\right)\left(1+\frac{a_{2}+a_{3}+a_{4}}{3}\right)\cdots\left(1+\frac{a_{N-3}+a_{N-2}+a_{N-1}}{3}\right)\times\\
 &\hspace{10.3mm} \vdots \\
 & \qquad\times\left(1+\frac{a_{1}+\cdots+a_{N-1}}{N-1}\right).
\end{aligned}
\ee
The quadratic invariant is
\be
I_2({\bf R}) = \frac{D({\bf R})C_2({\bf R})}{2(N+1)N(N-1)}  ,
\ee
where $C_2({\bf R})$ is the quadratic Casimir, defined via $2N T_{\bf R}^a T_{\bf R}^a = C_2({\bf R}) \mathds{1}$, or explicitly \cite{Dietrich:2006cm},
\be
C_2({\bf R}) = \sum_{m=1}^{N-1}\left[N(N-m)ma_{m}+m(N-m)a_{m}^{2}+\sum_{n=0}^{n-1}2n(N-m)a_{n}a_{m}\right]. \label{eq:C2}
\ee
One can show that $I_2(\mathbf{R}) \geq D(\mathbf{R})/2N$.
The cubic anomaly invariant is
\be 
I_3({\bf R}) = D({\bf R}) \sum_{p = 1}^{N-1} \sum_{q=1}^{N-1} \sum_{r=1}^{N-1} b_{pqr}(1+a_p)(1+a_q)(1+a_r),
\ee
where for $i\leq j\leq k$,
\be
b_{pqr} = \frac{2(N-3)!}{(N+2)!} p(N-2q)(N-r),
\ee
and all the other entries are determined by requiring that $b_{pqr}$ be totally symmetric~\cite{Banks:1976yg}.
Finally, the quartic coefficient $I_4(\bf{R})$ for general $N$ can be found in \Ref{Okubo:1981td}; since it vanishes for the SM gauge group, it does not impact the operators we consider.

\bibliographystyle{utphys-modified}
\bibliography{Standard_model_EFT}

\end{document}